\documentclass[review]{elsarticle}

\makeatletter
\def\ps@pprintTitle{%
    \let\@oddhead\@empty
    \let\@evenhead\@empty
    \def\@oddfoot{\reset@font\hfil\thepage\hfil}
    \let\@evenfoot\@oddfoot
}
\makeatother

\usepackage{graphicx}

\usepackage[a4paper]{geometry}

\usepackage{amsmath, amssymb, amsthm}
\usepackage{latexsym}
\usepackage{siunitx}

\usepackage{placeins}
\usepackage{tikz, tikz-uml}
\usetikzlibrary{decorations.pathreplacing}
\usetikzlibrary{shapes.misc}
\usetikzlibrary{positioning}

\definecolor{typecolor}{RGB}{0, 120, 255}
\definecolor{background}{gray}{0.95}
\definecolor{line}{gray}{0.7}

\tikzumlset{
    fill class = background,
    draw = line,
    font=\footnotesize\ttfamily
}

\usepackage{algorithm, algpseudocode}

\makeatletter
\algrenewcommand\ALG@beginalgorithmic{\footnotesize}
\makeatother

\usepackage{booktabs}

\usepackage{multirow}

\usepackage{subcaption}

\usepackage{url}
\usepackage{xcolor}
\definecolor{newcolor}{rgb}{.8,.349,.1}

\newcommand{\xvec}{\mathbf{x}}
\newcommand{\pvec}{\mathbf{p}}
\newcommand{\evec}{\mathbf{e}}
\newcommand{\vvec}{\mathbf{v}}
\newcommand{\Evec}{\mathbf{E}}
\newcommand{\Bvec}{\mathbf{B}}

\newcommand{\Figref}[1]{Fig.~\ref{#1}}
\newcommand{\Tabref}[1]{Tab.~\ref{#1}}
\newcommand{\Eqref}[1]{Eq.~\eqref{#1}}
\newcommand{\Algref}[1]{Alg.~\ref{#1}}
\newcommand{\Secref}[1]{Sec.~\ref{#1}}

\DeclareMathOperator*{\argmax}{arg\,max}

\begin{document}
    \newcommand{\opal}{\textsc{OPAL}}
\newcommand{\opalamr}{\textsc{OPAL-AMR}}
\newcommand{\ippl}{\textsc{IPPL}}
\newcommand{\amrmg}{\textsc{AMR-MG}}
\newcommand{\yt}{yt}

\newcommand{\boxlib}{\textsc{BoxLib}}
\newcommand{\amrex}{\textsc{AMReX}}
\newcommand{\trilinos}{\textsc{Trilinos}}
\newcommand{\muelu}{\textit{MueLu}}
\newcommand{\amesos}{\textit{Amesos2}}
\newcommand{\belos}{\textit{Belos}}
\newcommand{\ifpack}{\textit{Ifpack2}}
\newcommand{\tpetra}{\textit{Tpetra}}
\newcommand{\kokkos}{\textit{Kokkos}}

\begin{frontmatter}
    \title{On Architecture and Performance of Adaptive Mesh Refinement in an Electrostatics Particle-In-Cell Code}
    
    \author[]{Matthias Frey\corref{cor}} 
    \ead{matthias.frey@psi.ch}
    
    \author[]{Andreas Adelmann}
    \ead{andreas.adelmann@psi.ch}
    
    \author[]{Uldis Locans}

    \address{Paul Scherrer Institut, CH-5232 Villigen, Switzerland} 
    
    \cortext[cor]{Corresponding author}
    
    \begin{abstract}
    This article presents a hardware architecture independent implementation of an adaptive mesh refinement
    Poisson solver that is integrated into the electrostatic Particle-In-Cell beam dynamics code
    \opal{}.\ The Poisson solver is solely based on second generation \trilinos{} packages to
    ensure the desired hardware portability.\ Based on the massively parallel framework \amrex, formerly known as \boxlib,
    the new adaptive mesh refinement interface provides several refinement policies in order to enable precise large-scale neighbouring
    bunch simulations in high intensity cyclotrons.\ The solver is validated with a built-in multigrid solver
    of \amrex{} and a test problem with analytical solution.\ The parallel
    scalability is presented as well as an example of a neighbouring bunch simulation that covers the
    scale of the later anticipated physics simulation.\
    \end{abstract}
    
    
\end{frontmatter}

    \section{Introduction}
    In todays state-of-the-art beam dynamics codes the well-known Particle-In-Cell (PIC) \cite{Hockney:1988:CSU:62815} 
    technique has become indispensable.\ In contrast to the direct summation, where the force on a macro particle
    is obtained by the superposition of the forces due to all others, PIC models discretise a domain and deposit the charge of
    each macro particle onto a mesh in order to evaluate Coulomb's repulsion.\ In combination with the efficient parallelisation
    of such space-charge solvers using MPI (Message Passing Interface) or
    accelerators such as GPU (Graphics Processing Unit) and the MIC (Many Integrated Core) architecture, e.g.\ in 
    \cite{ADELMANN201683}, large-scale 
    simulations were enabled that are more realistic.\ Nevertheless,
    multi-bunch simulations of high intensity accelerators such as  cyclotrons require fine meshes in order to resolve the non-linear 
    effects in the evolution of the beams due to space-charge.\ A remedy to increase the resolution, reduce the computational effort 
    and also memory consumption is adaptive mesh refinement (AMR) \cite{BERGER1984484, BERGER198964}.\ In the context of 
    Vlasov-Poisson problems, AMR was applied by 
    \cite{HITTINGER2013118} using the Eulerian description for the coordinate and velocity space.\ Examples for a Lagrangian 
    formulation are the Unified Flow Solver (UFS) framework \cite{Kolobov} and WarpX \cite{VAY2018}.\\
    The diversity of today's computer architectures and the fast increase of emerging high performance computing technologies have 
    shown that
    it is getting more and more infeasible to design a scientific software to one specific hardware only.\ It is therefore obvious 
    that recent 
    source code developments reveal a trend towards architecture independent programming where the backend kernels exhibit the 
    hardware-specific implementation.\ An example are the second generation \trilinos{} packages that are built on top of the 
    \kokkos{} 
    library \cite{CarterEdwards20143202, Kokkos}.\\
    In this article the new AMR capability of the particle accelerator library \opal{} (Object-Oriented Particle Accelerator
    Library) \cite{opal:1, 2019arXiv190506654A} using \amrex{} \cite{AMReX} is 
    presented, as well as the built-in adaptive multigrid solver based on the algorithm in \cite{MartinPhdThesis} and the second 
    generation \trilinos{} packages \tpetra{} \cite{Tpetra}, \amesos{} and \belos{} \cite{AmesosBelos}, \muelu{}
    \cite{MueLu, MueLuURL} 
    and \ifpack{} \cite{Ifpack2}.\ The new implementation was benchmarked with the Poisson multigrid solver of 
    \amrex{} and the analytical example of a uniformly charged sphere.\\
    The new AMR feature of \opal{} will enable to study neighbouring bunch effects as they occur in high intensity cyclotrons due to
    the low turn separation in more detail.\ Previous investigations such as \cite{PhysRevSTAB.13.064201} for the PSI
    (Paul Scherrer Institut) Ring 
    cyclotron have already shown their existence but the PIC model was limited in resolution due to the high memory needs.\ It is
    hoped that the use of AMR will reduce the memory consumption for the mesh by decreasing the resolution in regions of void 
    while maintaining or even increasing the grid point density at locations of interest in order to resolve the neighbouring
    bunch interactions more precisely.\ In \cite{PhysRevSTAB.13.064201} was shown that the interaction of neighbouring bunches leads 
    to an increase at the tails of a particle distribution (i.e.\ increase of the number of halo particles) that usually causes
    particle losses and therefore an activation of the machine.\ Thus, it is essential to quantify this effect more precisely in order 
    to do predictions on further machine developments with higher beam current.\\
    Beside a short introduction to \opal{} in section \ref{sec:OPAL} and \amrex{} in section \ref{sec:AMReX}, section
    \ref{sec:interface} discusses the AMR interface in \opal.\ Section \ref{sec:AGM} explains the multigrid algorithm and
    its implementation using \trilinos{} with validation in section \ref{sec:benchmark}.\ A comparison of neighbouring 
    bunch simulations with either AMR turned on or off is shown in section \ref{sec:nbs}.\ The performance of the Poisson solver
    is discussed in section \ref{sec:scaling}.\ In the last section are conclusions and outlook.

    \section{The \opal{} Library}
    \label{sec:OPAL}
    The Object-Oriented Parallel Accelerator Library (\opal) \cite{opal:1, 2019arXiv190506654A} is an electrostatic PIC (ES-PIC) beam dynamics code for large-scale 
    particle accelerator simulations.\ Due to the general design its application ranges from high intensity cyclotrons to low 
    intensity proton therapy beamlines \cite{PhysRevAccelBeams.20.124702} with negligible space-charge.\ Beside the default FFT
    (Fast Fourier Transform) 
    Poisson solver for periodic and open boundary problems the built-in SAAMG (Smoothed Aggregation Algebraic Multigrid) solver 
    enables to simulate accelerators with arbitrary geometries \cite{ADELMANN20104554}.\ The time integration relies on the second 
    order Leapfrog, the fourth order Runge-Kutta (RK-4) or a multiple stepping Boris-Buneman method \cite{TOGGWEILER2014255}.\
    
    In beam dynamics the evolution of the density function $f(\xvec, \pvec, t)$ in time $t$ of the charged particle distribution in 
    phase space $(\xvec, \pvec)\in\mathbb{R}^{6}$ due to electromagnetic fields $\Evec(\xvec, t)$ and $\Bvec(\xvec, t)$ is described 
    by the Vlasov (or collisionless Boltzmann) equation
    \begin{equation}
    \label{eq:Vlasov}
    \frac{df(\xvec, \pvec, t)}{dt} = \gamma m_0\frac{\partial f}{\partial t}
                                   + \pvec\cdot\nabla_\xvec f
                                   + \frac{q}{\gamma m_0^2}\left(\gamma m_0\Evec(\xvec, t) + \pvec \times \Bvec(\xvec, t)\right) \cdot \nabla_\pvec f
                                   = 0,
    \end{equation}
    with particle charge $q$ and rest mass $m_0$.\ The relativistic momentum
    $\pvec = \gamma m_0\vvec$ with Lorentz factor $\gamma$ and particle velocity $\vvec$ is used together with the coordinate $\xvec$ 
    to specify the state of a particle in 
    the 6D phase space.\ Both, the electric and magnetic field, in \Eqref{eq:Vlasov} are a sum of an external and internal,
    i.e.\ space-charge, contribution
    \begin{align*}
        \Evec(\xvec, t) &= \Evec_{sc}(\xvec, t) + \Evec_{ext}(\xvec, t), \\
        \Bvec(\xvec, t) &= \Bvec_{sc}(\xvec, t) + \Bvec_{ext}(\xvec, t).
    \end{align*}
    The external fields are given by RF-cavities and by the magnetic field of the machine.\ In order to evaluate the
    electric self-field the beam is Lorentz transformed into its rest frame where the magnetic field induced by the motion of the 
    particles is negligible.\ Thus, the electric self-field is fully described by the electrostatic potential $\phi(\xvec, t)$, i.e.\
    \begin{equation*}
    \Evec_{sc}(\xvec, t) = -\nabla\phi(\xvec, t)
    \end{equation*}
    that is computed by Poisson's equation
    \begin{equation*}
    \Delta\phi(\xvec, t)= -\frac{\rho(\xvec, t)}{\varepsilon_0},
    \end{equation*}
    with charge density $\rho$ and vacuum permittivity $\varepsilon_0$.\ The magnetic self-field is afterwards restored by the inverse 
    Lorentz transform.\ This quasi-static approximation is known as Vlasov-Poisson equation.\
    
    \section{The \amrex{} Library}
    \label{sec:AMReX}
    The \amrex{} library \cite{AMReX} is a descendant of the parallel block-structured adaptive mesh refinement code named \boxlib.\  
    It is C++ based
    with an optional Fortran90 interface.\ Each level is distributed independently among MPI-processes in order to ensure load 
    balancing.\ The owned data is located either at nodes, faces, edges or centres of cells where the latter description is used in 
    the \opalamr{} implementation.\\
    In order to generate a level $l+1$ each cell of the underlying coarser level $l$ has to be marked to get refined or 
    not according to a user-defined criterion.\ In electrostatic problems natural choices are for example the charge density,
    the potential strength or the electric field (cf.\ \Secref{sec:amr_policies}).\ Subsequent AMR levels satisfy the relation
    \begin{equation}
    h_w^{l+1} = \frac{h_w^{l}}{r_w}\quad\forall w\in [x, y, z],
    \label{eq:mesh_spacing}
    \end{equation}
    where $r_w\in\mathbb{N}\setminus\{0\}$ is called the refinement ratio and $h_w^l$ specifies the mesh spacing of level $l$
    in direction of $w$.\ A sketch of a refined mesh is given in \Figref{fig:amr}.\ By definition, the coarsest level $(l=0)$ 
    covers the full domain $\Omega = \Omega^0$ whereas a fine level is defined by patches that may overlap several coarser grids.\ In 
    general, for a level $l>0$ with $n$ grids $g_i$ following holds
    \begin{align*}
    \Omega^{l} &= \left(\bigcup_{i=0}^{n-1} g_{i}^{l}\right) \subset \Omega^{l-1}, \\
     g_i^{l} \cap g_{j}^{l} &= \emptyset \quad \forall i, j\in\{0, 1, ..., n-1\} \mbox{ and } i\ne j.
    \end{align*}
    Although neighbouring grids aren't allowed to overlap they exchange data at interfaces via ghost cells.\ 
    
    \begin{figure}[!ht]
        \centering
        \begin{tikzpicture}[scale=1.0]
            \draw[step=5mm, gray!60!white] (0, 0) grid (6, 6);
            \draw[step=2.5mm, gray!60!white] (1, 2) grid (3.5, 5);
            \draw[step=2.5mm, gray!60!white] (3.5, 0.5) grid (6, 4);
            
            \draw[step=1.25mm, gray!60!white] (1.5, 2.5) grid (5, 3.5);
            
            \draw[step=1.25mm, gray!60!white] (4, 1) grid (5.5, 2);
            
            \draw[step=1.25mm, gray!60!white] (1.25, 4) grid (3, 4.75);
            
            \draw[very thick] (1, 2) -- (1, 5) -- (3.5, 5) -- (3.5, 4);
            \draw[very thick] (1, 2) -- (3.5, 2);
            
            \draw[very thick] (3.5, 0.5) -- (3.5, 2.5);
            \draw[very thick] (3.5, 3.5) -- (3.5, 4) -- (6, 4) -- (6, 0.5) -- (3.5, 0.5);
            
            \draw[very thick] (1.5, 2.5) -- (1.5, 3.5) -- (5, 3.5) -- (5, 2.5) -- (1.5, 2.5);
            
            \draw[very thick] (0, 0) rectangle (6, 6);
            
            \draw[very thick] (4, 1) rectangle (5.5, 2);
            
            \draw[very thick] (1.25, 4) rectangle (3, 4.75);
            
            \node[red] at (5.75, 5.75) {${\bf \Omega_0}$};
            \node[red] at (5.75, 3.75) {${\bf \Omega_1}$};
            \node[red] at (4.75, 3.25) {${\bf \Omega_2}$};
            
            \node[red] at (5.25, 1.75) {${\bf \Omega_2}$};
            
            \node[red] at (2.75, 4.5) {${\bf \Omega_2}$};
            
        \end{tikzpicture}
        \caption{Sketch of a block-structured mesh refinement of a Cartesian grid $\Omega_0$ in 2D with \amrex.\ Fine levels denoted
        by $\Omega_1$ and $\Omega_2$ may span
        multiple coarser grids as indicated.\ At interfaces among grids of same level ghost cells allow exchanging data.}
        \label{fig:amr}
    \end{figure}
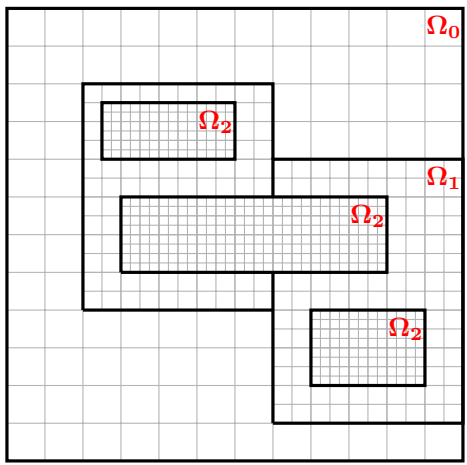\FloatBarrier\parindent 0pt
    
    \section{Adaptive Mesh Refinement in the \opal{} Library}
    \label{sec:interface}
    In order to allow AMR and uniform mesh PIC algorithms, the interface in \opal{} is implemented in a lightweight fashion where
    an AMR library is used as a black box.\ The AMR functionality is provided by concrete implementations of the abstract 
    base class that defines common requirements on AMR libraries such as refinement strategies and mesh update functions.\ The
    actual AMR implementation is therefore hidden allowing multiple AMR dependencies.\
    
    In AMR mode the allocation of work among MPI-processes is controlled by \amrex.\ In contrast to \opal{} where load balancing
    is optimised w.r.t.\ the macro particles, \amrex{} aims to achieve a uniform workload of grid operations.\ These two 
    parallelisation paradigms are contradictory and cause
    additional MPI-communication for every PIC operation if both, grids and particles, are kept evenly distributed among the
    MPI-processes.\ In order to reduce communication effort at the expense of possible particle load imbalances the developed AMR
    interface distributes the particles according to their grids.\ For this purpose a new particle layout manager is
    created that stores further AMR specific attributes, i.e.\ the level and the grid a particle lives on.\
    
    A peculiarity of the PIC model in \opal{} is the adjustment of the grid $\Omega_0$ (cf.\ \Figref{fig:amr}) to the particle
    bunch.\ The mesh that is co-moving
    with the macro particles adapts dynamically to the dimension of the bunch in rest frame, keeping the number
    of grid points per dimension constant, with the consequence of a constantly changing grid spacing.\ In longitudinal 
    direction, i.e.\ the direction of travel, this change includes the correction of relativistic length contraction in laboratory
    frame.\ In AMR mode instead the macro particles are mapped to a fixed domain since the problem geometry has to be predefined in 
    \amrex.\ This linear transformation includes the Lorentz transform of the particles.\ Adaptive mesh refinement, particle 
    partitioning and the calculation of the electrostatic potential (cf.\ 
    \Secref{sec:domain_transform}) are carried out there.\
    
    Spurious self-forces on particles close by coarse-fine 
    grid interfaces that occur in AMR due to image charges are corrected by buffer cells as described in 
    \cite{1749-4699-5-1-014019}.\ Another solution as depicted in \cite{COLELLA2010947} would be the modification of the charge 
    deposition algorithm using a convolution of Green's function for particles near a refinement boundary.
    
    \subsection{Domain Transform}
    \label{sec:domain_transform}
    In order to prevent particles leaving the predefined domain of the mesh where the AMR hierarchy is built, they are mapped into
    a  computation space denoted by $\mathcal{S}_{c}$ for the evaluation of Poisson's equation, the 
    repartition of the particles to MPI-processes and the mesh refinement.\ Therefore, the geometry can be 
    kept at $\delta\mathcal{S}_{c}$ where $\delta$ specifies a constant box increment in percent to increase the margin of the 
    mesh.\ In the co-moving frame the natural choice of the computation space is $\mathcal{S}_{c} = [-1, 1]^3$
    since the bunch is located around the design trajectory with the reference particle at $(x, y, z) = (0, 0, 0)$.\ In order to
    consider an inhomogeneous problem domain, the box dimension of $\mathcal{S}_{c}$ can be adjusted by the user at the beginning.\ 
    After
    solving Poisson's equation the electrostatic potential and the electric field have to be rescaled properly.\ Instead of
    rescaling the fields at the location of the particles, it is directly done on the grid as depicted in
    \Figref{fig:sc_workflow}.\ The mapping of the particle coordinates in co-moving space $\mathcal{S}_{p}$ to
    computation space $\mathcal{S}_{c}$ includes also the Lorentz transform.\
    
    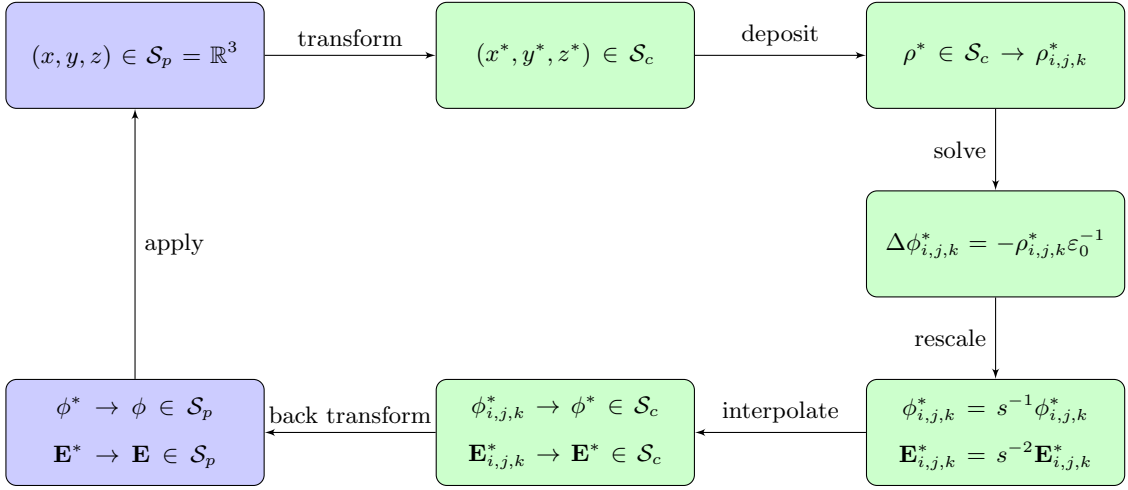
\begin{figure}[!ht]
        \centering
        \tikzstyle{Sp_space} = [rectangle, draw, fill=blue!20,
                             text width=9em, text centered, rounded corners, minimum height=4em]
        \tikzstyle{Sc_space} = [rectangle, draw, fill=green!20,
                                text width=9em, text centered, rounded corners, minimum height=4em]
        \tikzstyle{line} = [draw, -latex']
        \begin{tikzpicture}[node distance=2.5cm, auto]
            \node [Sp_space] (Sp) {\small $(x,y,z)\in\mathcal{S}_{p}=\mathbb{R}^3$};
            
            \node [Sc_space, right=2.25 of Sp] (Sc) {\small $(x^*,y^*,z^*)\in\mathcal{S}_{c}$};
            
            \node [Sc_space, right=2.25cm of Sc] (deposit) {\small $\rho^*\in\mathcal{S}_{c}\rightarrow\rho_{i, j, k}^*$};
            
            \node [Sc_space, below of=deposit] (solve) {\small $\Delta\phi_{i,j,k}^*=-\rho_{i, j, k}^*\varepsilon_0^{-1}$};

            \node [Sc_space, below of=solve] (scale) {\small $\phi_{i,j,k}^* = s^{-1}\phi_{i,j,k}^*$ \\
                                                           $\mathbf{E}_{i,j,k}^* = s^{-2}\mathbf{E}_{i,j,k}^*$};
            
            \node [Sc_space, left=2.25cm of scale] (interp) {\small $\phi_{i,j,k}^*\rightarrow\phi^*\in\mathcal{S}_{c}$ \\
                                                            $\mathbf{E}_{i,j,k}^*\rightarrow\mathbf{E}^*\in\mathcal{S}_{c}$};
            
            \node [Sp_space, left=2.25cm of interp] (bt) {\small $\phi^*\rightarrow\phi\in\mathcal{S}_{p}$ \\
                                                          $\mathbf{E}^*\rightarrow\mathbf{E}\in\mathcal{S}_{p}$};

            \path [line] (Sp) -- node[above]{\small transform} (Sc);
            \path [line] (Sc) -- node[above]{\small deposit} (deposit);
            \path [line] (deposit) -- node[left]{\small solve} (solve);
            \path [line] (solve) -- node[left]{\small rescale} (scale);
            \path [line] (scale) -- node[above]{\small interpolate} (interp);
            \path [line] (interp) -- node[above]{\small back transform} (bt);
            \path [line] (bt) -- node[right]{\small apply} (Sp);
        \end{tikzpicture}
        \caption{Workflow of the space-charge calculation.\ Poisson's equation is solved in the computation domain
        and rescaled afterwards.\ All steps in particle space $\mathcal{S}_{p}$ and computation space
        $\mathcal{S}_{c}$ are marked in blue and green, respectively.\ The mapping of the particle coordinates in space
        $\mathcal{S}_{p}$ to $\mathcal{S}_{c}$ involves also the Lorentz transform.}
        \label{fig:sc_workflow}
    \end{figure}
    
    \subsubsection{Particle Coordinate}
    Let $\mathbf{x} = (x_0, x_1, x_2)\in\mathcal{S}_p$ be a coordinate of some particle in the particle space $\mathcal{S}_p$ and
    let $\mathbf{l} = (l_0, l_1, l_2) > \mathbf{0}$, then we define
    \begin{equation*}
        \Gamma(\mathbf{x}, \mathbf{l}) := \max_{i=\{1, 2, 3\}}\left|\frac{x_i}{l_i}\right|.
    \end{equation*}
    The transform of an individual particle at position $\mathbf{x}\in\mathcal{S}_{p}$ into computation space
    $\mathbf{x}^*\in\mathcal{S}_{c} = [-l_0, l_0]\times [-l_1, l_1]\times [-l_2, l_2]$ is therefore given by
    \begin{equation*}
        \mathbf{x}^* = \frac{\mathbf{x}}{s}\quad\mbox{with}\quad
        s = \argmax_{\mathbf{x}\in\mathcal{S}_{p}}\sum_{i=0}^{N-1}\Gamma(\mathbf{x}_i, \mathbf{l}),
    \end{equation*}
    where $N$ is the number of particles.
    
    \subsubsection{Electrostatic Potential}
    Let $\phi\in\mathcal{S}_p$ be the electrostatic potential in particle space $\mathcal{S}_p$ and $\phi^*\in\mathcal{S}_c$ the
    corresponding potential value in computation space $\mathcal{S}_c$, then they relate as
    \begin{equation}
        \phi = \frac{1}{s}\phi^*.
        \label{eq:potential_transform}
        \end{equation}
    \begin{proof}
        Let the discrete charge density of $N$ particles be described by \cite[eq.\ 1.6]{Jackson99}
        \begin{equation*}
            \rho(\xvec) = \sum_{i=1}^{N}q_i\delta(\xvec - \xvec_i) \quad\quad \xvec\in\mathbb{R}^{d},
        \end{equation*}
        in $d$ dimensions and the coordinates being transformed as denoted above then
        \begin{equation*}
        \rho = s^{-d}\rho^{*}
        \end{equation*}
        with $s>0$ and
        \begin{equation*}
        \frac{\partial}{\partial w} = \frac{\partial w^{*}}{\partial w}\frac{\partial}{\partial w^{*}}
                                    = s^{-1}\frac{\partial}{\partial w^{*}}
        \end{equation*}
        where $w = x_1, x_2, ..., x_d$.\ Thus,
        \begin{align*}
            \Delta\phi          &= -\frac{\rho}{\epsilon_0} \\
            s^{-2}\Delta^{*}\phi   &= -s^{-d}\frac{1}{\varepsilon_0}\rho^{*} \\
            s^{-2}\Delta^{*}\phi   &= s^{-d}\Delta^{*}\phi^{*} \\
                         \phi   &= s^{2-d}\phi^{*}.
        \end{align*}
        Therefore, the potential transforms in $3$ dimensions as denoted in \Eqref{eq:potential_transform}.\ In 2 dimensions the
        electrostatic potential remains.
    \end{proof}
    
    \subsubsection{Electric  Field}
    Let $\mathbf{E}\in\mathcal{S}_p$ be the electric field in particle space $\mathcal{S}_p$ and
    $\mathbf{E}^*\in\mathcal{S}_c$ the corresponding electric field vector in computation space $\mathcal{S}_c$, then they relate as
    \begin{equation}
        \mathbf{E} = \frac{1}{s^2}\mathbf{E}^*.
        \label{eq:efield_transform}
    \end{equation}
    \begin{proof}
    According to Gauss' law the electric field is the derivative of the electrostatic potential.\ Thus, an additional $s^{-1}$
    contributes to the transformation, therefore,
    \begin{equation*}
        \Evec = s^{1-d}\Evec^{*}
    \end{equation*}
    that coincides with \eqref{eq:efield_transform} in $3$ dimensions.
    \end{proof}
    
    \subsection{Adaptive Mesh Refinement Policies}
    \label{sec:amr_policies}
    Beside the regrid function each AMR module implements the charge deposition, the particle-to-core
    (re-)distribution and various refinement strategies.\ There are currently six refinement policies available.\
    Most refinement strategies are directly connected to particle properties since it is desirable to increase the spatial
    resolution at their location.\
    Natural choices of refinement criteria are the charge density per cell, the electrostatic potential and the electric field.\
    They are explained in more detail below.\ Other methods limit the minimum or maximum number of particles within a cell.\
    The last tagging option refines cells based on the momentum of particles.\ While the first three methods refine the mesh
    based on the grid data, the latter methods use particle information directly.\ All methods apply a user-defined
    threshold $\lambda$ in order to control the mesh refinement.\ This threshold 
    denotes either the minimum charge density per cell
    \begin{equation}
        |\rho^{l}_{i,j,k}|\ge \lambda,
        \label{eq:tag_charge}
    \end{equation}
    or a scale factor $\lambda\in[0, 1]$ in order
    to refine every grid cell $(i,j,k)$ on a level $l$ that satisfies
    \begin{equation*}
        |\phi^{l}_{i,j,k}|\ge \lambda \max_{i,j,k} |\phi^{l}|
    \end{equation*}
    or 
    \begin{equation*}
        |E^{l}_{w; i,j,k}| \ge \lambda\max_{i,j,k}|E_{w}^{l}|,
    \end{equation*}
    in case of the electrostatic potential $\phi$ or the electric field components $E_{w}$ with $w\in\{x, y, z\}$, respectively.\
    The charge
    density in \Eqref{eq:tag_charge} is scaled in order to account for the domain transformation as previously mentioned and
    explained in detail in \Secref{sec:domain_transform}.\
    Examples of AMR based on the charge density, potential and electric field with various thresholds are shown in 
    \Figref{fig:tag_charge}, \Figref{fig:tag_pot} and \Figref{fig:tag_ef}, respectively.\
    
    \begin{figure}[htp]
        \centering
        \includegraphics[width=0.49\textwidth]{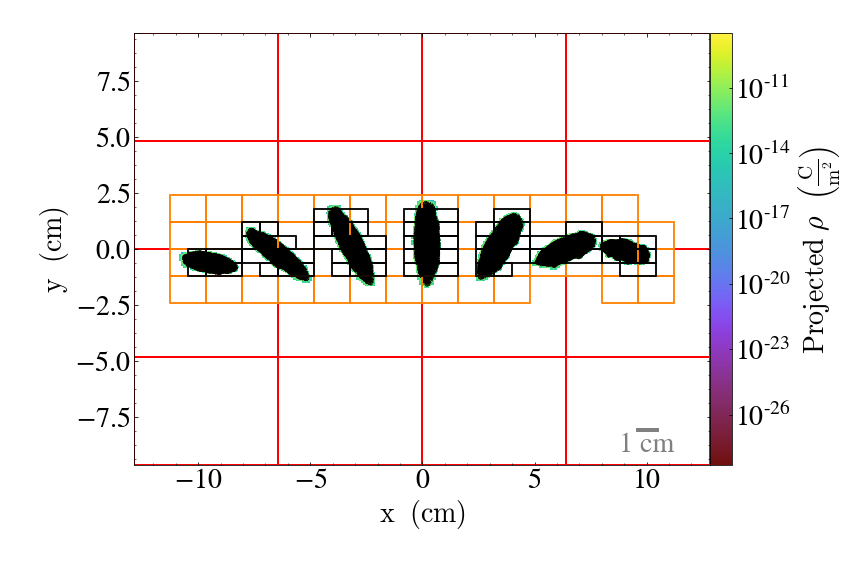}
        \hfill
        \includegraphics[width=0.49\textwidth]{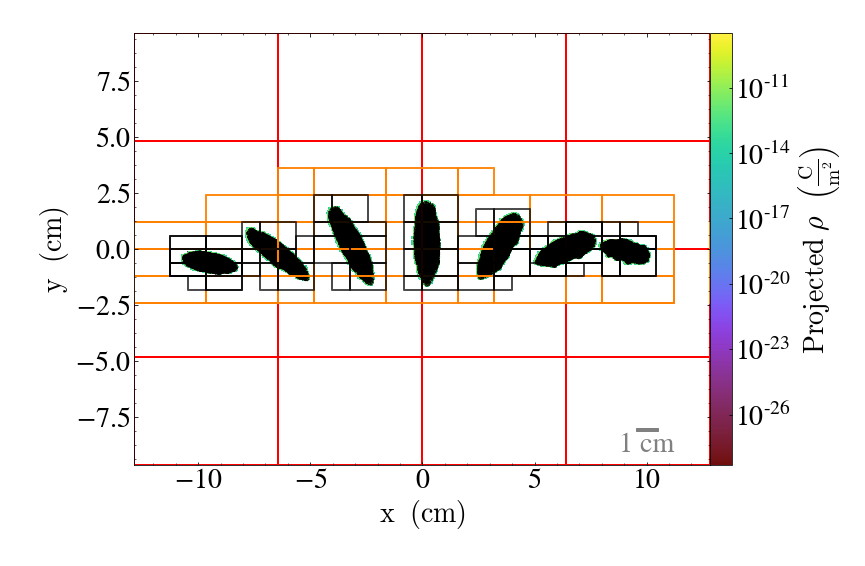}
        \hfill
        \includegraphics[width=0.49\textwidth]{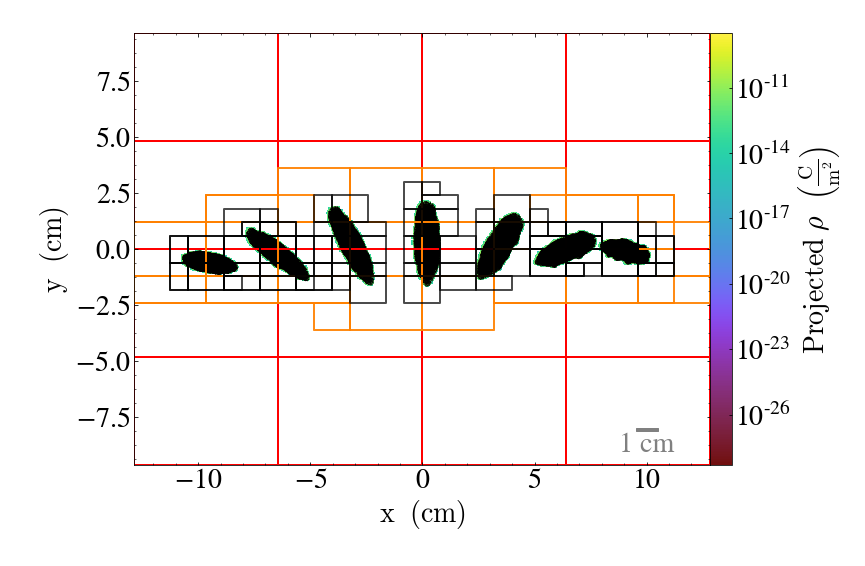}
        \hfill
        \includegraphics[width=0.49\textwidth]{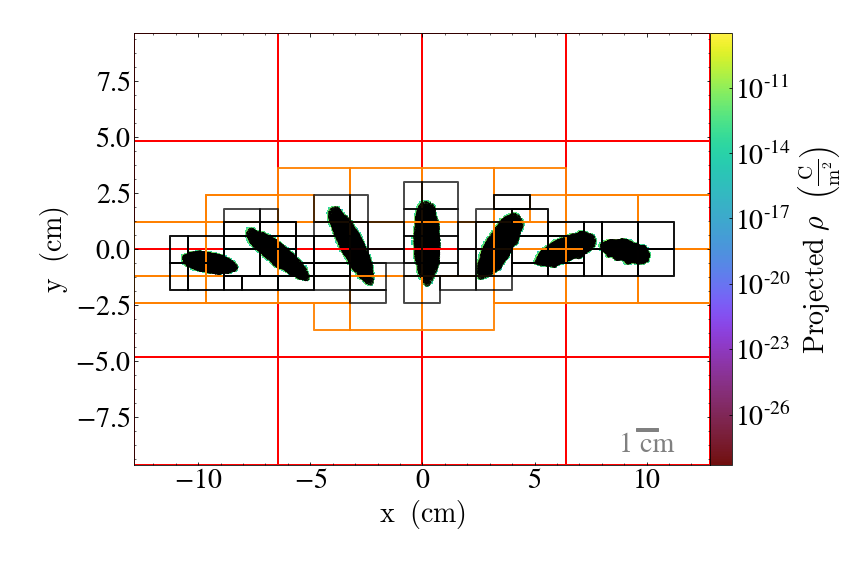}
        \caption{Integrated projection of the charge denstiy onto the $xy$-plane showing 7 adjacent particle bunches.\
        Adaptive mesh refinement with charge density threshold $\SI{1e-6}{C/m^3}$ (top left),
        $\SI{1e-7}{C/m^3}$ (top right), $\SI{1e-8}{C/m^3}$ (bottom left), $\SI{1e-9}{C/m^3}$ (bottom right).\ Plotted
        with an own extension of the \yt{} package \cite{2011ApJS..192....9T}.}
        \label{fig:tag_charge}
    \end{figure}
    
    \begin{figure}[htp]
        \centering
        \includegraphics[width=0.49\textwidth]{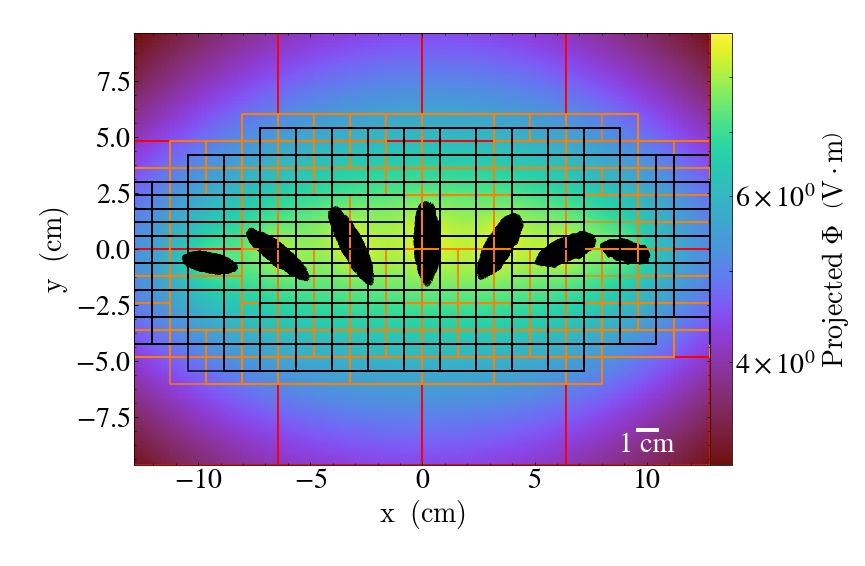}\hfill
        \includegraphics[width=0.49\textwidth]{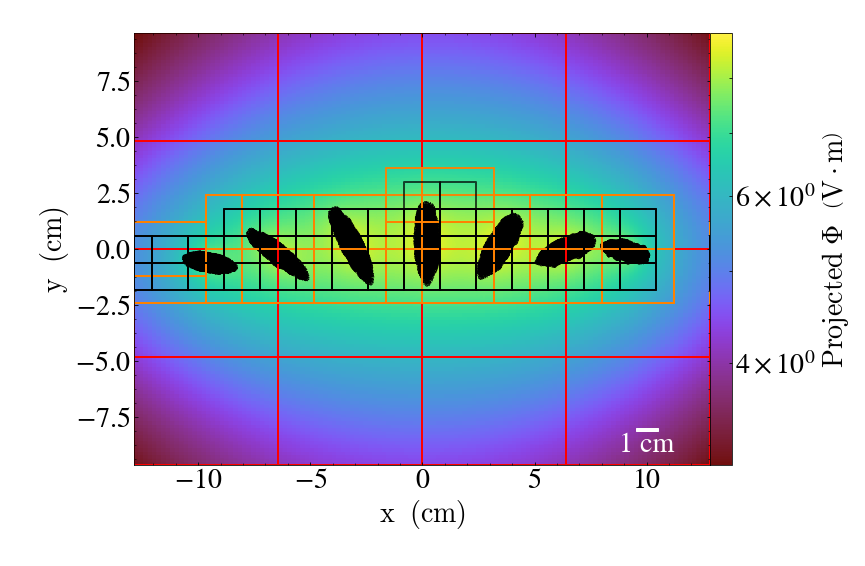}\hfill
        \includegraphics[width=0.49\textwidth]{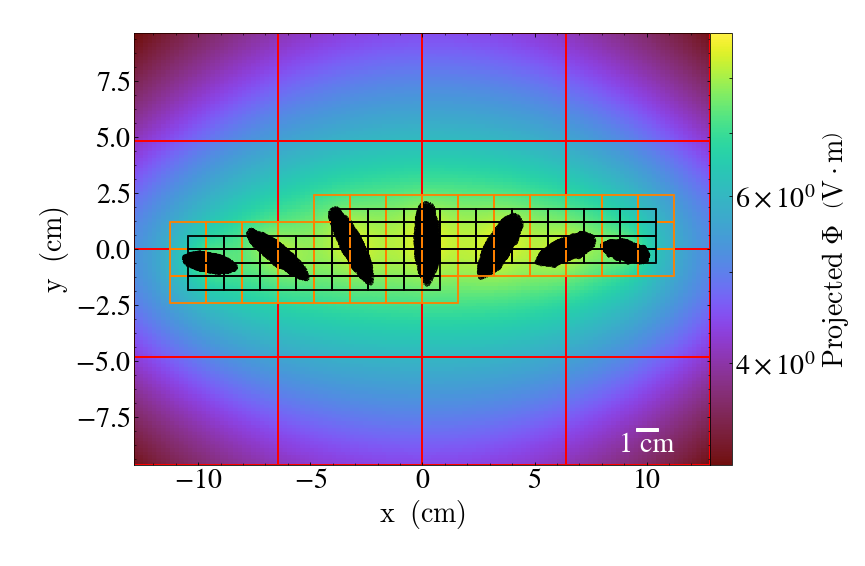}\hfill
        \includegraphics[width=0.49\textwidth]{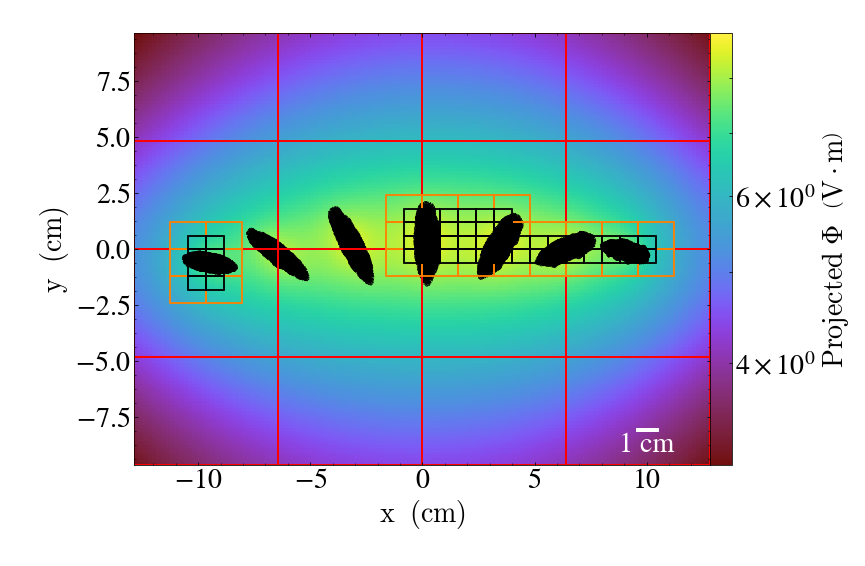}
        \caption{Integrated projection of the electrostatic potential onto the $xy$-plane showing 7 adjacent particle bunches.\
        Adaptive mesh refinement based on the electrostatic potential with
        thresholds $\lambda$ from left to right and top to bottom: $0.25$, $0.5$, $0.75$ and $0.95$.\ Plotted
        with an own extension of the \yt{} package \cite{2011ApJS..192....9T}.}
        \label{fig:tag_pot}
    \end{figure}
    
    \begin{figure}[htp]
        \centering
        \includegraphics[width=0.49\textwidth]{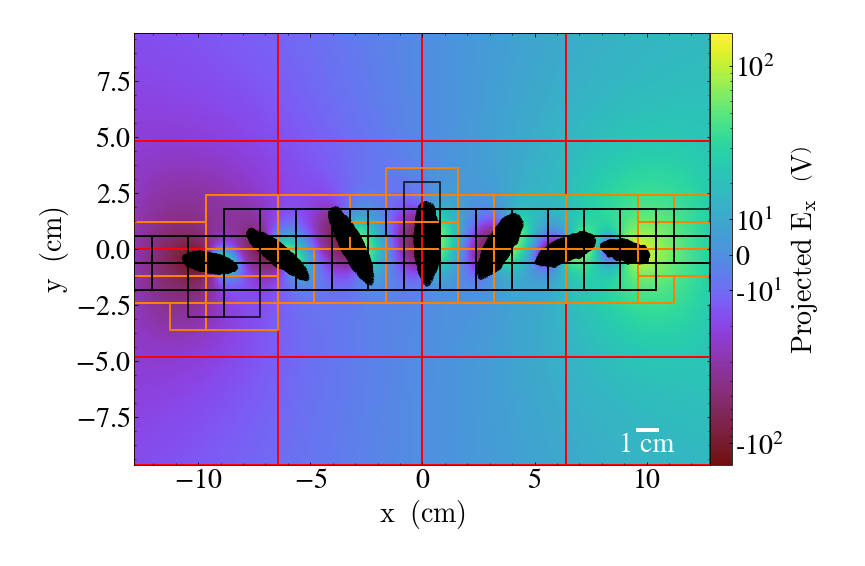}\hfill
        \includegraphics[width=0.49\textwidth]{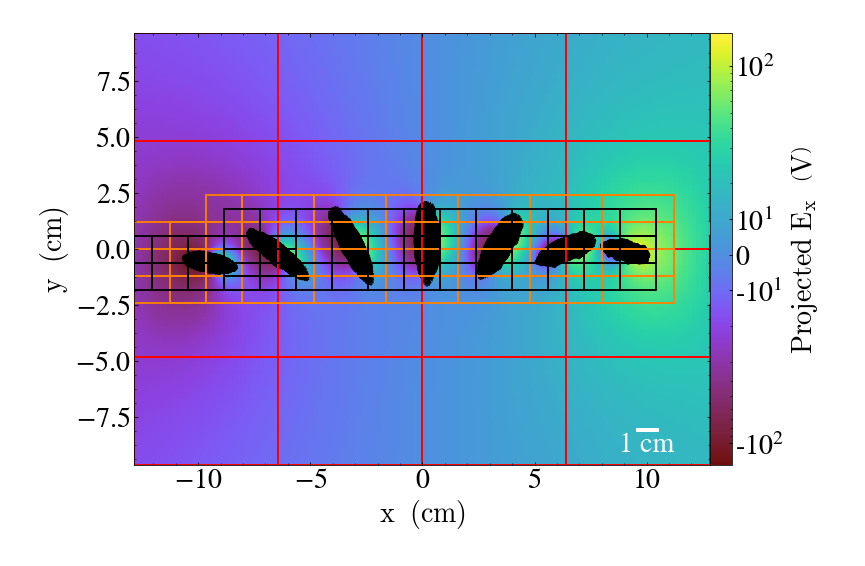}\hfill
        \includegraphics[width=0.49\textwidth]{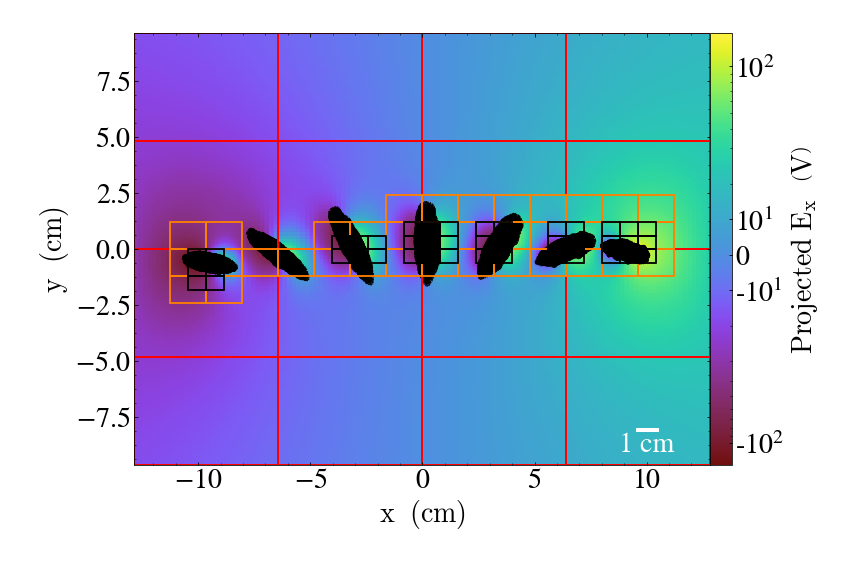}\hfill
        \includegraphics[width=0.49\textwidth]{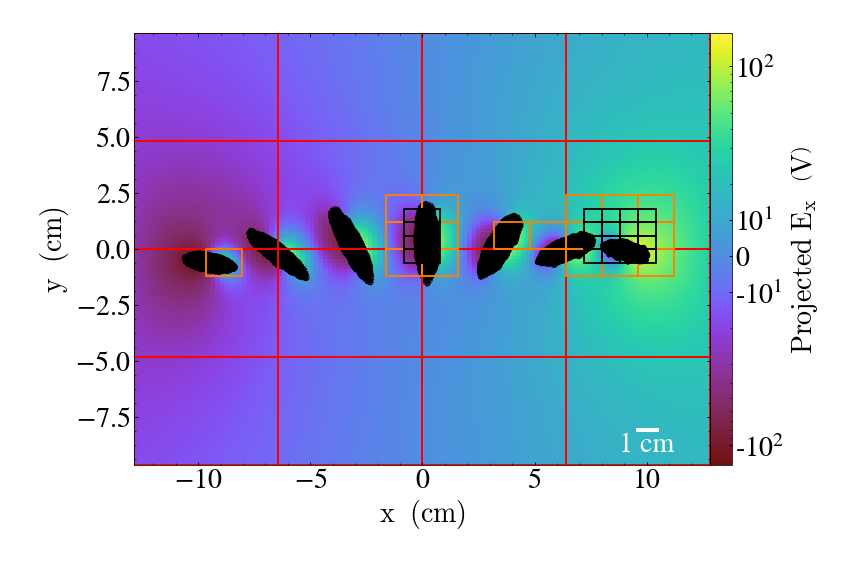}
        \caption{Integrated projection of the electric field component $E_x$ onto the $xy$-plane showing 7 adjacent particle bunches.\
        Adaptive mesh refinement based on the electric field components with
        thresholds $\lambda$ from left to right and top to bottom: $0.25$, $0.5$, $0.75$ and $0.95$.\ Plotted
        with an own extension of the \yt{} package \cite{2011ApJS..192....9T}.}
        \label{fig:tag_ef}
    \end{figure}
    
    \section{Adaptive Geometric Multigrid}
    \label{sec:AGM}
    This section describes the algorithm of the adaptive geometric multigrid (AGMG) according to \cite{Martin96, MartinPhdThesis}
    and its implementation with the second generation packages of \trilinos, i.e.\ \tpetra{} \cite{Tpetra}, \amesos{} and \belos{}
    \cite{AmesosBelos}, \muelu{} \cite{MueLu, MueLuURL} and \ifpack{} \cite{Ifpack2}.\
    A cell-centred implementation is also presented in \cite{ALMGREN19981}.\ In opposite to previous implementations
    the one presented here is hardware independent thanks to the aforementioned \trilinos{} packages that have the \kokkos{}
    \cite{CarterEdwards20143202, Kokkos} library as backend.\ Another benefit is the convenient exchange of kernels such as
    smoothers (e.g.\ Gauss-Seidel or Jacobi) provided by \ifpack{} or linear solvers of \belos, \amesos{} and \muelu.\ The sparse 
    matrices and vectors are instances of \tpetra{} classes.\
    
    \subsection{Coarse-Fine Interface}
    AGMG is a special variant of the classical geometric multigrid since not all levels cover the full domain $\Omega = \Omega^0$
    (cf.\ \Figref{fig:amr}).\
    At interfaces between
    subsequent levels $\partial\Omega^{l,l+1}$ the \textit{elliptic matching condition} (i.e.\ Neumann and Dirichlet boundary 
    condition) must be satisfied in order to ensure continuity of the solution.\ This condition is met by \textit{flux differencing}
    \begin{equation}
    \mathcal{L}^{l}\phi(\xvec)           = \sum_{d=1}^{3}
                                           \frac{f(\xvec+\frac{1}{2}h_{d}^{l}\evec_d) - f(\xvec-\frac{1}{2}h_{d}^{l}\evec_d)}{h_d^l}
                                         + \mathcal{O}\left((h_d^l)^2\right),
    \label{eq:lap_flux_differencing}
    \end{equation}
    with mesh spacing $h_d$ an unit vector $\evec_d$ where either
    \begin{equation}
        \begin{aligned}
        f(\xvec+\frac{1}{2}h_{d}^{l}\evec_d) &= \frac{\phi(\xvec+h_{d}^{l}\evec_d) - \phi(\xvec)}{h_d^l}, \\
        f(\xvec-\frac{1}{2}h_{d}^{l}\evec_d) &= \frac{\phi(\xvec) - \phi(\xvec-h_{d}^{l}\evec_d)}{h_d^l}
        \end{aligned}
        \label{eq:coarse_fluxes}
    \end{equation}
    on $\Omega^{l}$ or 
    \begin{equation}
        \begin{aligned}
        f(\xvec+\frac{1}{2}h_{d}^{l}\evec_d) &= \sum_{i,j\in\{\pm\frac{1}{4}\}}
                                                \frac{\phi(\xvec+\frac{3}{4}h_{d}^{l}\evec_d + ih_{d^+}^l\evec_{d^+}
                                              + jh_{d^-}^l\evec_{d^-})
                                              - \phi(\xvec+\frac{1}{4}h_{d}^{l}\evec_d + ih_{d^+}^l\evec_{d^+}
                                              + jh_{d^-}^l\evec_{d^-})}{4h_d^{l+1}}, \\
        f(\xvec-\frac{1}{2}h_{d}^{l}\evec_d) &= \sum_{i,j\in\{\pm\frac{1}{4}\}}
                                                \frac{\phi(\xvec-\frac{1}{4}h_{d}^{l}\evec_d + ih_{d^+}^l\evec_{d^+}
                                              + jh_{d^-}^l\evec_{d^-})
                                              - \phi(\xvec-\frac{3}{4}h_{d}^{l}\evec_d + ih_{d^+}^l\evec_{d^+}
                                              + jh_{d^-}^l\evec_{d^-})}{4h_d^{l+1}}
        \end{aligned}
        \label{eq:flux_differencing}
    \end{equation}
    with $d^+, d^-\in\{1, 2, 3\}\setminus\{d\}$ and $d^+ \ne d^-$ at the interface $\partial\Omega^{l,l+1}$, i.e.\ the average flux
    across the boundary where a mesh refinement ratio (cf.\ \Eqref{eq:mesh_spacing}) of $r_d = 2\ \forall d\in\{1, 2, 3\}$ is
    assumed.\ In case
    of a cell without adjacent finer cells the flux differencing 
    reduces to the usual second order Laplacian discretisation
    \begin{equation}
    \mathcal{L}^{l}\phi(\xvec) = \sum_{d=1}^{3}\frac{\phi(\xvec+h_d^l\evec_d) - 2\phi(\xvec) + \phi(\xvec - h_d^l\evec_d)}{(h_d^l)^2}
                               + \mathcal{O}\left((h_d^l)^2\right).
    \end{equation}
    An illustration of the stencil of \Eqref{eq:lap_flux_differencing} with fluxes computed either by \Eqref{eq:coarse_fluxes} or
    \Eqref{eq:flux_differencing} is shown in \Figref{fig:flux_differencing}.\ In order to simplify the representation 
    the example is in 2D with only one coarse-fine interface on the left side.\ Hence, the corresponding finite difference stencil is 
    given by
    \begin{equation*}
            \mathcal{L}^{l}\phi(\xvec) = \frac{f(\xvec + \frac{1}{2}h_{x}^{l}\evec_{x})
                                             - f(\xvec - \frac{1}{2}h_{x}^{l}\evec_{x})}{h_{x}^{l}}
                                       + \frac{f(\xvec + \frac{1}{2}h_{y}^{l}\evec_{y})
                                             - f(\xvec - \frac{1}{2}h_{y}^{l}\evec_{y})}{h_{y}^{l}},
    \end{equation*}
    where
    \begin{equation*}
        \begin{aligned}
            f(\xvec + \frac{1}{2}h_{x}^{l}\evec_{x}) &= \frac{\phi(\xvec + h_{x}^{l}\evec_{x}) - \phi(\xvec)}{h_{x}^{l}}, \\
            f(\xvec - \frac{1}{2}h_{x}^{l}\evec_{x}) &= \frac{1}{2h_{x}^{l+1}}\left(\phi_{high}^{ghost}-\phi_{high} +
                                                                                    \phi_{low}^{ghost}-\phi_{low}\right), \\
            f(\xvec + \frac{1}{2}h_{y}^{l}\evec_{y}) &= \frac{\phi(\xvec + h_{y}^{l}\evec_{y}) - \phi(\xvec)}{h_{y}^{l}}, \\
            f(\xvec - \frac{1}{2}h_{y}^{l}\evec_{y}) &= \frac{\phi(\xvec) - \phi(\xvec - h_{y}^{l}\evec_{y})}{h_{y}^{l}}.
        \end{aligned}
    \end{equation*}
    
    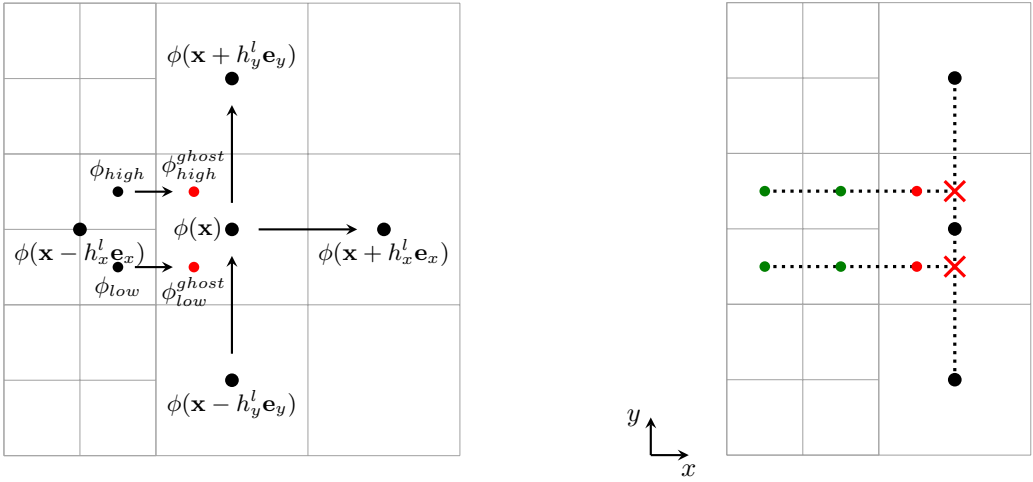
\begin{figure}[!ht]
        \centering
        \begin{subfigure}[t]{0.45\textwidth}
            \begin{tikzpicture}[>=stealth]
                    \draw[step=20mm, gray!70!white] (0, 0) grid (6, 6);
                    \draw[step=10mm, gray!70!white] (0, 0) grid (2, 6);
                    
                    \draw[fill=black, draw=none] (3, 3) circle (0.09) node[left]{\small $\phi(\xvec)$};
                    
                    \draw[fill=black, draw=none] (1, 3) circle (0.09) node[below]{\small $\phi(\xvec - h_{x}^{l}\evec_{x})$};
                    \draw[fill=black, draw=none] (5, 3) circle (0.09) node[below]{\small $\phi(\xvec + h_{x}^{l}\evec_{x})$};
                    
                    \draw[fill=black, draw=none] (3, 5) circle (0.09) node[above]{\small $\phi(\xvec + h_{y}^{l}\evec_{y})$};
                    \draw[fill=black, draw=none] (3, 1) circle (0.09) node[below]{\small $\phi(\xvec - h_{y}^{l}\evec_{y})$};
                        
                    \draw[fill=red, draw=none] (2.5, 3.5) circle (0.07) node[above]{\small $\phi_{high}^{ghost}$};
                    \draw[fill=red, draw=none] (2.5, 2.5) circle (0.07) node[below]{\small $\phi_{low}^{ghost}$};
                
                    \draw[fill=black, draw=none] (1.5, 3.5) circle (0.07) node[above]{\small $\phi_{high}$};
                    \draw[fill=black, draw=none] (1.5, 2.5) circle (0.07) node[below]{\small $\phi_{low}$};
                    
                    \draw[thick, ->, shorten <=10pt, shorten >=10pt] (3, 3) -- (3, 5);
                    \draw[thick, ->, shorten <=10pt, shorten >=10pt] (3, 3) -- (5, 3);
                    \draw[thick, ->, shorten <=10pt, shorten >=10pt] (3, 1) -- (3, 3);
                    
                    \draw[thick, ->, shorten <=15pt, shorten >=15pt] (2.25, 3.5) -- (2.75, 3.5);
                    \draw[thick, ->, shorten <=15pt, shorten >=15pt] (2.25, 2.5) -- (2.75, 2.5);

                    \draw[] (3.5, 0) node[below] {$\textcolor{white}{x}$};
                    \draw[] (4, 0.5) node[left] {$\textcolor{white}{y}$};
            \end{tikzpicture}
            \caption{The red nodes indicate ghost cells that need to be interpolated.}
        \end{subfigure}
        \hfill
        \begin{subfigure}[t]{0.45\textwidth}
            \begin{tikzpicture}[>=stealth]
                    \tikzset{cross/.style={cross out, draw=black, minimum size=2*(#1-\pgflinewidth), inner sep=0pt, outer sep=0pt},
                    cross/.default={5pt}}
                    
                    \draw[thick, ->] (-1, 0) -- (-0.5, 0) node[below] {$x$};
                    \draw[thick, ->] (-1, 0) -- (-1, 0.5) node[left] {$y$};

                    \draw[step=20mm, gray!70!white] (0, 0) grid (4, 6);
                    \draw[step=10mm, gray!70!white] (0, 0) grid (2, 6);
                    
                    \draw[very thick, dotted] (3, 1) -- (3, 5);
                    \draw[very thick, dotted] (0.5, 3.5) -- (3, 3.5);
                    \draw[very thick, dotted] (0.5, 2.5) -- (3, 2.5);
                    
                    \draw[very thick] (3, 3.5) node[cross,red] {};
                    \draw[very thick] (3, 2.5) node[cross,red] {};
                    
                    \draw[fill=black, draw=none] (3, 3) circle (0.09) node[right]{};
                    
                    \draw[fill=black, draw=none] (3, 5) circle (0.09) node[above]{};
                    \draw[fill=black, draw=none] (3, 1) circle (0.09) node[below]{};
                        
                    \draw[fill=red, draw=none] (2.5, 3.5) circle (0.07) node[above]{};
                    \draw[fill=red, draw=none] (2.5, 2.5) circle (0.07) node[above]{};
                
                    \draw[fill=green!50!black, draw=none] (1.5, 3.5) circle (0.07) node[above]{};
                    \draw[fill=green!50!black, draw=none] (0.5, 3.5) circle (0.07) node[above]{};

                    \draw[fill=green!50!black, draw=none] (1.5, 2.5) circle (0.07) node[above]{};
                    \draw[fill=green!50!black, draw=none] (0.5, 2.5) circle (0.07) node[above]{};
            \end{tikzpicture}
            \caption{The red crosses specify the intermediate interpolation points using coarse cells.}
            \label{fig:flux_differencing_b}
        \end{subfigure}
        \caption{Illustration of flux differencing in 2D at a coarse-fine interface on the left side.\ In 2D the coarse-fine
        interface is 1D.}
        \label{fig:flux_differencing}
    \end{figure}
    In 3D ghost cells are expressed in terms of valid coarse and fine cells where a two-step second order Lagrange interpolation in 2D
    \begin{equation}
        \phi^{interpolated}(u, v) = \sum_{i, j = 0}^{2}L_{i}(u)L_{j}(v)\phi(u_i, v_j)
        \label{eq:2d_lagrange}
    \end{equation}
    with
    \begin{equation*}
        L_i(x) = \frac{(x-x_k)(x-x_l)}{(x_i-x_k)(x_i-x_l)}\quad\quad (l\ne i\ne k\ne l)
    \end{equation*}
    is performed.\ In 2D this corresponds to 1D Lagrange interpolations.\ First, the intermediate points symbolised as
    red crosses in \Figref{fig:flux_differencing_b} are computed with \Eqref{eq:2d_lagrange} where 
    only non-covered coarse cells parallel to the interface are taken.\
    Second, the fine cells normal to the boundary are used together with the intermediate locations to obtain the ghost cells
    with \Eqref{eq:2d_lagrange}.\
    
    In 3D the interface is surface perpendicular to the current coarse-fine boundary.\ Depending on the
    surrounding cells this surface distinguishes nine configurations to evaluate the 2D quadratic Lagrange interpolation as shown in 
    \Figref{fig:Lagrange_configurations}.\ The current location of the interface is denoted by the black dot.\
    According to \Eqref{eq:2d_lagrange} nine non-refined coarse cells are required for second order interpolation denoted by the
    cells highlighted in red.\ For this purpose a surface consisting of 25 cells is checked perpendicular to the coarse-fine
    interface of interest.\ Ideally none of the surrounding coarse cells is refined such that the interpolation pattern shown in 
    \Figref{subfig:case0}
    is applied.\ The cases in \Figref{subfig:case1} to \Figref{subfig:case4} indicate a mesh refinement on a single side of this 
    surface perpendicular to the coarse-fine interface.\ In case fine cells form a corner one of the patterns 
    \Figref{subfig:case5} to \Figref{subfig:case8} is appropriate.\ The selection of the interpolation pattern follows a list ordered
    according to \Figref{fig:Lagrange_configurations}, i.e.\ from left to right and top to bottom.\ In order to simplify the 
    evaluation of the interpolation scheme an integer value is assigned to each configuration obtained by its
    representation as a bit pattern (see \Tabref{tab:bit_pattern}).\ For this purpose all 25 cells are given a number denoting
    the position of the bits.\ A bit is flipped to one if the corresponding cell is not covered by fine cells.\
    In case none of the nine patterns is applicable the interpolation order is reduced and thus one of 
    the four first order Lagrange interpolation configurations of \Figref{fig:Lagrange_2d_linear} is taken instead.\ The 
    implementation follows exactly the same scheme with conversion shown in \Tabref{tab:Lagrange_2d_linear}.
    \begin{figure}[!ht]
        \centering
        \begin{tikzpicture}[scale=0.8, >=stealth]
            \newcommand{\cell}[3]{
                \fill[#3, opacity=0.6] (#1, #2) -- +(0, .5) -- +(.5, .5) -- +(.5,0) -- cycle;
            }
            
            \colorlet{cred}{red!70!white}
            \colorlet{cblue}{blue!50!white}
        
            \begin{scope}
            \draw[step=5mm, gray] (0, 0) grid (2.5, 2.5);
            \cell{1}{1}{cred}
            \cell{1}{1.5}{cred}
            \cell{1}{0.5}{cred}
            \cell{0.5}{1}{cred}
            \cell{1.5}{1}{cred}
            
            \cell{1.5}{1.5}{cred}
            \cell{1.5}{0.5}{cred}
            \cell{0.5}{1.5}{cred}
            \cell{0.5}{0.5}{cred}
            
            \draw[fill=black, draw=none] (1.25, 1.25) circle (0.04) node[]{};
            
            \node[] at (1.2, -0.5) {\parbox{0.3\linewidth}{\subcaption{case 0}\label{subfig:case0}}};
            
            \end{scope}
            
            \begin{scope}[xshift=3cm]
            \draw[step=5mm, gray] (0, 0) grid (2.5, 2.5);
            \draw[step=2.5mm, gray] (0, 1.5) grid (2.5, 2.5);
            
            \cell{1}{0.5}{cred}
            \cell{1}{1.}{cred}
            \cell{1}{0}{cred}
            \cell{1.5}{1}{cred}
            \cell{0.5}{1}{cred}
            
            \cell{1.5}{0}{cred}
            \cell{1.5}{0.5}{cred}
            \cell{0.5}{0.5}{cred}
            \cell{0.5}{0}{cred}
            
            \draw[fill=black, draw=none] (1.25, 1.25) circle (0.04) node[]{};
            
            \node[] at (1.2, -0.5) {\parbox{0.3\linewidth}{\subcaption{case 1}\label{subfig:case1}}};
            
            \end{scope}
            
            \begin{scope}[xshift=6cm]
            \draw[step=5mm, gray] (0, 0) grid (2.5, 2.5);
            \draw[step=2.5mm, gray] (1.5, 0) grid (2.5, 2.5);
            
            \cell{0.5}{1}{cred}
            \cell{1}{0.5}{cred}
            \cell{1}{1.5}{cred}
            \cell{1}{1}{cred}
            \cell{0}{1}{cred}
            
            \cell{0.5}{1.5}{cred}
            \cell{0}{0.5}{cred}
            \cell{0.5}{0.5}{cred}
            \cell{0}{1.5}{cred}
            
            \draw[fill=black, draw=none] (1.25, 1.25) circle (0.04) node[]{};
            
            \node[] at (1.2, -0.5) {\parbox{0.3\linewidth}{\subcaption{case 2}\label{subfig:case2}}};
            
            \end{scope}
            
            \begin{scope}[xshift=9cm]
            \draw[step=5mm, gray] (0, 0) grid (2.5, 2.5);
            \draw[step=2.5mm, gray] (0, 0) grid (2.5, 1);
            
            \cell{1}{1.5}{cred}
            \cell{1}{2}{cred}
            \cell{1}{1}{cred}
            \cell{1.5}{1}{cred}
            \cell{0.5}{1}{cred}
            
            \cell{1.5}{2}{cred}
            \cell{1.5}{1.5}{cred}
            \cell{0.5}{1.5}{cred}
            \cell{0.5}{2}{cred}
            
            \draw[fill=black, draw=none] (1.25, 1.25) circle (0.04) node[]{};
            
            \node[] at (1.2, -0.5) {\parbox{0.3\linewidth}{\subcaption{case 3}\label{subfig:case3}}};
            
            \end{scope}
            
            \begin{scope}[xshift=12cm]
            \draw[step=5mm, gray] (0, 0) grid (2.5, 2.5);
            \draw[step=2.5mm, gray] (0, 0) grid (1, 2.5);
            
            \cell{1.5}{1}{cred}
            \cell{1}{0.5}{cred}
            \cell{1}{1.5}{cred}
            \cell{1}{1}{cred}
            \cell{2}{1}{cred}
            
            \cell{1.5}{1.5}{cred}
            \cell{2}{0.5}{cred}
            \cell{1.5}{0.5}{cred}
            \cell{2}{1.5}{cred}
            
            \draw[fill=black, draw=none] (1.25, 1.25) circle (0.04) node[]{};
            
            \node[] at (1.2, -0.5) {\parbox{0.3\linewidth}{\subcaption{case 4}\label{subfig:case4}}};
            
            \end{scope}
            
            \begin{scope}[yshift=-3.5cm]
            \draw[step=5mm, gray] (0, 0) grid (2.5, 2.5);
            \draw[step=2.5mm, gray] (0, 0) grid (1, 1.5);
            \draw[step=2.5mm, gray] (0, 1.5) grid (2.5, 2.5);
            
            \cell{1}{0.5}{cred}
            \cell{1}{1}{cred}
            \cell{1}{0}{cred}
            \cell{1.5}{1}{cred}
            \cell{2}{1}{cred}
            
            \cell{1.5}{0}{cred}
            \cell{1.5}{0.5}{cred}
            \cell{2}{0.5}{cred}
            \cell{2}{0}{cred}
            
            \draw[fill=black, draw=none] (1.25, 1.25) circle (0.04) node[]{};
            
            \node[] at (1.2, -0.5) {\parbox{0.3\linewidth}{\subcaption{case 5}\label{subfig:case5}}};
            
            \end{scope}
            
            \begin{scope}[xshift=3cm, yshift=-3.5cm]
            \draw[step=5mm, gray] (0, 0) grid (2.5, 2.5);
            \draw[step=2.5mm, gray] (1.5, 0) grid (2.5, 1.5);
            \draw[step=2.5mm, gray] (0, 1.5) grid (2.5, 2.5);
            
            \cell{1}{0.5}{cred}
            \cell{1}{1}{cred}
            \cell{1}{0}{cred}
            \cell{0.5}{1}{cred}
            \cell{0}{1}{cred}
            
            \cell{0.5}{0}{cred}
            \cell{0.5}{0.5}{cred}
            \cell{0}{0.5}{cred}
            \cell{0}{0}{cred}
            
            \draw[fill=black, draw=none] (1.25, 1.25) circle (0.04) node[]{};
            
            \node[] at (1.2, -0.5) {\parbox{0.3\linewidth}{\subcaption{case 6}\label{subfig:case6}}};
            
            \end{scope}
            
            \begin{scope}[xshift=6cm, yshift=-3.5cm]
            \draw[step=5mm, gray] (0, 0) grid (2.5, 2.5);
            
            \draw[step=2.5mm, gray] (1.5, 0) grid (2.5, 2.5);
            \draw[step=2.5mm, gray] (0, 0) grid (1.5, 1);
            
            \cell{0.5}{1}{cred}
            \cell{1}{2}{cred}
            \cell{1}{1.5}{cred}
            \cell{1}{1}{cred}
            \cell{0}{1}{cred}
            
            \cell{0.5}{1.5}{cred}
            \cell{0}{2}{cred}
            \cell{0.5}{2}{cred}
            \cell{0}{1.5}{cred}
            
            \draw[fill=black, draw=none] (1.25, 1.25) circle (0.04) node[]{};
            
            \node[] at (1.2, -0.5) {\parbox{0.3\linewidth}{\subcaption{case 7}\label{subfig:case7}}};
            
            \end{scope}
            
            \begin{scope}[xshift=9cm, yshift=-3.5cm]
            \draw[step=5mm, gray] (0, 0) grid (2.5, 2.5);
            
            \draw[step=2.5mm, gray] (0, 1) grid (1, 2.5);
            \draw[step=2.5mm, gray] (0, 0) grid (2.5, 1);
            
            \cell{1.5}{1}{cred}
            \cell{1}{2}{cred}
            \cell{1}{1.5}{cred}
            \cell{1}{1}{cred}
            \cell{2}{1}{cred}
            
            \cell{1.5}{1.5}{cred}
            \cell{2}{2}{cred}
            \cell{1.5}{2}{cred}
            \cell{2}{1.5}{cred}
            
            \draw[fill=black, draw=none] (1.25, 1.25) circle (0.04) node[]{};
            
            \node[] at (1.2, -0.5) {\parbox{0.3\linewidth}{\subcaption{case 8}\label{subfig:case8}}};
            
            \end{scope}

            \begin{scope}[xshift=12cm, yshift=-3.5cm]
                \draw[thick, ->] (0.25, 0.5) -- (1.75, 0.5) node[below right] {$x$};
                \draw[thick, ->] (0.25, 0.5) -- (0.25, 2) node[above left] {$y$};
                \draw[thick, ->] (0.25, 0.5) -- (0.75, 1) node[above right] {$z$};
            \end{scope}
        \end{tikzpicture}
        \caption{All possible configurations for 2D quadratic Lagrange interpolation where the red cells are used for the 
        interpolation.\ The coarse-fine interface is perpendicular to the shown cell layer (i.e.\ in $z$-direction).\
        The black dot indicates the cell at the current coarse-fine-interface.}
        \label{fig:Lagrange_configurations}
    \end{figure}
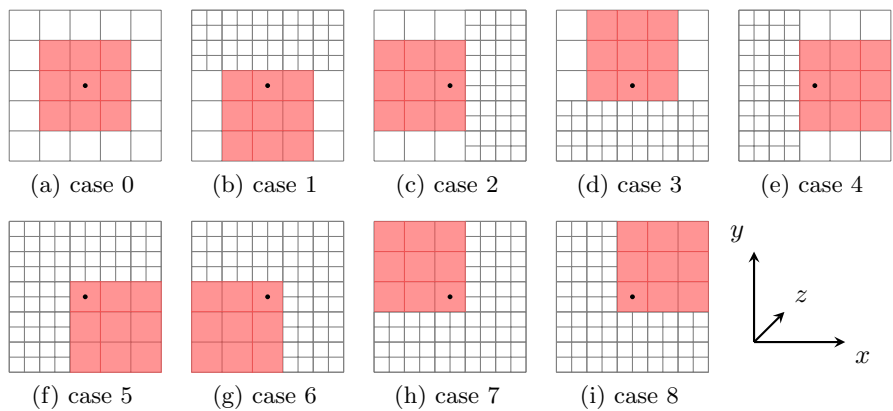\FloatBarrier\parindent 0pt
    
    \begin{table}[!ht]
         \centering
        \colorlet{cred}{red!70!white}
        \colorlet{cblue}{blue!50!white}
        \begin{minipage}{0.15\textwidth}
            \centering
            \begin{tikzpicture}
                \newcommand{\cell}[3]{
                    \fill[#3, opacity=0.6] (#1, #2) -- +(0, .5) -- +(.5, .5) -- +(.5,0) -- cycle;
                }
                
                \begin{scope}
                \draw[step=5mm, gray] (0, 0) grid (2.5, 2.5);
                \cell{1}{1}{cred}
                \cell{1}{1.5}{cred}
                \cell{1}{0.5}{cred}
                \cell{0.5}{1}{cred}
                \cell{1.5}{1}{cred}
                
                \cell{1.5}{1.5}{cred}
                \cell{1.5}{0.5}{cred}
                \cell{0.5}{1.5}{cred}
                \cell{0.5}{0.5}{cred}
                
                \foreach \i in {0,...,4}
                    \foreach \j in {0,...,4}
                    {
                        \pgfmathtruncatemacro{\lab}{\i + 5 * \j}
                        \draw[fill=black, draw=none] (0.5 * \i + 0.25, 0.5 * \j + 0.25) node[]{\lab};
                    }
                \end{scope}
            \end{tikzpicture}
        \end{minipage}
        \hfill
        \begin{minipage}{0.8\textwidth}\centering
            \begin{tabular}{crc}
                 \toprule
                bit pattern & unsigned long & case\\
                 \midrule
                \textcolor{red}{$0~0~0~0~0~0~1~1~1~0~0~1~1~1~0~0~1~1~1~0~0~0~0~0~0$} & $473'536$ & $0$ \\ 
                $0~0~0~0~0~0~0~0~0~0~0~1~1~1~0~0~1~1~1~0~0~1~1~1~0$ & $14'798$ & $1$  \\ 
                $0~0~0~0~0~0~0~1~1~1~0~0~1~1~1~0~0~1~1~1~0~0~0~0~0$ & $236'768$ & $2$ \\ 
                $0~1~1~1~0~0~1~1~1~0~0~1~1~1~0~0~0~0~0~0~0~0~0~0~0$ & $15'153'152$ & $3$ \\ 
                $0~0~0~0~0~1~1~1~0~0~1~1~1~0~0~1~1~1~0~0~0~0~0~0~0$ & $947'072$ & $4$ \\ 
                $0~0~0~0~0~0~0~0~0~0~1~1~1~0~0~1~1~1~0~0~1~1~1~0~0$ & $29'596$ & $5$ \\ 
                $0~0~0~0~0~0~0~0~0~0~0~0~1~1~1~0~0~1~1~1~0~0~1~1~1$ & $7'399$ & $6$ \\ 
                $0~0~1~1~1~0~0~1~1~1~0~0~1~1~1~0~0~0~0~0~0~0~0~0~0$ & $7'576'576$ & $7$ \\ 
                $1~1~1~0~0~1~1~1~0~0~1~1~1~0~0~0~0~0~0~0~0~0~0~0~0$ & $30'306'304$ & $8$ \\ 
                 \bottomrule
             \end{tabular}
        \end{minipage}
        \caption{Bit patterns of the second order Lagrange interpolation schemes with ordering according to 
        \Figref{fig:Lagrange_configurations}.\ The second column contains the corresponding number used to
        detect a pattern.\ An example of the conversion between grid and bits is indicated for the
        pattern on the right side with bit string highlighted in red.}
        \label{tab:bit_pattern}
     \end{table}
    
    \begin{figure}[!ht]
        \centering
        \begin{tikzpicture}[scale=0.8]
            \newcommand{\cell}[3]{
                \fill[#3, opacity=0.6] (#1, #2) -- +(0, .5) -- +(.5, .5) -- +(.5,0) -- cycle;
            }
            
            \colorlet{cred}{red!70!white}
            \colorlet{cblue}{blue!50!white}
        
            \begin{scope}
            \draw[step=5mm, gray] (0, 0) grid (1.5, 1.5);
            
            \cell{0.5}{0.5}{cred}
            \cell{0}{0.5}{cred}
            \cell{0.5}{0}{cred}
            
            \cell{0}{0}{cred}
            
            \draw[fill=black, draw=none] (0.75, 0.75) circle (0.04) node[]{};
            
            \node[] at (0.72, -0.5) {\parbox{0.3\linewidth}{\subcaption{case 0}\label{subfig:case0_1order}}};
            \end{scope}
            
            \begin{scope}[xshift=3cm]
            \draw[step=5mm, gray] (0, 0) grid (1.5, 1.5);
            
            \cell{0.5}{0.5}{cred}
            \cell{0}{0.5}{cred}
            \cell{0.5}{1}{cred}
            
            \cell{0}{1}{cred}
            
            \draw[fill=black, draw=none] (0.75, 0.75) circle (0.04) node[]{};
            
            \node[] at (0.72, -0.5) {\parbox{0.3\linewidth}{\subcaption{case 1}\label{subfig:case1_1order}}};
            \end{scope}
            
            \begin{scope}[xshift=6cm]
            \draw[step=5mm, gray] (0, 0) grid (1.5, 1.5);
            
            \cell{0.5}{0.5}{cred}
            \cell{1}{0.5}{cred}
            \cell{0.5}{1}{cred}
            
            \cell{1}{1}{cred}
            
            \draw[fill=black, draw=none] (0.75, 0.75) circle (0.04) node[]{};
            
            \node[] at (0.72, -0.5) {\parbox{0.3\linewidth}{\subcaption{case 2}\label{subfig:case2_1order}}};
            \end{scope}
            
            \begin{scope}[xshift=9cm]
            \draw[step=5mm, gray] (0, 0) grid (1.5, 1.5);
            
            \cell{0.5}{0.5}{cred}
            \cell{1}{0.5}{cred}
            \cell{0.5}{0}{cred}
            
            \cell{1}{0}{cred}
            
            \draw[fill=black, draw=none] (0.75, 0.75) circle (0.04) node[]{};
            
            \node[] at (0.72, -0.5) {\parbox{0.3\linewidth}{\subcaption{case 3}\label{subfig:case3_1order}}};
            \end{scope}
            
        \end{tikzpicture}
        \caption{All possible configurations for 2D linear Lagrange interpolation at which the red cells are used to build the
        Lagrange coefficients.\ The black dot indicates the cell at the current coarse-fine-interface.\ The interface is
        perpendicular to the shown cell layer.}
        \label{fig:Lagrange_2d_linear}
    \end{figure}
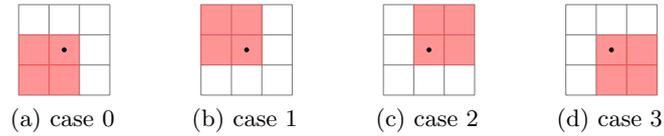\FloatBarrier\parindent 0pt
    
    \begin{table}[!ht]
        \centering
        \colorlet{cred}{red!70!white}
        \colorlet{cblue}{blue!50!white}
        \begin{minipage}{0.15\textwidth}
            \centering
            \begin{tikzpicture}
                \newcommand{\cell}[3]{
                    \fill[#3, opacity=0.6] (#1, #2) -- +(0, .5) -- +(.5, .5) -- +(.5,0) -- cycle;
                }
                
                \begin{scope}
                \draw[step=5mm, gray] (0, 0) grid (1.5, 1.5);
                \cell{0.5}{0.5}{cred}
                \cell{0}{0.5}{cred}
                \cell{0.5}{0}{cred}
                
                \cell{0}{0}{cred}
                
                \foreach \i in {0,...,2}
                    \foreach \j in {0,...,2}
                    {
                        \pgfmathtruncatemacro{\lab}{\i + 3 * \j}
                        \draw[fill=black, draw=none] (0.5 * \i + 0.25, 0.5 * \j + 0.25) node[]{\lab};
                    }
                \end{scope}
            \end{tikzpicture}
        \end{minipage}
        \begin{minipage}{0.5\textwidth}\centering
            \begin{tabular}{crc}
                \toprule
                bit pattern & unsigned long & case\\
                \midrule
                $\textcolor{red}{0~0~0~0~1~1~0~1~1}$ & $27$ & $0$ \\ 
                $0~1~1~0~1~1~0~0~0$ & $216$ & $1$ \\ 
                $1~1~0~1~1~0~0~0~0$ & $432$ & $2$ \\ 
                $0~0~0~1~1~0~1~1~0$ & $54$ & $3$ \\ 
                \bottomrule
            \end{tabular}
        \end{minipage}
        \caption{Bit patterns for 2D first order Lagrange interpolation (cf.\ Fig.~\ref{fig:Lagrange_2d_linear}).\
        The first row highlighted in red indicates the example pattern on the right side.}
        \label{tab:Lagrange_2d_linear}
    \end{table}

    \subsection{Boundary Conditions}
    Assuming the beam in vacuum and neglecting any beam pipes the electrostatic potential converges to zero at infinity.\
    In order to resemble this behaviour in \textit{finite difference} a common approximation is the \textit{Asymptotic Boundary 
    Condition} (ABC) presented in \cite{doi:10.1002/cpa.3160330603, 10.2307/2101222} that is also denoted as
    \textit{radiative} or \textit{open} boundary condition (BC).\ The first order approximation ABC-1 is given by
    \begin{equation}
    \frac{\partial\phi(r)}{\partial r} + \frac{1}{r}\phi(r) = \mathcal{O}(r^{-3}).
    \label{eq:ABC-1}
    \end{equation}
    Instead to spherical coordinates a formulation in Cartesian coordinates is applied for example in
    \cite{58681, 241666, doi:10.1063/1.4931738}.\ In spherical coordinates the $n$-th order approximation (ABC-$n$) is easily 
    evaluated by
    \begin{equation*}
    \left(\prod_{j=1}^{n}\left(\frac{\partial}{\partial r} + \frac{2j - 1}{r}\right)\right)\phi(r) = \mathcal{O}(r^{1-2n}),
    \end{equation*}
    where the product is computed in decreasing order and $n\in\mathbb{N}$.\ The implementation presented in this article 
    uses Robin boundary 
    conditions to approximate open boundaries.\ The formula looks similar to \Eqref{eq:ABC-1} except that the radial derivative
    is replaced by a normal derivative w.r.t.\ the mesh boundary, i.e.\ \cite{ADELMANN20104554}
    \begin{equation}
    \frac{\partial\phi}{\partial n} + \frac{1}{d}\phi = 0
    \label{eq:robin_bc}
    \end{equation}
    where $d>0$ is an artificial distance.\ The condition is discretised using
    \textit{central difference}.\ In addition to open BCs 
    according to \Eqref{eq:robin_bc} the solver presented here allows to impose homogeneous Dirichlet and periodic BCs at the mesh (or 
    physical) boundaries.
    
    \subsection{Algorithm and Implementation Details}
    \label{sec:multigrid}
    Following the notation of \cite{Martin96, MartinPhdThesis}, the full domain $\Omega$ is given by
    \begin{equation*}
    \Omega = \sum_{l=0}^{l_{max}}\Omega^l - \mathcal{P}(\Omega^{l+1}),
    \end{equation*}
    where the projection $\mathcal{P}$ from level $l+1$ to level $l$ satisfies $\mathcal{P}(\Omega^{l+1})\subset\Omega^l$.\ Due to
    the properties of the refinement Poisson's equation is described by
    \begin{equation*}
    \mathcal{L}^{comp}\phi = -\frac{\rho}{\varepsilon_0}\mbox{ on }\Omega
    \end{equation*}
    with composite Laplacian operator $\mathcal{L}^{comp}$ that considers only non-refined regions of each level.\ The full algorithm
    is illustrated in matrix notation in \Algref{alg:main} to \Algref{alg:v_cycle}.\ It performs a V-cycle in the residual correction 
    formulation with pre- and post-smoothing of the error.\ The iterative procedure stops when the $l_{p}$-norm of the residual of all 
    levels with $p\in\{1, 2, \infty\}$ is smaller than the corresponding right-hand side norm.\ Since \amrex{} assigns the grids
    to cores independent of the underlying level distribution, the implementation provides special matrices, i.e.\ $B_{crse}^{l}$ and
    $B_{fine}^{l}$, to handle the coarse-fine-interfaces.\ Thus, each AMR level stores up to ten matrices and four vectors 
    represented by \tpetra{} objects.\ These are the composite Laplacian matrix $A_{comp}^{l}$, the Laplacian matrix assuming no-finer
    grids $A_{nf}^{l}$, the coarse boundary matrix $B_{crse}^{l}$ and fine boundary matrix $B_{fine}^{l}$, the restriction and
    interpolation matrices $R^{l}$ and $I^{l}$, respectively, the gradient matrices $\mathbf{G}^{l}$ and the matrix to get all
    uncovered cells $U^{l}$.\ The vectors per level are the charge density $\rho^{l}$, electrostatic potential $\phi^{l}$, residual
    $r^{l}$ and error $e^{l}$.\ Whereas the vectors span the whole level domain, some matrices only cover a subdomain or carry 
    additional information for the coarse-fine interfaces as shown in \Figref{fig:matrices_and_domain}.\ The coarse and fine boundary
    matrices encompass one side of the Lagrange interpolation stencil that is completed by the Laplacian matrices.\ In case of the
    finest level the composite and no-fine Laplacian matrices coincide.
    
    The pre- and post-relaxation steps on line 8 and 16, respectively, of \Algref{alg:v_cycle} use the algorithms provided by
    \ifpack{} (e.g.\ Gauss-Seidel, Jacobi, etc.).\
    The linear system of equations on the coarsest level (\Algref{alg:v_cycle}, line 20) is either solved by direct solvers available 
    via \amesos{} or iterative solvers of \belos.\ Furthermore, an interface to \muelu{} allows Smoothed Aggregation Algebraic 
    Multigrid (SAAMG) as bottom solver.
    
    \begin{algorithm}[H]
        \begin{algorithmic}[1]
            \Require Level $l\ge 0$
            \Ensure  Updated residual $r^{l}$ on the composite domain
            \Function{Residual}{$l$}
                \If{$l = l_{max}$}
                    \State $r^{l} \gets \rho^{l} - A_{nf}^{l}\phi^{l} - B_{crse}^{l}\phi^{l-1}$
                \Else
                    \State $r^{l} \gets U^{l}\rho^{l}  - U^{l}\cdot
                            \left(A_{comp}^{l}\phi^{l} + B_{crse}^{l}\phi^{l-1} + B_{fine}^{l}\phi^{l+1}\right)$
                \EndIf
            \EndFunction
        \end{algorithmic}
        \caption{Residual evaluation on the composite domain}
        \label{alg:residual}
    \end{algorithm}
    
    \begin{algorithm}[H]
        \begin{algorithmic}[1]
            \Require Charge density $\rho$, electrostatic potential $\phi$, electric field $\mathbf{E}$
                     and finest level $l_{max}$
            \Ensure  Electrostatic potential $\phi$ and electric field $\mathbf{E}$
            \Function{Solve}{$\rho, \phi, \mathbf{E}, l_{max}$}
            
                \For{$l = 0$ to $l_{max}$}
                    \State \Call{Residual}{$l$}\Comment{Initialise residual}
                \EndFor
            
                \State $i\gets 0$
                
                \While{$i < i_{max}\wedge \exists l\in[0, l_{max}]:\ ||r^{l}||_{p} > \varepsilon||\rho^{l}||_{p}$}
                \Comment{$p\in\{1, 2, \infty\}$}
                    \State \Call{Relax}{$l_{max}$}\Comment{Start of V-cycle}
                    
                    \For{$l = 0$ to $l_{max}$}
                        \State \Call{Residual}{$l$}\Comment{Update residual}
                    \EndFor
                    
                    \State $i\gets i + 1$
                \EndWhile
                
                \For{$l=l_{max}-1$ to $0$}
                    \State $\phi^{l} \gets U^{l}\phi^{l} + R^{l}\phi^{l+1}$\Comment{Average down}
                \EndFor
                
                \For{$l=0$ to $l_{max}$}
                    \For{$d = 0$ to $3$}
                        \State $\mathbf{E}_{d}^{l} \gets -\mathbf{G}_{d}^{l}\phi^{l}$\Comment{Evaluate electric field}
                    \EndFor
                \EndFor
                
            \EndFunction
        \end{algorithmic}
        \caption{Main loop of AGMG}
        \label{alg:main}
    \end{algorithm}
    
    \begin{algorithm}[H]
        \begin{algorithmic}[1]
            \Require Level $l \ge 0$
            \Ensure  Electrostatic potential $\phi$
            \Function{Relax}{$l$}
                \If{$l = l_{max}$}
                    \State $r^{l}\gets \rho^{l} - A_{nf}^{l}\phi^{l} - B_{crse}^{l}\phi^{l-1}$
                \EndIf
                \If{$l > 0$}
                    \State $\phi_{save}^{l}\gets \phi^{l}$
                    \State $e^{l-1}\gets 0$
                    \State \Call{Smooth}{$e^{l}, r^{l}$}\Comment{Pre-smooth: Gauss-Seidel, Jacobi, ...}
                    \State $\phi^{l}\gets\phi^{l}+e^{l}$
                    \State $r^{l-1} \gets R^{l}\cdot\left(r^{l} - A_{nf}^{l}e^{l} - B_{crse}^{l}e^{l-1}\right)$
                           \Comment{Restrict on covered domain}
                    \State $r^{l-1} \gets U^{l-1}r^{l-1}
                                          - A_{comp}^{l-1}\phi^{l-1}
                                          - B_{crse}^{l-1}\phi^{l-2}
                                          - B_{fine}^{l-1}\phi^{l}$
                           \Comment{Residual update on uncovered domain}
                    \State \Call{Relax}{$l-1$}
                    \State $e^{l}\gets I^{l-1} e^{l-1}$\Comment{Prolongation / Interpolation}
                    \State $r^{l}\gets r^{l} - A_{nf}^{l}e^{l} - B_{crse}^{l}e^{l-1}$
                    \State $\delta e^{l}\gets 0$
                    \State \Call{Smooth}{$\delta e^{l}, r^{l}$}\Comment{Post-smooth: Gauss-Seidel, Jacobi, ...}
                    \State $e^{l}\gets e^{l}+\delta e^{l}$
                    \State $\phi^{l}\gets\phi_{save}^{l}+e^{l}$
                \Else
                    \State $Ae^{0} = r^{0}$\Comment{Solve linear system of equations}
                    \State $\phi^{0}\gets\phi^{0} + e^{0}$
                \EndIf
            \EndFunction
        \end{algorithmic}
        \caption{Residual correction V-Cycle}
        \label{alg:v_cycle}
    \end{algorithm}
    
    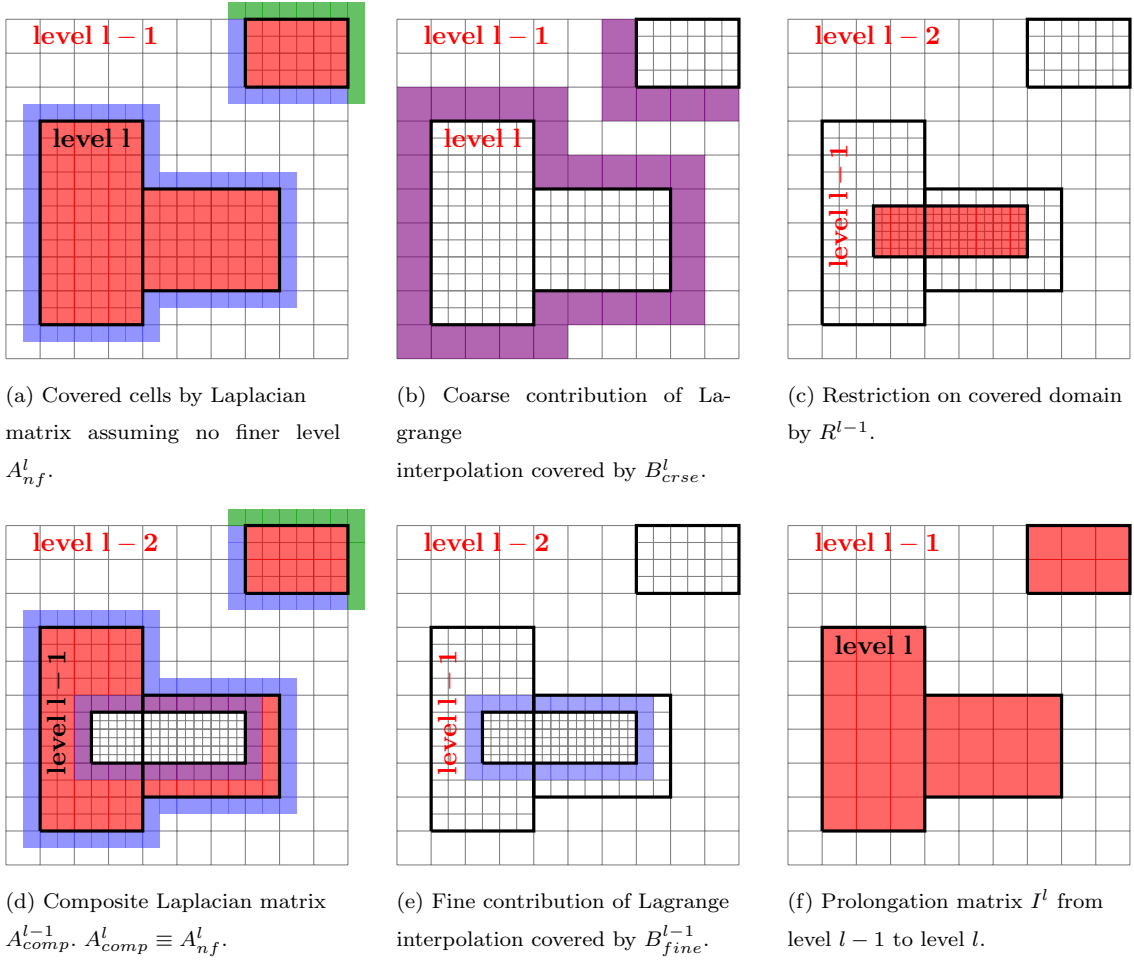
\begin{figure}[!ht]
        \centering
        \begin{subfigure}[t]{0.3\textwidth}
            \centering
            \newcommand{\cell}[3]{
                \fill[#3, opacity=0.6] (#1, #2) -- +(0, 1) -- +(1, 1) -- +(1,0) -- cycle;
            }
            \newcommand{\halfcell}[3]{
                \fill[#3, opacity=0.6] (#1, #2) -- +(0, 0.5) -- +(0.5, 0.5) -- +(0.5,0) -- cycle;
            }
            \begin{tikzpicture}[scale=0.45]
                \begin{scope}
                    \draw[step=10mm, white] (0, 0) grid (11, 11);
                    \draw[step=10mm, gray] (0, 0) grid (10, 10);
                    
                    \draw[step=5mm, gray] (1, 1) grid (4, 7);
                    
                    \draw[step=5mm, gray] (4, 2) grid (8, 5);
                    
                    \draw[step=5mm, gray] (7, 8) grid (10, 10);
                    \foreach \i in {1,...,6}
                        \foreach \j in {1,...,3}
                        {
                            \cell{\j}{\i}{red}
                        }
                    \foreach \i in {2,...,4}
                        \foreach \j in {4,...,7}
                        {
                            \cell{\j}{\i}{red}
                        }
                    
                    \foreach \i in {8,...,9}
                        \foreach \j in {7,...,9}
                        {
                            \cell{\j}{\i}{red}
                        }
                
                    \foreach \i in {0.5,1,...,7}
                    {
                        \halfcell{0.5}{\i}{blue!70!white}
                    }
                    
                    \foreach \j in {1,1.5,...,4}
                    {
                        \halfcell{\j}{0.5}{blue!70!white}
                    }
                    
                    \halfcell{4}{1}{blue!70!white}
                    
                    \foreach \j in {4,4.5,...,8}
                    {
                        \halfcell{\j}{1.5}{blue!70!white}
                    }
                    
                    \foreach \j in {1,1.5,...,4}
                    {
                        \halfcell{\j}{7}{blue!70!white}
                    }
                
                    \foreach \i in {5,5.5,...,6.5}
                    {
                        \halfcell{4}{\i}{blue!70!white}
                    }
                    
                    \foreach \j in {4.5,5,...,8}
                    {
                        \halfcell{\j}{5}{blue!70!white}
                    }
                
                    \foreach \i in {2,2.5,...,4.5}
                    {
                        \halfcell{8}{\i}{blue!70!white}
                    }
                
                    \foreach \i in {7.5,8,...,9.5}
                    {
                        \halfcell{6.5}{\i}{blue!70!white}
                    }
                    
                    \foreach \j in {7,7.5,...,9.5}
                    {
                        \halfcell{\j}{7.5}{blue!70!white}
                    }
                    
                    \foreach \j in {6.5,7,...,10}
                    {
                        \halfcell{\j}{10}{green!60!black}
                    }
                    
                    \foreach \i in {7.5,8,...,9.5}
                    {
                        \halfcell{10}{\i}{green!60!black}
                    }
                
                    \draw[very thick] (1, 1) -- (4, 1) -- (4, 2);
                    \draw[very thick] (4, 5) -- (4, 7) -- (1, 7) -- (1, 1);
                    \draw[very thick] (4, 2) -- (8, 2) -- (8, 5) -- (4, 5);
                    \draw[very thick] (7, 8) -- (10, 8) -- (10, 10) -- (7, 10) -- (7, 8);
                    \draw[very thick] (4, 2) -- (4, 5);
                    
                    \node[red, right] at (0.5, 9.5) {$\bf level\ l-1$};
                    \node at (2.5, 6.5) {$\bf level\ l$};
                \end{scope}
            \end{tikzpicture}
            \captionsetup{font=footnotesize}
            \caption{Covered cells by Laplacian \\ matrix assuming
            no finer level $A_{nf}^{l}$.}
        \end{subfigure}
        \hfill
        \begin{subfigure}[t]{0.3\textwidth}
            \centering
            \newcommand{\cell}[3]{
                \fill[#3, opacity=0.6] (#1, #2) -- +(0, 1) -- +(1, 1) -- +(1,0) -- cycle;
            }
            \newcommand{\halfcell}[3]{
                \fill[#3, opacity=0.6] (#1, #2) -- +(0, 0.5) -- +(0.5, 0.5) -- +(0.5,0) -- cycle;
            }
            \begin{tikzpicture}[scale=0.45]
                \begin{scope}
                    \draw[step=10mm, white] (0, 0) grid (11, 11);
                    \draw[step=10mm, gray] (0, 0) grid (10, 10);
                    \draw[step=5mm, gray] (1, 1) grid (4, 7);
                    
                    \draw[step=5mm, gray] (4, 2) grid (8, 5);
                    
                    \draw[step=5mm, gray] (7, 8) grid (10, 10);
                
                    \foreach \i in {0,...,7}
                    {
                        \cell{0}{\i}{violet}
                    }
                    
                    \foreach \j in {1,...,4}
                    {
                        \cell{\j}{0}{violet}
                    }
                    
                    \foreach \j in {4,...,8}
                    {
                        \cell{\j}{1}{violet}
                    }
                    
                    \foreach \j in {1,...,4}
                    {
                        \cell{\j}{7}{violet}
                    }

                    \foreach \i in {5,...,6}
                    {
                        \cell{4}{\i}{violet}
                    }
                    
                    \foreach \j in {5,...,8}
                    {
                        \cell{\j}{5}{violet}
                    }
                    
                    \foreach \i in {2,...,4}
                    {
                        \cell{8}{\i}{violet}
                    }

                    \foreach \i in {7,...,9}
                    {
                        \cell{6}{\i}{violet}
                    }
                    
                    \foreach \j in {7,...,9}
                    {
                        \cell{\j}{7}{violet}
                    }
                    \draw[very thick] (1, 1) -- (4, 1) -- (4, 2);
                    \draw[very thick] (4, 5) -- (4, 7) -- (1, 7) -- (1, 1);
                    \draw[very thick] (4, 2) -- (8, 2) -- (8, 5) -- (4, 5);
                    \draw[very thick] (7, 8) -- (10, 8) -- (10, 10) -- (7, 10) -- (7, 8);
                    \draw[very thick] (4, 2) -- (4, 5);
                    
                    \node[red, right] at (0.5, 9.5) {$\bf level\ l-1$};
                    \node[red] at (2.5, 6.5) {$\bf level\ l$};
                \end{scope}
            \end{tikzpicture}
            \captionsetup{font=footnotesize}
            \caption{Coarse contribution of Lagrange \\ interpolation covered by $B_{crse}^{l}$.}
        \end{subfigure}
        \hfill
        \begin{subfigure}[t]{0.3\textwidth}
            \centering
            \newcommand{\cell}[3]{
                \fill[#3, opacity=0.6] (#1, #2) -- +(0, 1) -- +(1, 1) -- +(1,0) -- cycle;
            }
            \newcommand{\halfcell}[3]{
                \fill[#3, opacity=0.6] (#1, #2) -- +(0, 0.5) -- +(0.5, 0.5) -- +(0.5,0) -- cycle;
            }
            \begin{tikzpicture}[scale=0.45]
                \begin{scope}
                    \draw[step=10mm, white] (0, 0) grid (11, 11);
                    \draw[step=10mm, gray] (0, 0) grid (10, 10);
                    \draw[step=5mm, gray] (1, 1) grid (4, 7);
                    
                    \draw[step=5mm, gray] (4, 2) grid (8, 5);
                    
                    \draw[step=5mm, gray] (7, 8) grid (10, 10);
                    \draw[step=2.5mm, gray] (2.5, 3) grid (7, 4.5);
                    \foreach \i in {3,3.5,4}
                        \foreach \j in {2.5,3,...,6.5}
                        {
                            \halfcell{\j}{\i}{red}
                        }
                    \draw[very thick] (1, 1) -- (4, 1) -- (4, 2);
                    \draw[very thick] (4, 5) -- (4, 7) -- (1, 7) -- (1, 1);
                    \draw[very thick] (4, 2) -- (8, 2) -- (8, 5) -- (4, 5);
                    \draw[very thick] (7, 8) -- (10, 8) -- (10, 10) -- (7, 10) -- (7, 8);
                    \draw[very thick] (4, 2) -- (4, 3);
                    \draw[very thick] (4, 4.5) -- (4, 5);
                    \draw[very thick] (4, 2) -- (4, 5);
                    \draw[very thick] (2.5, 3) -- (7, 3) -- (7, 4.5) -- (2.5, 4.5) -- (2.5, 3);
                    
                    \node[red, right] at (0.5, 9.5) {$\bf level\ l-2$};
                    \node[red,rotate=90] at (1.5, 4.5) {$\bf level\ l-1$};
                \end{scope}
            \end{tikzpicture}
            \captionsetup{font=footnotesize}
            \caption{Restriction on covered domain \\ by $R^{l-1}$.}
        \end{subfigure}
        \hfill
        \begin{subfigure}[t]{0.3\textwidth}
            \centering
            \newcommand{\cell}[3]{
                \fill[#3, opacity=0.6] (#1, #2) -- +(0, 1) -- +(1, 1) -- +(1,0) -- cycle;
            }
            \newcommand{\halfcell}[3]{
                \fill[#3, opacity=0.6] (#1, #2) -- +(0, 0.5) -- +(0.5, 0.5) -- +(0.5,0) -- cycle;
            }
            \begin{tikzpicture}[scale=0.45]
                \begin{scope}
                    \draw[step=10mm, white] (0, 0) grid (11, 11);
                    \draw[step=10mm, gray] (0, 0) grid (10, 10);
                    \draw[step=5mm, gray] (1, 1) grid (4, 7);
                    
                    \draw[step=5mm, gray] (4, 2) grid (8, 5);
                    
                    \draw[step=5mm, gray] (7, 8) grid (10, 10);
                    \draw[step=2.5mm, gray] (2.5, 3) grid (7, 4.5);
                    
                    \foreach \i in {1,...,6}
                        \foreach \j in {1,...,1}
                        {
                            \cell{\j}{\i}{red}
                        }
                    
                    \foreach \i in {1,5,6}
                        \foreach \j in {2,3}
                        {
                            \cell{\j}{\i}{red}
                        }
                    
                    \foreach \j in {2,2.5,...,7}
                    {
                        \halfcell{\j}{4.5}{violet}
                        \halfcell{\j}{2.5}{violet}
                        \halfcell{\j}{2}{red}
                    }
                    \foreach \i in {3,3.5,4}
                    {
                        \halfcell{2}{\i}{violet}
                        \halfcell{7}{\i}{violet}
                    }
                    
                    \foreach \i in {2,2.5,...,4.5}
                    {
                        \halfcell{7.5}{\i}{red}
                    }
                    
                    \foreach \i in {8,...,9}
                        \foreach \j in {7,...,9}
                        {
                            \cell{\j}{\i}{red}
                        }
                    
                    \foreach \i in {0.5,1,...,7}
                    {
                        \halfcell{0.5}{\i}{blue!70!white}
                    }
                    
                    \foreach \j in {1,1.5,...,4}
                    {
                        \halfcell{\j}{0.5}{blue!70!white}
                    }
                    
                    \halfcell{4}{1}{blue!70!white}
                    
                    \foreach \j in {4,4.5,...,8}
                    {
                        \halfcell{\j}{1.5}{blue!70!white}
                    }
                    
                    \foreach \j in {1,1.5,...,4}
                    {
                        \halfcell{\j}{7}{blue!70!white}
                    }

                    \foreach \i in {5,5.5,...,6.5}
                    {
                        \halfcell{4}{\i}{blue!70!white}
                    }
                    
                    \foreach \j in {4.5,5,...,8}
                    {
                        \halfcell{\j}{5}{blue!70!white}
                    }
                    
                    \foreach \i in {2,2.5,...,4.5}
                    {
                        \halfcell{8}{\i}{blue!70!white}
                    }

                    \foreach \i in {7.5,8,...,9.5}
                    {
                        \halfcell{6.5}{\i}{blue!70!white}
                    }
                    
                    \foreach \j in {7,7.5,...,9.5}
                    {
                        \halfcell{\j}{7.5}{blue!70!white}
                    }

                    \foreach \j in {6.5,7,...,10}
                    {
                        \halfcell{\j}{10}{green!60!black}
                    }
                    
                    \foreach \i in {7.5,8,...,9.5}
                    {
                        \halfcell{10}{\i}{green!60!black}
                    }
                    
                    \draw[very thick] (1, 1) -- (4, 1) -- (4, 2);
                    \draw[very thick] (4, 5) -- (4, 7) -- (1, 7) -- (1, 1);
                    \draw[very thick] (4, 2) -- (8, 2) -- (8, 5) -- (4, 5);
                    \draw[very thick] (7, 8) -- (10, 8) -- (10, 10) -- (7, 10) -- (7, 8);
                    \draw[very thick] (4, 2) -- (4, 3);
                    \draw[very thick] (4, 4.5) -- (4, 5);
                    \draw[very thick] (4, 2) -- (4, 5);
                    \draw[very thick] (2.5, 3) -- (7, 3) -- (7, 4.5) -- (2.5, 4.5) -- (2.5, 3);
                    \node[red, right] at (0.5, 9.5) {$\bf level\ l-2$};
                    \node[black,rotate=90] at (1.5, 4.5) {$\bf level\ l-1$};
                \end{scope}
            \end{tikzpicture}
            \captionsetup{font=footnotesize}
            \caption{Composite Laplacian matrix \\ $A_{comp}^{l-1}$.\
            $A_{comp}^{l} \equiv A_{nf}^{l}$.}
        \end{subfigure}
        \hfill
        \begin{subfigure}[t]{0.3\textwidth}
            \centering
            \newcommand{\cell}[3]{
                \fill[#3, opacity=0.6] (#1, #2) -- +(0, 1) -- +(1, 1) -- +(1,0) -- cycle;
            }
            \newcommand{\halfcell}[3]{
                \fill[#3, opacity=0.6] (#1, #2) -- +(0, 0.5) -- +(0.5, 0.5) -- +(0.5,0) -- cycle;
            }
            \begin{tikzpicture}[scale=0.45]
                \begin{scope}
                    \draw[step=10mm, white] (0, 0) grid (11, 11);
                    \draw[step=10mm, gray] (0, 0) grid (10, 10);
                    \draw[step=5mm, gray] (1, 1) grid (4, 7);
                    
                    \draw[step=5mm, gray] (4, 2) grid (8, 5);
                    
                    \draw[step=5mm, gray] (7, 8) grid (10, 10);
                    \draw[step=2.5mm, gray] (2.5, 3) grid (7, 4.5);
                    
                    \foreach \i in {2.5, 3,...,4.5}
                    {
                        \halfcell{2}{\i}{blue!60!white}
                    }
                    
                    \foreach \j in {2.5, 3,...,7}
                    {
                        \halfcell{\j}{2.5}{blue!60!white}
                    }
                    
                    \foreach \j in {2.5, 3,...,7}
                    {
                        \halfcell{\j}{4.5}{blue!60!white}
                    }
                    
                    \foreach \i in {3, 3.5, 4}
                    {
                        \halfcell{7}{\i}{blue!60!white}
                    }
                    
                    \draw[very thick] (1, 1) -- (4, 1) -- (4, 2);
                    \draw[very thick] (4, 5) -- (4, 7) -- (1, 7) -- (1, 1);
                    \draw[very thick] (4, 2) -- (8, 2) -- (8, 5) -- (4, 5);
                    \draw[very thick] (7, 8) -- (10, 8) -- (10, 10) -- (7, 10) -- (7, 8);
                    \draw[very thick] (4, 2) -- (4, 3);
                    \draw[very thick] (4, 4.5) -- (4, 5);
                    \draw[very thick] (4, 2) -- (4, 5);
                    \draw[very thick] (2.5, 3) -- (7, 3) -- (7, 4.5) -- (2.5, 4.5) -- (2.5, 3);
                    
                    \node[red, right] at (0.5, 9.5) {$\bf level\ l-2$};
                    \node[red,rotate=90] at (1.5, 4.5) {$\bf level\ l-1$};
                \end{scope}
            \end{tikzpicture}
            \captionsetup{font=footnotesize}
            \caption{Fine contribution of Lagrange \\ interpolation covered by $B_{fine}^{l-1}$.}
        \end{subfigure}
        \hfill
        \begin{subfigure}[t]{0.3\textwidth}
            \centering
            \newcommand{\cell}[3]{
                \fill[#3, opacity=0.6] (#1, #2) -- +(0, 1) -- +(1, 1) -- +(1,0) -- cycle;
            }
            \newcommand{\halfcell}[3]{
                \fill[#3, opacity=0.6] (#1, #2) -- +(0, 0.5) -- +(0.5, 0.5) -- +(0.5,0) -- cycle;
            }
            \begin{tikzpicture}[scale=0.45]
                \begin{scope}
                    \draw[step=10mm, white] (0, 0) grid (11, 11);
                    \draw[step=10mm, gray] (0, 0) grid (10, 10);
                    
                    
                    
                    \foreach \i in {1,...,6}
                        \foreach \j in {1,2,3}
                        {
                            \cell{\j}{\i}{red}
                        }
                    
                    \foreach \i in {2,...,4}
                        \foreach \j in {4,...,7}
                        {
                            \cell{\j}{\i}{red}
                        }
                    
                    \foreach \i in {8,...,9}
                        \foreach \j in {7,...,9}
                        {
                            \cell{\j}{\i}{red}
                        }
                    
                    \draw[very thick] (1, 1) -- (4, 1) -- (4, 2);
                    \draw[very thick] (4, 5) -- (4, 7) -- (1, 7) -- (1, 1);
                    \draw[very thick] (4, 2) -- (8, 2) -- (8, 5) -- (4, 5);
                    \draw[very thick] (7, 8) -- (10, 8) -- (10, 10) -- (7, 10) -- (7, 8);
                    \draw[very thick] (4, 2) -- (4, 5);
                    
                    \node[red, right] at (0.5, 9.5) {$\bf level\ l-1$};
                    \node[black] at (2.5, 6.5) {$\bf level\ l$};
                \end{scope}
            \end{tikzpicture}
            \captionsetup{font=footnotesize}
            \caption{Prolongation matrix $I^{l}$ from \\ level $l-1$ to level $l$.}
        \end{subfigure}
        \caption{Cell domain occupied by matrices.\ Red: \textit{Usual} cell domain; Green: Physical / mesh boundary; Blue: Fine 
        contribution of Lagrange interpolation; Violet: Coarse contribution of Lagrange interpolation.}
        \label{fig:matrices_and_domain}
    \end{figure}\FloatBarrier\parindent 0pt
    
    \section{Poisson Solver Validation}
    \label{sec:benchmark}
    The Poisson solver is validated using three different examples.\ First, the preservation of symmetry is tested.\
    Second, a comparison with the analytical solution of a uniformly charged sphere in free space is shown.\ Although
    AMR is not turned on for a single-bunch simulation in the real application, it is nevertheless a good mini-app to check for any 
    discontinuities at the coarse-fine interfaces among levels.\
    In a third example the solver is validated by means of the built-in Poisson multi-level (ML) solver of \amrex{} where 11
    Gaussian-shaped bunches are placed in a chain using Dirichlet boundary conditions in the computation domain mimicking a
    multi-bunch simulation in high intensity cyclotrons as studied in \cite{PhysRevSTAB.13.064201}.\
    The last two tests use the charge density to obtain the mesh refinements with threshold
    $\lambda = \SI{1}{fC/m^3}$ (cf.\ \Eqref{eq:tag_charge} in \Secref{sec:amr_policies}).\\
    All line and projection plots are generated with an own extension of the \yt{} package \cite{2011ApJS..192....9T}.\ In the 
    following, a regular PIC model with a uniform single-level mesh, i.e.\ without refinement,
    is an AMR simulation of at most level zero.\
    
    \subsection{Symmetry Conservation}
    In order to check symmetry preservation we initialise a three level problem where each level covers the centred 
    region as shown below in \Figref{fig:artificial_charge_density}.\ At each level, the grid cells are assigned to the same charge 
    density value, starting at \SI{1}{C/m^3} on level zero and increasing by \SI{0.5}{C/m^3} on each subsequent level.\ Therefore,
    cutting a line through the centre of the domain yields a perfectly symmetric electrostatic potential and anti-symmetric
    electric field components mirrored at the centre.\ According to \Figref{fig:artificial_absolute_error}, the symmetry is
    preserved with absolute errors in the order of magnitude of machine precision and thus negligible.\
    
    \begin{figure}[htp]
        \centering
        \begin{subfigure}[t]{0.47\textwidth}
            \includegraphics[width=1.0\textwidth]{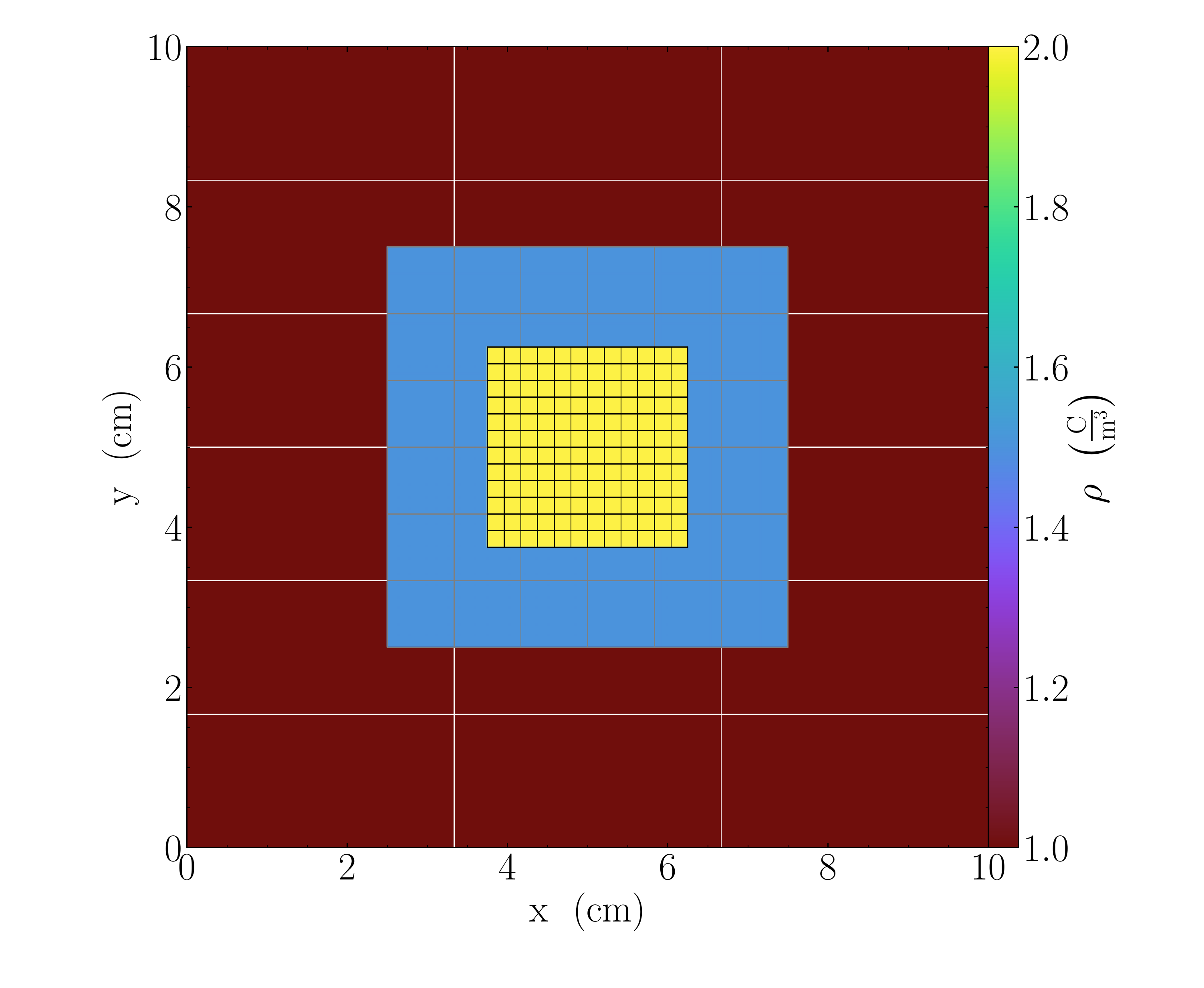}
        \caption{Slice through the domain showing the charge density per level.}
        \label{fig:artificial_charge_density}
        \end{subfigure}
        \hfill
        \begin{subfigure}[t]{0.47\textwidth}
            \includegraphics[width=1.0\textwidth]{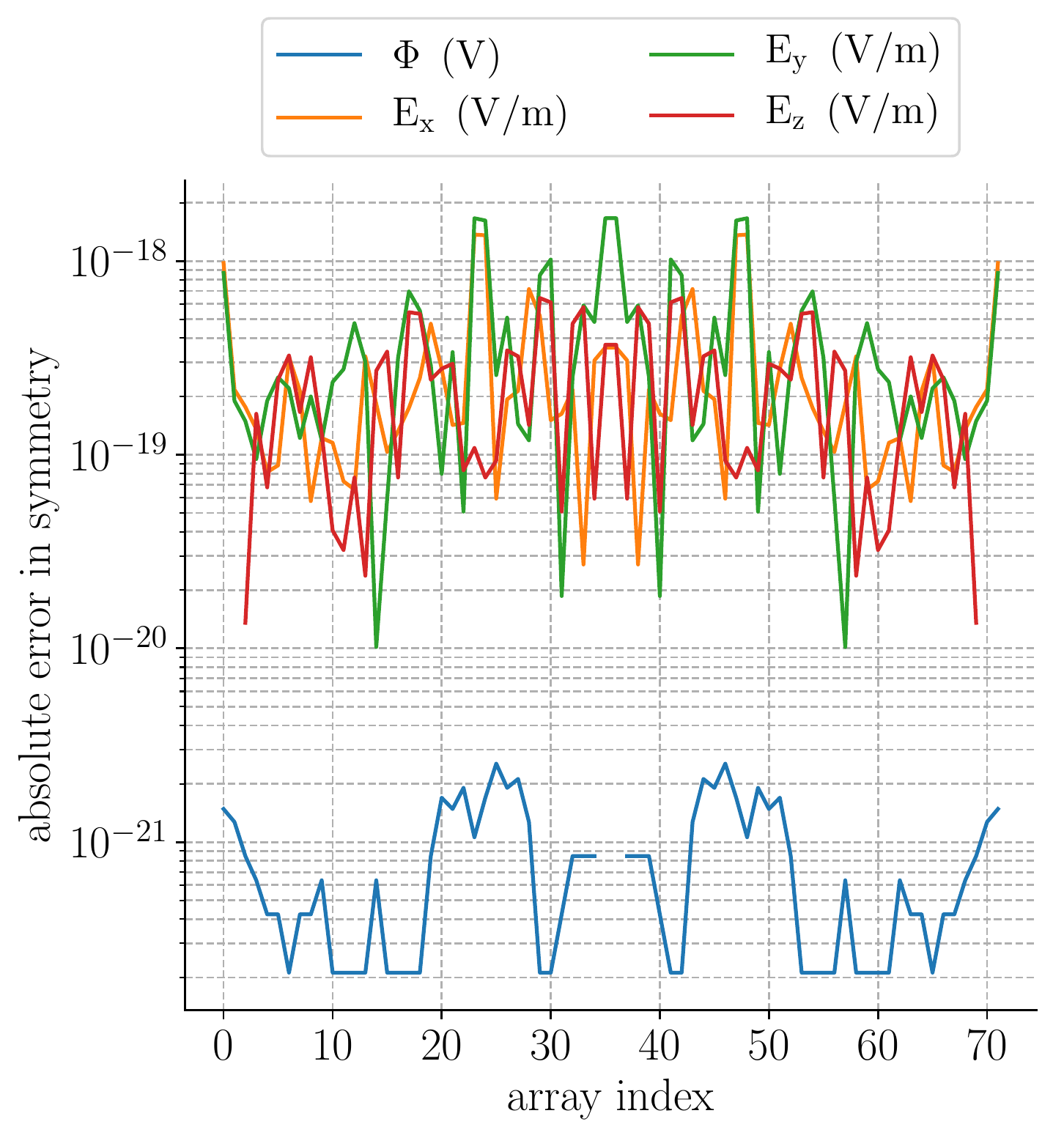}
        \caption{Symmetry errors when cutting a line through the centre of the domain.}
        \label{fig:artificial_absolute_error}
        \end{subfigure}
        \caption{Charge density and absolute error in symmetry of electric field components and electrostatic potential.\
        Starting at a charge density of \SI{1}{C/m^3} on level zero (full domain), it is incremented by \SI{0.5}{C/m^3} on
        subsequent higher levels.}
    \end{figure}
    
    \subsection{Uniformly Charged Sphere in Free Space}
    In this mini-app $10^6$ particles are randomly picked within a sphere of radius $R = 5\ \si{mm}$ centred at origin.\
    In order to simplify comparison to the analytical solution
    \begin{equation*}
        E(r) = \frac{Q}{4\pi\epsilon_0}
        \begin{cases}
            r^{-2}, & r > R \\
            R^{-3}r, & r \le R
        \end{cases},
    \end{equation*}

    \begin{equation*}
    \phi(r) = \frac{Q}{4\pi\epsilon_0}
        \begin{cases}
            r^{-1}, & r > R \\
            (2R)^{-1}\cdot(3 - r^2 R^{-2}), & r \le R
        \end{cases},
    \end{equation*}
    each particle carries a charge of $q = 4\pi\epsilon_0 R^2\cdot10^{-2}\ \si{C}$.\ Thus, the peak value of the electric field is
    $10^4\ \si{V/m}$ and  $75\ \si{V}$ for the potential.\ The computation is performed using a base grid of $36^3$ grid points
    and 2 refined levels.\ The mesh is increased by $\delta = \SI{20}{\percent}$ compared to the computation domain (cf.\
    \Secref{sec:domain_transform}).\ The line plots of \Figref{fig:comparison_uniform_sphere} show the results for various artificial
    distances $d$ of \Eqref{eq:robin_bc}.\ The solution with distance $d=1.7$ agrees well with the analytical 
    solution.\ As expected the potential deviates at the boundaries from the analytical solution due to the numerical approximation
    of the open boundaries.

    \begin{figure}[!ht]
        \centering
        \begin{subfigure}[b]{0.49\textwidth}
            \centering
            \includegraphics[width=1\textwidth]
                {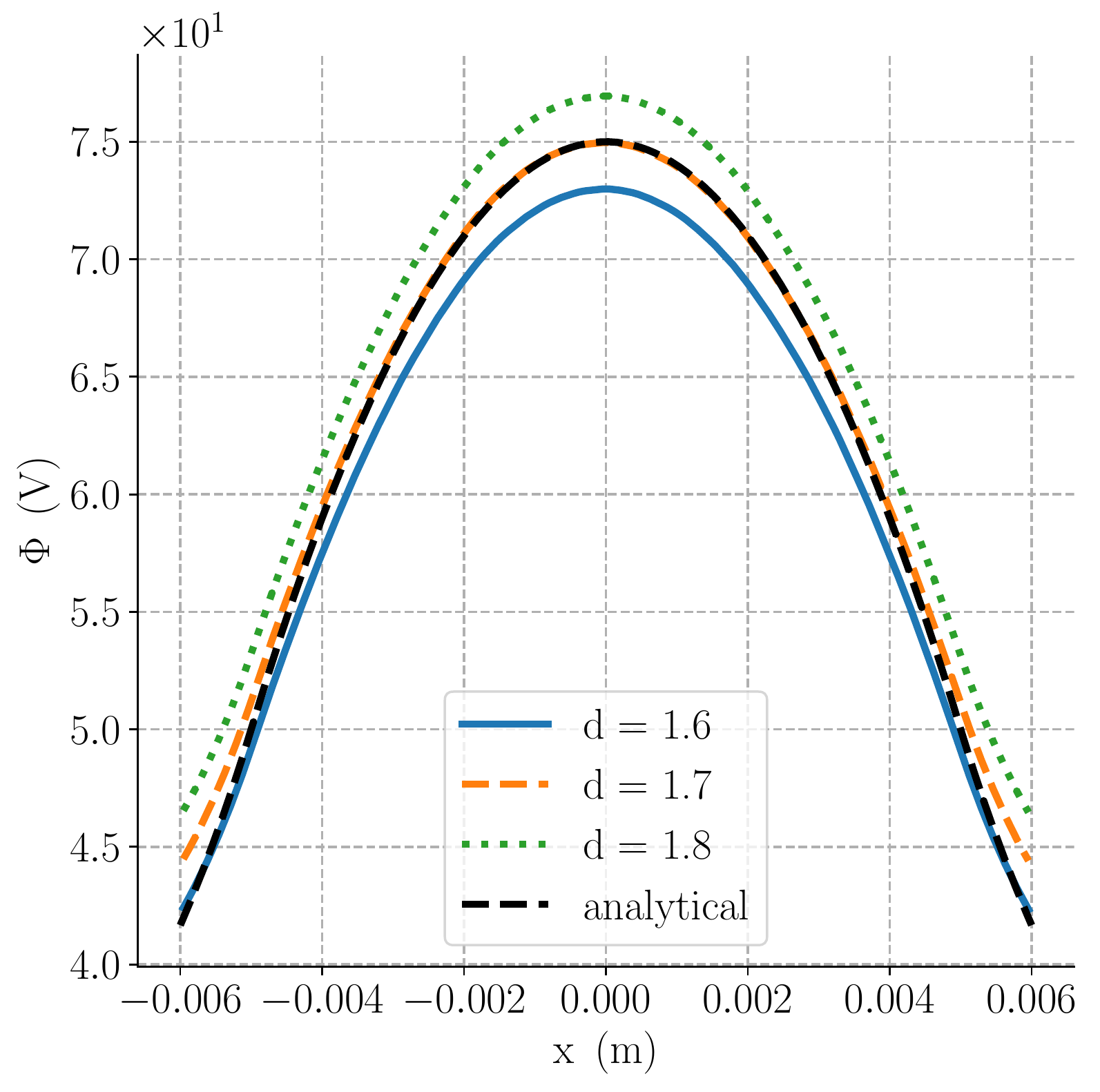}
            \caption{Electrostatic potential in $x$-direction.}
        \end{subfigure}
        \hfill
        \begin{subfigure}[b]{0.49\textwidth}
            \centering
            \includegraphics[width=1\textwidth]
                {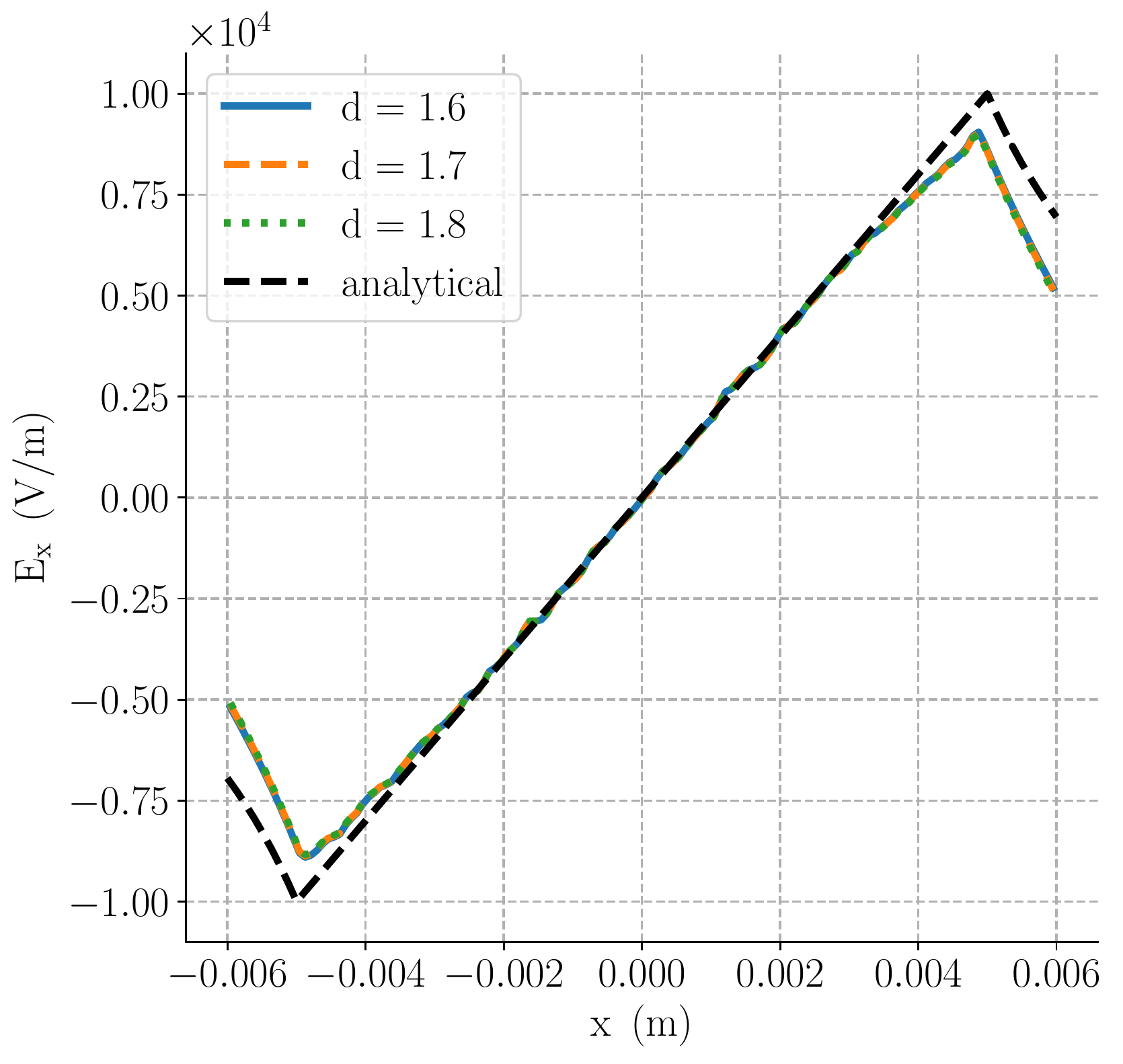}
            \caption{Electric field component in $x$-direction.}
            \label{fig:unif_b}
        \end{subfigure}
        \caption{Comparison of the analytical and numerical solution of a uniformly charged sphere in free space with
        various artificial distances $d$ of the open boundary condition (cf.\ \Eqref{eq:robin_bc}).\ The lines in
        (\subref{fig:unif_b}) coincide.}
        \label{fig:comparison_uniform_sphere}
    \end{figure}\FloatBarrier\parindent 0pt
    
    \begin{figure}[!ht]
        \centering
        \begin{subfigure}[t]{0.49\textwidth}
            \centering
            \includegraphics[width=1\textwidth]
                {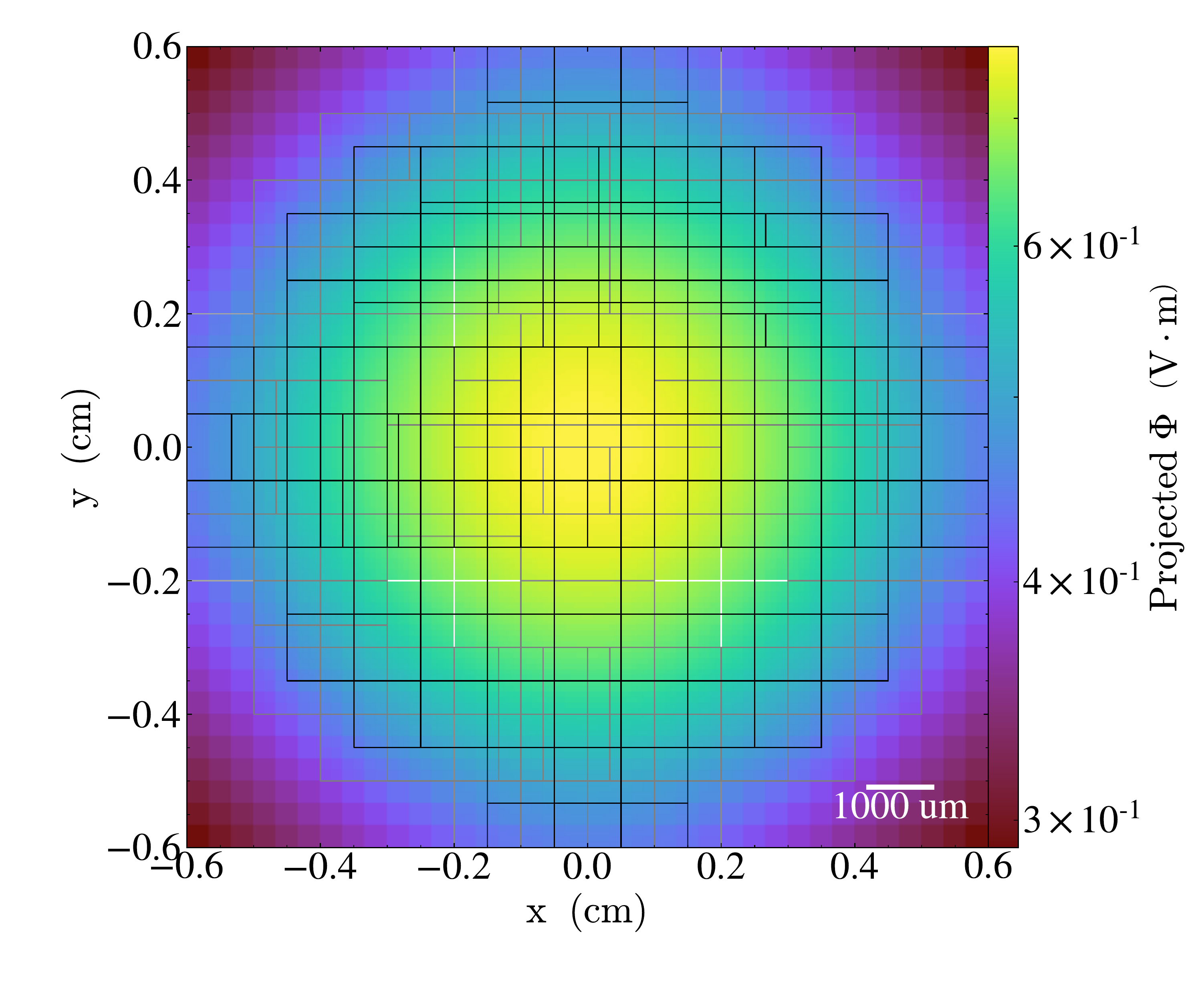}
            \caption{Electrostatic potential.}
        \end{subfigure}
        \hfill
        \begin{subfigure}[t]{0.49\textwidth}
            \centering
            \includegraphics[width=1\textwidth]
                {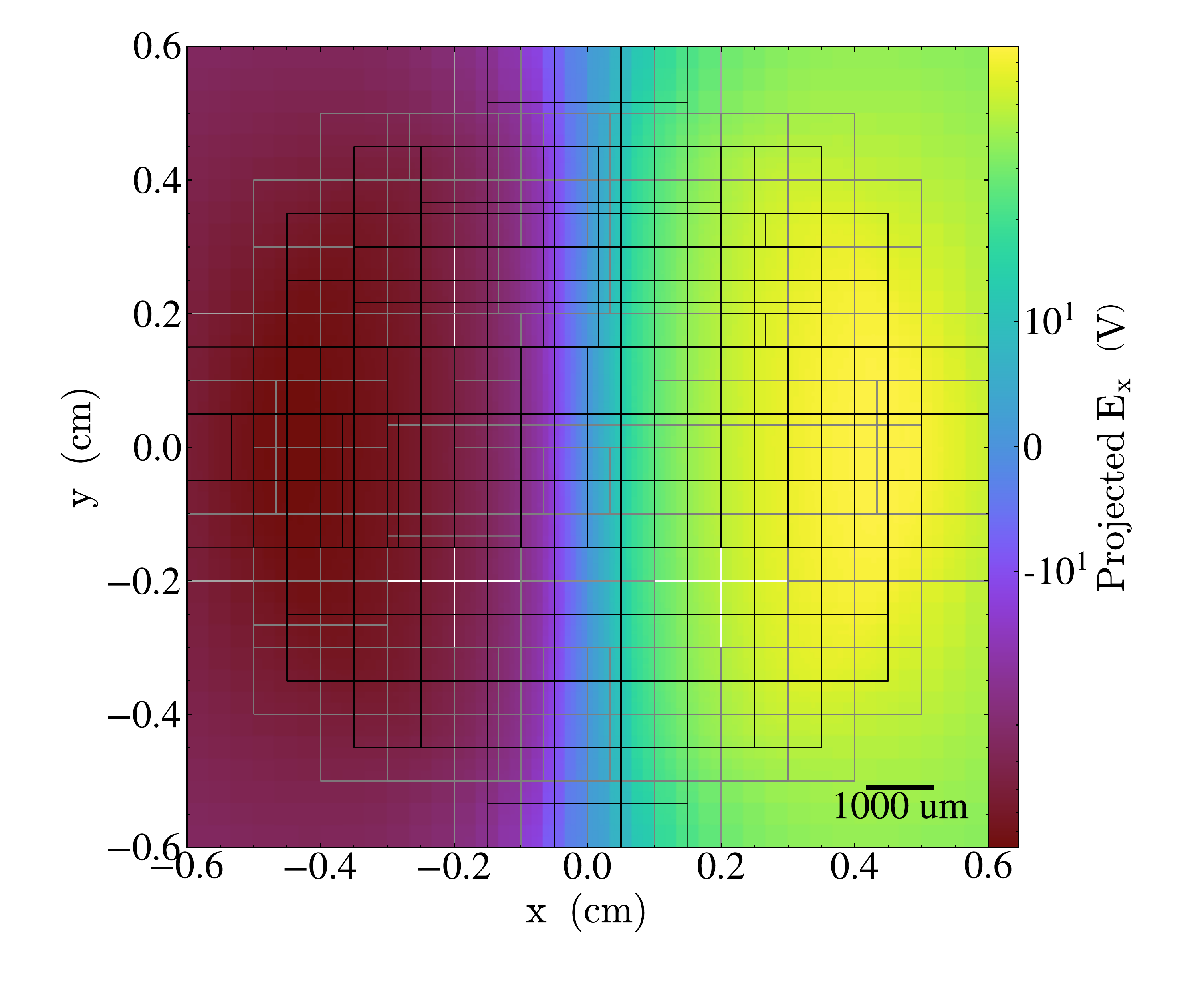}
            \caption{Electric field component in $x$-direction.}
        \end{subfigure}
        \caption{Integrated projection plots onto the $xy$-plane of the electrostatic potential and its electric field
        component $E_x$.}
        \label{fig:sliceplot_potential_uniform_sphere}
    \end{figure}\FloatBarrier\parindent 0pt
    
    \subsection{11 Gaussian-Shaped Bunches}
    \label{sec:N_Gaussian_Shaped_Bunches}
    In this mini-app the newly implemented solver is compared to the Poisson solver of \amrex{}.\ Each bunch is
    initialised with $10^6$ macro particles of charge \SI{0.1}{fC}.\ The particles per bunch are picked using a
    one-dimensional Gaussian distribution per dimension with mean $\mu_y = \mu_z = 0\ \si{m}$ and standard deviation
    $\sigma_y = \sigma_z = 5\ \si{mm}$.\ In horizontal direction the standard deviation is $\sigma_x = 1.5\ \si{mm}$ with a mean 
    shift of $4\ \si{cm}$ to the neighbouring bunches.\ The problem is solved on a $144^3$ base grid and 2 levels of refinement.\ 
    At the mesh boundaries the Dirichlet boundary condition $\partial\phi = 0$ is imposed.\
    The mesh is increased by $\delta = \SI{10}{\percent}$ as explained in \Secref{sec:domain_transform}.\
    As indicated by the line plots of \Figref{fig:lineplot} both solutions agree.\ The potential has a maximum absolute error
    \SI{0.022}{V} that corresponds to a maximum relative error of \SI{0.51}{\percent}.\
    
    \begin{figure}[!ht]
        \centering
        \begin{subfigure}[t]{0.49\textwidth}
            \includegraphics[width=\textwidth]
                {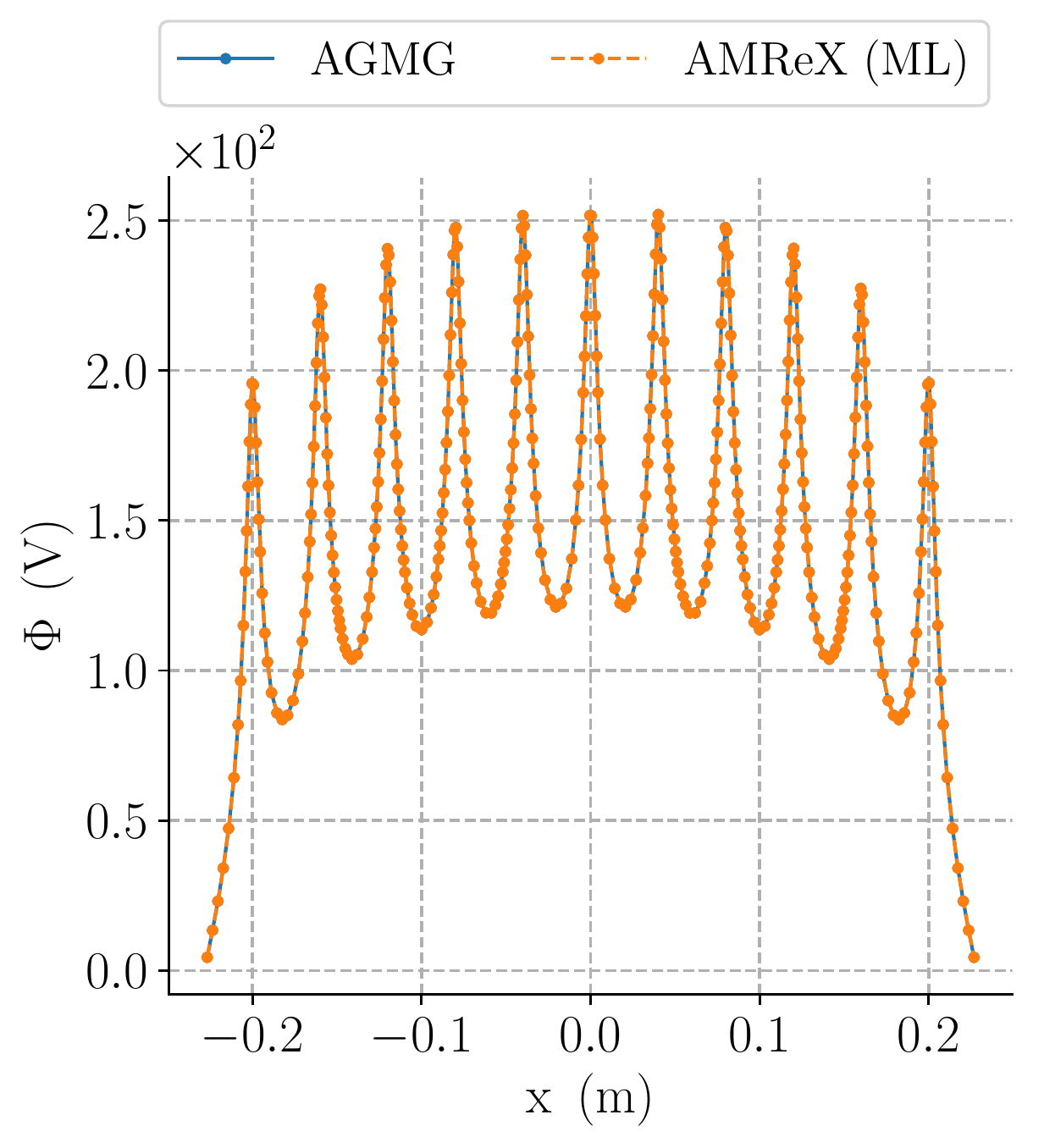}
            \caption{Electrostatic potential in $x$-direction.}
        \end{subfigure}
        \hfill
        \begin{subfigure}[t]{0.49\textwidth}
            \includegraphics[width=\textwidth]
                {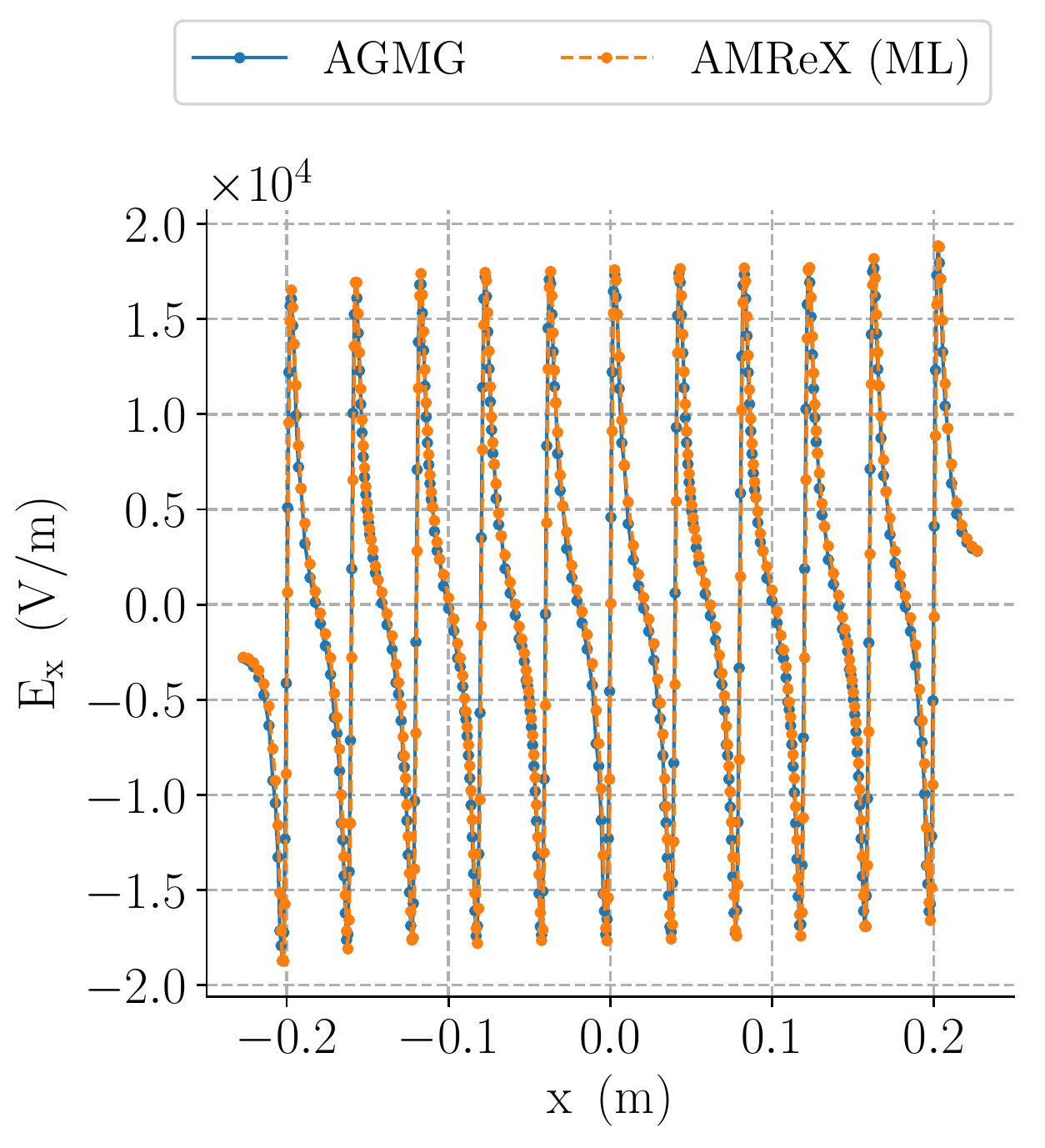}
            \caption{Electric field component in $x$-direction.}
        \end{subfigure}
        \caption{Line plots of the electrostatic potential and electric field of the multi-bunch
        test example.}
        \label{fig:lineplot}
    \end{figure}\FloatBarrier\parindent 0pt

    \section{Neighbouring Bunch Simulation}
    \label{sec:nbs}
    As initially stated the new AMR feature in \opal{} is mainly developed to study neighbouring bunch simulations
    (cf.\ \Figref{fig:tag_charge}) in high intensity cyclotrons \cite{PhysRevSTAB.13.064201}.\ This type of simulation
    injects
    a new particle bunch after every turn.\ The computation domain therefore increases over time, resulting in a
    decrease in resolution in regular PIC.\ To overcome this issue the domain must be extremely finely discretised, at the expense
    of a high memory consumption and a waste of computing resources in regions without particles.\
    In this section we illustrate the benefit of AMR over regular PIC w.r.t.\ memory and accuracy using a simplified
    model of the PSI Ring cyclotron.\
    The simulation integrates either 5, 7, 9 or 11 
    neighbouring bunches each with $10^5$ or $10^6$ particles over one turn using 360 steps.\ In AMR mode the charge density per cell
    is used as refinement criterion (cf.\ 
    \Secref{sec:amr_policies}) with cell threshold $\lambda = \SI{1}{nC/m^3}$.\ The AMR hierarchy is updated after every tenth 
    integration step.\ All simulations have an enlarged mesh of $\delta = \SI{20}{\percent}$ compared to the computation
    domain.\ Poisson's equation is solved in a box with dimension $[-1,1]\times[-0.75, 0.75]\times[-0.75, 0.75]$ to
    take into account the inhomogeneity of the problem.\ At its boundaries we apply Robin BC
    (cf.\ \Eqref{eq:robin_bc}) with $d=1.7$.\
    
    The results are compared to the single-level execution where we use the root mean square (rms) beam size, i.e.\
    \begin{equation}
        \sigma_w = \sqrt{\langle w^2\rangle}
        \label{eq:rms_sigma}
    \end{equation}
    and the beam-profile parameter \cite{Wangler}, which is a statistical measure to determine the
    proportion of halo particles in a beam, i.e.\
    \begin{equation}
        \xi_w = \frac{\langle w^4\rangle}{\langle w^2\rangle^2},
        \label{eq:halo}
    \end{equation}
    where $\langle w^n\rangle$ denotes the $n$-th moment of the particle distribution in coordinate $w\in\{x, y, z\}$.\
    In \Figref{fig:rms_5_bunches} and \Figref{fig:rms_11_bunches} are the rms beam sizes and in 
    \Figref{fig:halo_5_bunches} and \Figref{fig:halo_11_bunches} the beam-profile parameters of the centre bunch in
    a simulation of 5 and 11 adjacent bunches, respectively.\ The result of regular PIC with
    $512^3$ grid points is compared to two AMR simulations with either $64^3$ grid points on the coarsest level
    and three levels of refinement or $128^3$ grid points on the coarsest level and two levels of refinement.\ All three
    simulations have therefore the same resolution on the finest grid.\ The halo parameters and rms beam sizes have an absolute error
    below $\mathcal{O}(10^{-5})$ compared to the regular PIC model.\ As observed in \Figref{fig:memory} and
    \Tabref{tab:memory_usage}, however, the average resident set size (RSS),
    i.e.\ the amount of occupied physical memory, per MPI-process is on average at least four times smaller 
    with AMR than FFT PIC.\ All simulations ran with 36 MPI-processes.\
    
    Beside the memory benefit, AMR reduces also the time to solution as visualised in \Figref{fig:timing}.\ The detailed
    timing results of the Poisson solver and fourth order Runge-Kutta integration for 5 and 11 neighbouring bunches are shown in 
    \Tabref{tab:detailed_timing_1e5} and \Tabref{tab:detailed_timing_1e6}.\ As expected, the particle integration grows in
    proportion to the increase in particles per bunch.\ The timings indicate that possible particle load imbalances do not harm
    the performance of the AMR PIC models significantly since the computation of the potential and electric field consume at least
    \SI{87.7}{\percent} and \SI{63.2}{\percent} in case of $10^5$ and $10^6$ particles per bunch, respectively.\ Overall,
    the runtime of the shown AMR configurations is at least \SI{62.5}{\percent} shorter compared to FFT PIC. 
    
    \begin{table}[htp]
        \centering
        \def\p#1{~(\SI{#1}{\percent})~}
        \begin{tabular}{l
                        S[scientific-notation=false, round-mode=places, round-precision=0, table-format=5]@{}
                        S[round-mode=places, round-precision=1, table-format=2.1]@{}
                        S[round-mode=places, round-precision=0, table-format=5]@{}
                        S[round-mode=places, round-precision=1, table-format=2.1]@{}
                        S[round-mode=places, round-precision=1, table-format=2.1]@{}
                        S[round-mode=places, round-precision=1, table-format=2.1]@{}
                        S[round-mode=places, round-precision=1, table-format=2.1]@{}
                        S[round-mode=places, round-precision=1, table-format=2.1]@{}
                        S[round-mode=places, round-precision=0, table-format=5]@{}
                        S[round-mode=places, round-precision=0, table-format=5]}
            \toprule
            \bf PIC model       & \multicolumn{4}{c}{\bf Poisson timing (s)}
                                & \multicolumn{4}{c}{\bf RK-4 timing (s)}
                                & \multicolumn{2}{c}{\bf total timing (s)} \\
                                  \cmidrule(lr){2-5} \cmidrule(lr){6-9} \cmidrule(lr){10-11}
                                & \multicolumn{2}{c}{5 nbs} & \multicolumn{2}{c}{11 nbs}
                                & \multicolumn{2}{c}{5 nbs} & \multicolumn{2}{c}{11 nbs}
                                & \multicolumn{1}{c}{5 nbs} & \multicolumn{1}{c}{11 nbs} \\
                                \midrule
            Amr-64          & 1103  & \p{95.33}  & 762.1 & \p{87.698} & 7.03  & \p{0.605}   & 15.48 & \p{1.783}   & 1157 & 868.8 \\
            Amr-128         & 1659  & \p{96.79}  & 1296  & \p{91.98}  & 8.53  & \p{0.49591} & 18.47 & \p{1.312}   & 1714 & 1409 \\
            Uniform-512     & 20420 & \p{99.66}  & 19100 & \p{99.117} & 33.02 & \p{0.1610}  & 76.31 & \p{0.3959}  & 20490 & 19270 \\
            FFT-512         & 9325  & \p{98.157} & 9142  & \p{97.827} & 5.35  & \p{0.0568}  & 11.75 & \p{0.12627} & 9500 & 9345 \\ 
            \bottomrule
        \end{tabular}
        \caption{Detailed timing results (max.\ CPU time) of the Poisson solver and time integration with fourth order
        Runge-Kutta (RK-4) for 5 and 11 neighbouring bunches (nbs) of $10^5$ macro particles each.\ The percentages are
        w.r.t.\ the total runtimes shown in the last two columns.}
        \label{tab:detailed_timing_1e5}
    \end{table}
    
    \begin{table}[htp]
        \centering
        \def\p#1{~(\SI{#1}{\percent})~}
        \begin{tabular}{l
                        S[round-mode=places, round-precision=0, table-format=5]@{}
                        S[round-mode=places, round-precision=1, table-format=2.1]@{}
                        S[round-mode=places, round-precision=0, table-format=5]@{}
                        S[round-mode=places, round-precision=1, table-format=2.1]@{}
                        S[round-mode=places, round-precision=1, table-format=3.1]@{}
                        S[round-mode=places, round-precision=1, table-format=2.1]@{}
                        S[round-mode=places, round-precision=1, table-format=3.1]@{}
                        S[round-mode=places, round-precision=1, table-format=2.1]@{}
                        S[round-mode=places, round-precision=0, table-format=5]@{}
                        S[round-mode=places, round-precision=0, table-format=5]}
            \toprule
            \bf PIC model       & \multicolumn{4}{c}{\bf Poisson timing (s)}
                                & \multicolumn{4}{c}{\bf RK-4 timing (s)}
                                & \multicolumn{2}{c}{\bf total timing (s)} \\
                                  \cmidrule(lr){2-5} \cmidrule(lr){6-9} \cmidrule(lr){10-11}
                                & \multicolumn{2}{c}{5 nbs} & \multicolumn{2}{c}{11 nbs}
                                & \multicolumn{2}{c}{5 nbs} & \multicolumn{2}{c}{11 nbs}
                                & \multicolumn{1}{c}{5 nbs} & \multicolumn{1}{c}{11 nbs} \\
                                \midrule
            Amr-64          & 1777  & \p{76.926} & 2225  & \p{63.936} & 73.89 & \p{3.199} & 164.7 & \p{4.732} & 2310 & 3480 \\
            Amr-128         & 2070  & \p{78.290} & 2300  & \p{63.221} & 71.5  & \p{2.704} & 161.9 & \p{4.450} & 2644 & 3638 \\
            Uniform-512     & 20750 & \p{96.332} & 19240 & \p{90.159} & 334.3 & \p{0.155} & 765.8 & \p{3.588} & 21540 & 21340 \\
            FFT-512         & 8978  & \p{96.093} & 9032  & \p{92.998} & 52.72 & \p{0.564} & 118.0 & \p{1.214} & 9343 & 9712 \\ 
            \bottomrule
        \end{tabular}
        \caption{Detailed timing results (max.\ CPU time) of the Poisson solver and time integration with fourth order
        Runge-Kutta (RK-4)
        for 5 and 11 neighbouring bunches (nbs) of $10^6$ macro particles each.\ The percentages are
        w.r.t.\ the total runtimes shown in the last two columns.}
        \label{tab:detailed_timing_1e6}
    \end{table}
    
    \begin{table}[htp]
        \centering
        \begin{tabular}{l
                        S[round-mode=places, round-precision=4, table-format=1.4]
                        S[round-mode=places, round-precision=4, table-format=1.4]
                        S[round-mode=places, round-precision=4, table-format=1.4]
                        S[round-mode=places, round-precision=4, table-format=1.4]}
            \toprule
            \bf PIC model       & \multicolumn{2}{c}{\bf Avg.\ RSS with 5 nbs (GiB)}
                                & \multicolumn{2}{c}{\bf Avg.\ RSS with 11 nbs (GiB)} \\
                                  \cmidrule(lr){2-3} \cmidrule(lr){4-5}
                                & \multicolumn{1}{c}{$10^5$ ppb} & \multicolumn{1}{c}{$10^6$ ppb}
                                & \multicolumn{1}{c}{$10^5$ ppb} & \multicolumn{1}{c}{$10^6$ ppb} \\
                                \midrule
            Amr-64          & 0.282870695326 & 0.438596828779 & 0.250065981017 & 0.568327318297 \\
            Amr-128         & 0.321462098757 & 0.459891067611 & 0.284363561206 & 0.590016651154 \\
            Uniform-512     & 4.05242522558  & 4.13070490625 & 4.05336374707 & 4.19315369924 \\
            FFT-512         & 2.15717821651  & 2.28764022721 & 2.17569025358 & 2.38899207115 \\
            \bottomrule
        \end{tabular}
        \caption{Average resident size (RSS) in Gibibyte (GiB) per MPI-process over all 360 integration
        steps with 5 or 11 neighbouring bunches (nbs) and $10^5$ or $10^6$ macro particles per bunch (ppb).}
        \label{tab:memory_usage}
    \end{table}
    
    The particle load balancing is quantified as the average number of particles per MPI-process $\langle N_p\rangle_{s}$ over all
    integration steps $s$ divided by the total number of particles in simulation $N_{t}$, i.e.\
    \begin{equation*}
        \frac{\langle N_p \rangle_{s}}{N_t}.
    \end{equation*}
    In the best case all MPI-processes have $N_t / P_t$ particles during integration where $P_t$ is the total number of
    processes.\ \Figref{fig:particle_lbal_1e5} and \Figref{fig:particle_lbal_1e6} show the number of cores
    that deviate from the optimum particle count within a few percent.\ The load balancing between $10^5$ and $10^6$ particles per 
    bunch does not differ significantly.\ A similar observation is done in \Figref{fig:grid_lbal_1e5} and \Figref{fig:grid_lbal_1e6}
    where the optimal number of grid points among the MPI-processes is evaluated.
    
    \begin{figure}[htp]
        \centering
        \includegraphics[width=\textwidth]{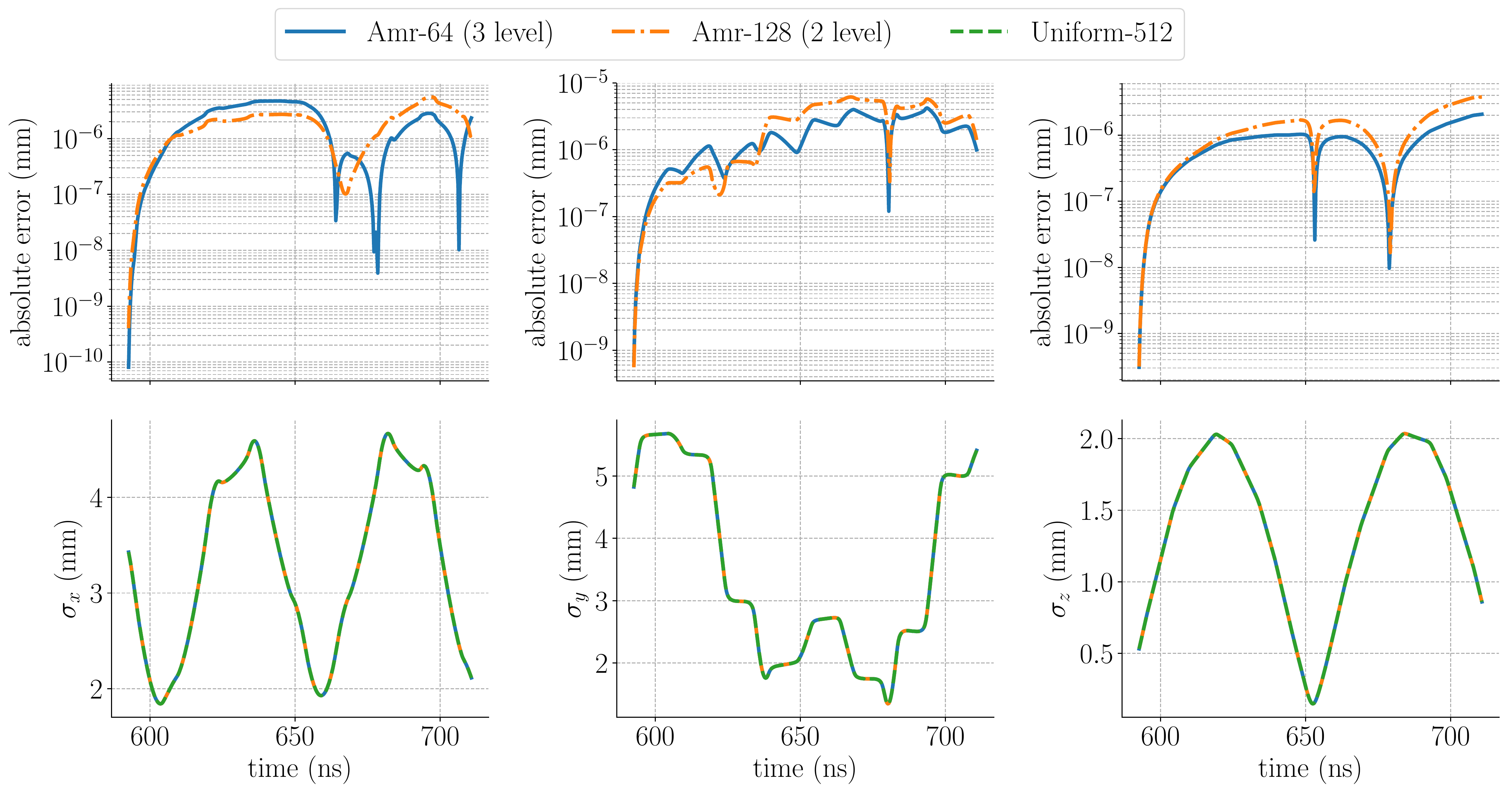}
        \caption{Evolution of the rms beam size (cf.\ \Eqref{eq:rms_sigma}) of the centre bunch in a simulation of
        5 adjacent bunches and
        the absolute error of AMR models to the reference simulation with uniform mesh of $512^3$ grid points (Uniform-512).\
        On the finest level all three simulations have the same mesh resolution.}
        \label{fig:rms_5_bunches}
    \end{figure}
    
    \begin{figure}[htp]
        \centering
        \includegraphics[width=\textwidth]{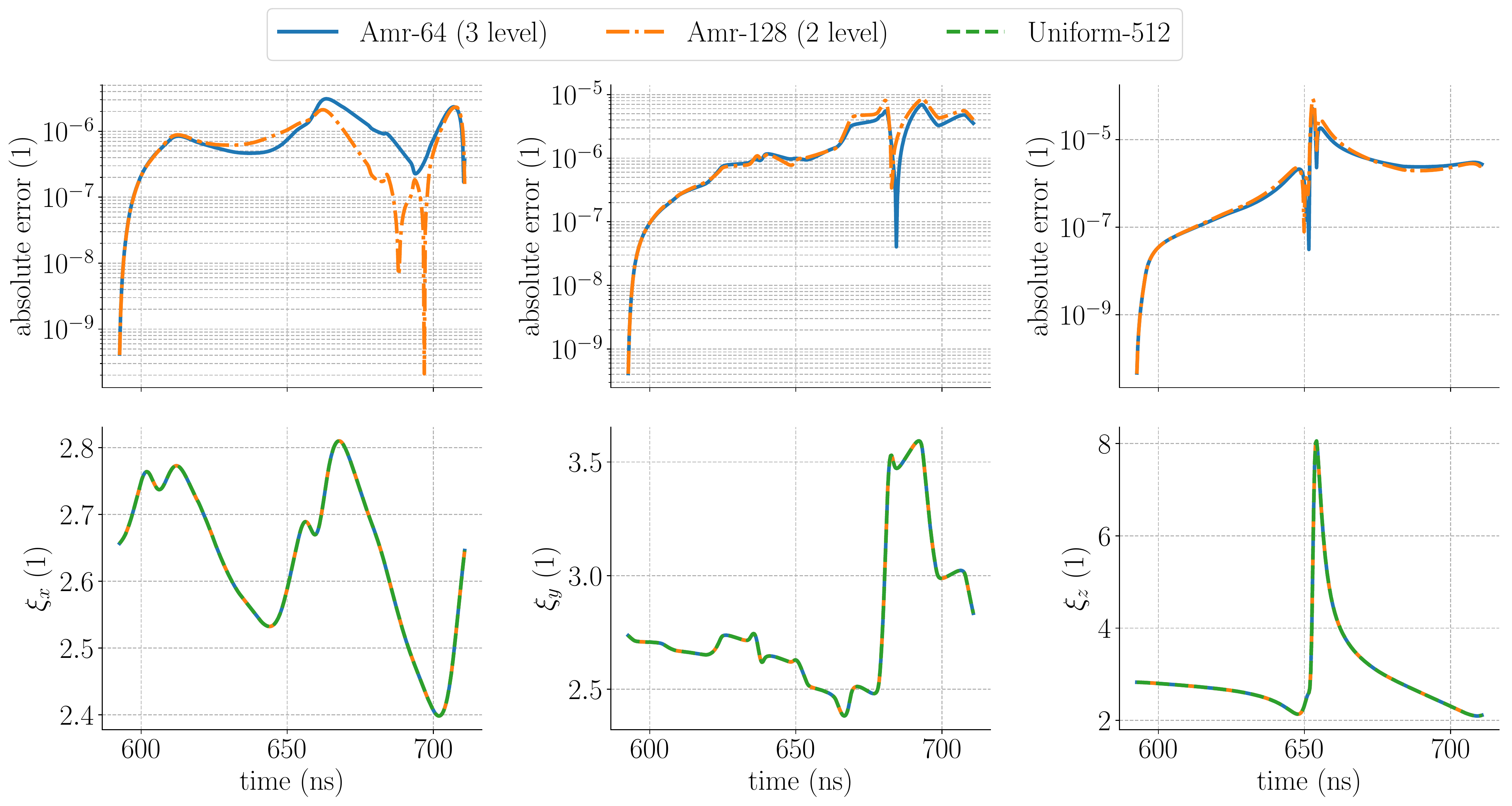}
        \caption{Evolution of the beam-profile parameters (cf.\ \Eqref{eq:halo}) of the centre bunch in a simulation of 5 adjacent 
        bunches and the absolute error of AMR models to the reference simulation with uniform mesh of $512^3$ grid points
        (Uniform-512).\ On the finest level all three simulations have the same mesh resolution.}
        \label{fig:halo_5_bunches}
    \end{figure}
    
    \begin{figure}[htp]
        \centering
        \includegraphics[width=\textwidth]{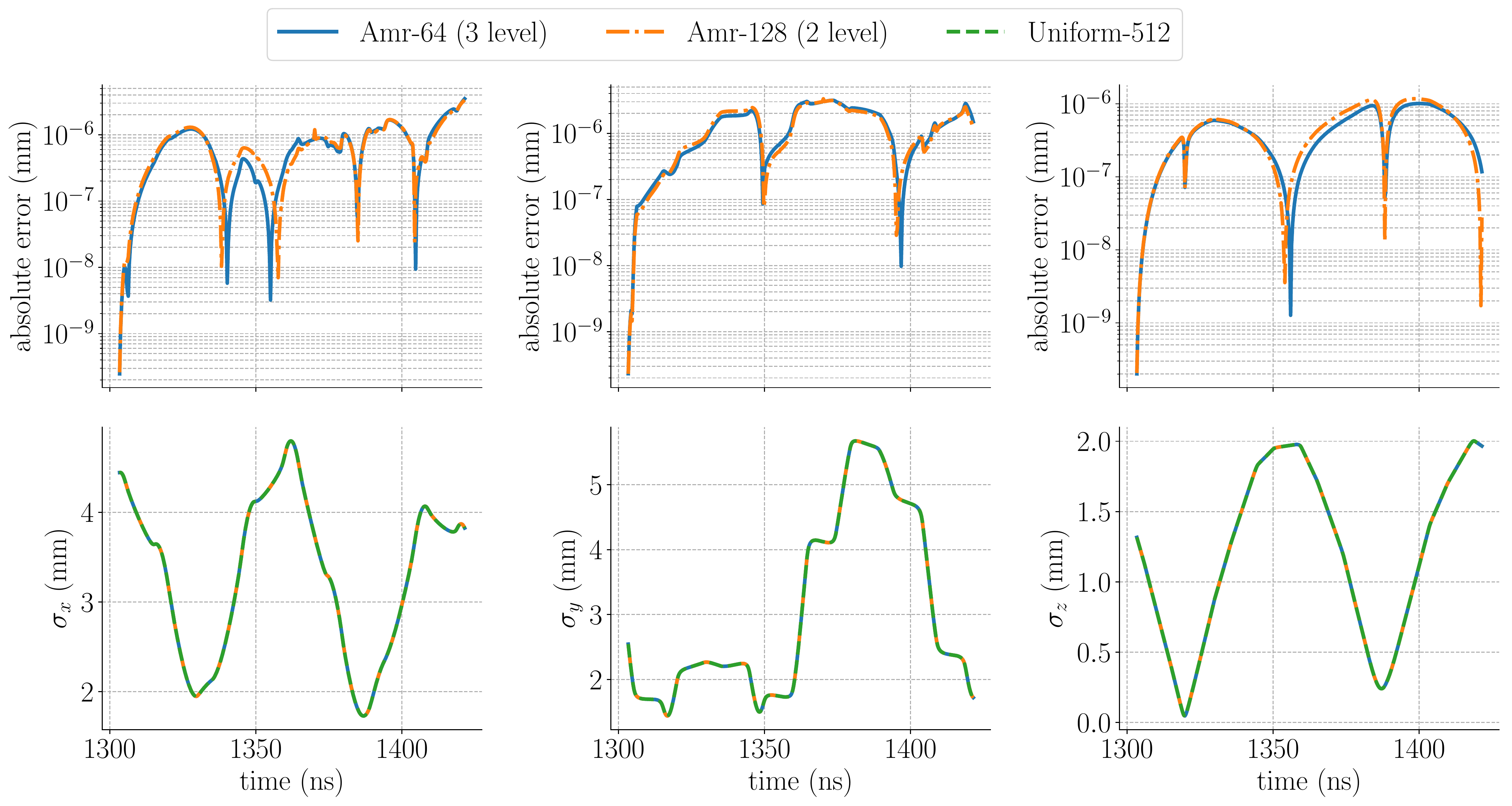}
        \caption{Evolution of the rms beam size (cf.\ \Eqref{eq:rms_sigma}) of the centre bunch in a simulation of 11
        adjacent bunches and
        the absolute error of AMR models to the reference simulation with uniform mesh of $512^3$ grid points (Uniform-512).\
        On the finest level all three simulations have the same mesh resolution.}
        \label{fig:rms_11_bunches}
    \end{figure}
    
    \begin{figure}[htp]
        \centering
        \includegraphics[width=\textwidth]{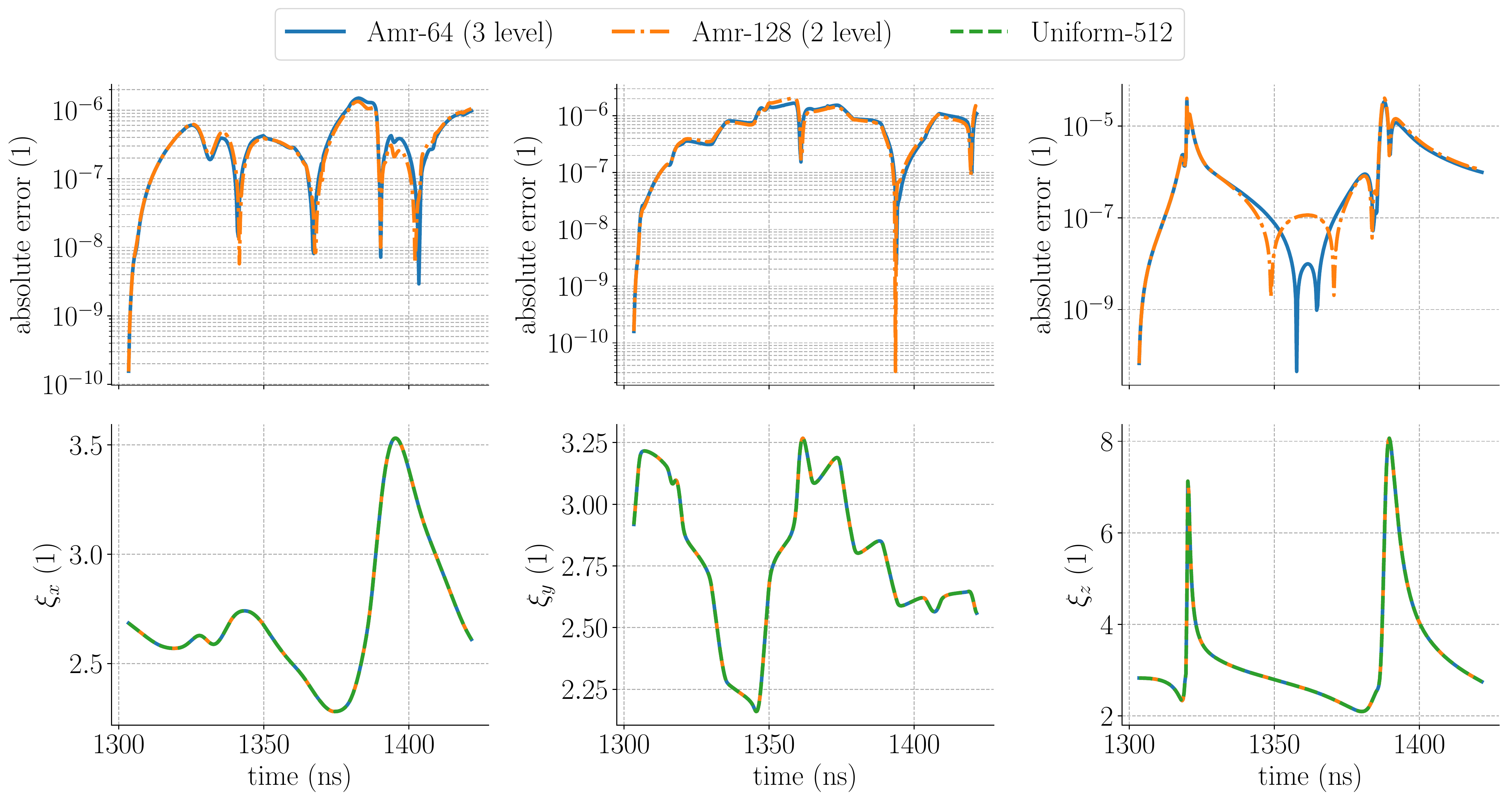}
        \caption{Evolution of the beam-profile parameters (cf.\ \Eqref{eq:halo})
        of the centre bunch in a simulation of 11 adjacent bunches and
        the absolute error of AMR models to the reference simulation with uniform mesh of $512^3$ grid points (Uniform-512).\
        On the finest level all three simulations have the same mesh resolution.}
        \label{fig:halo_11_bunches}
    \end{figure}
    
    \begin{figure}[htp]
        \centering
        \includegraphics[width=\textwidth]{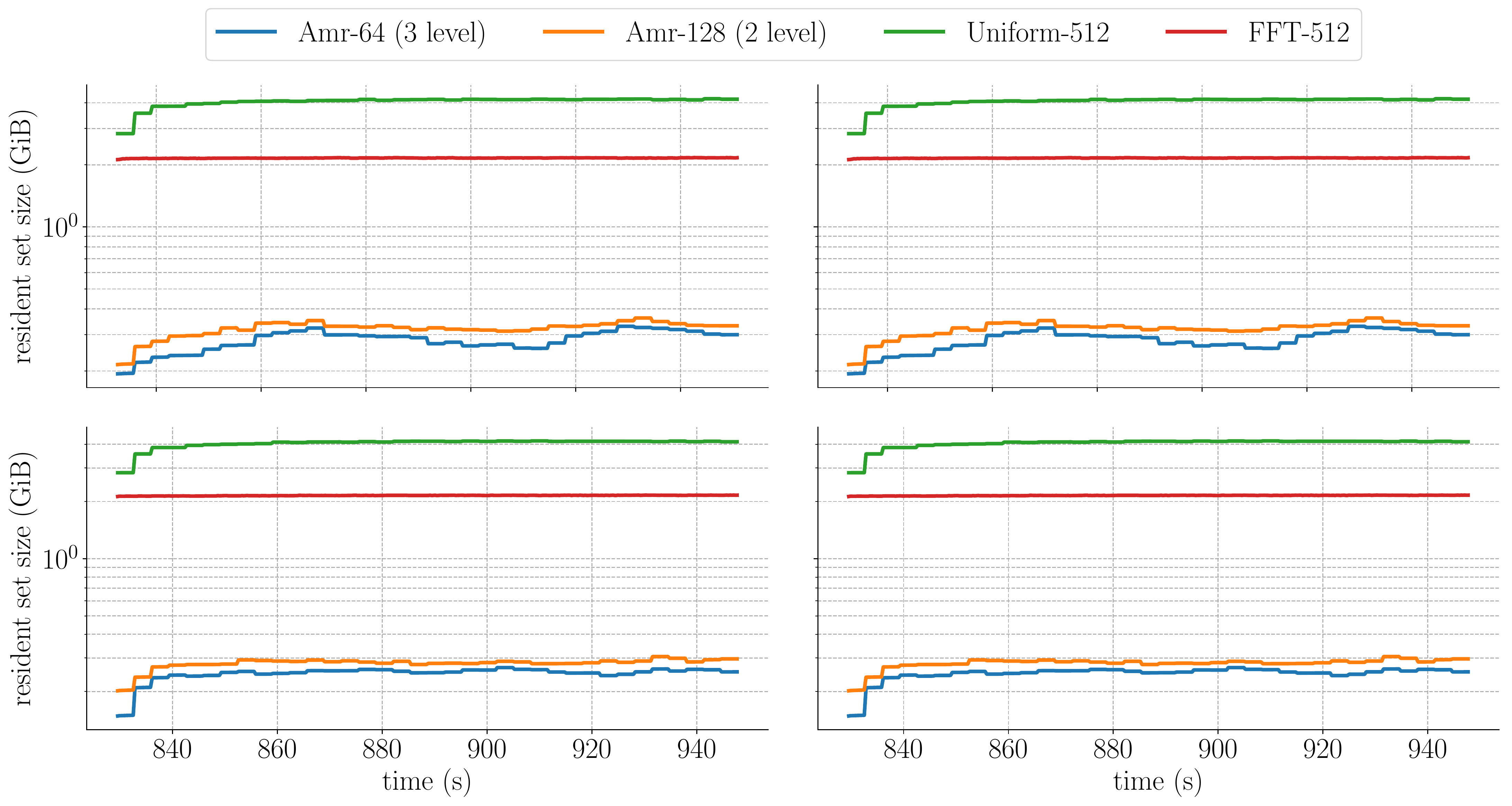}
        \caption{Average resident set size (RSS) in Gibibyte (GiB) per MPI-process with 5 (top left), 7 (top right), 9
        (bottom left) and 11 (bottom right) neighbouring bunches.\ Each bunch consists of $10^5$ macro particles.\
        All simulations were run with $36$ MPI-processes.}
        \label{fig:memory}
    \end{figure}
    
    \begin{figure}[htp]
        \centering
        \includegraphics[width=\textwidth]{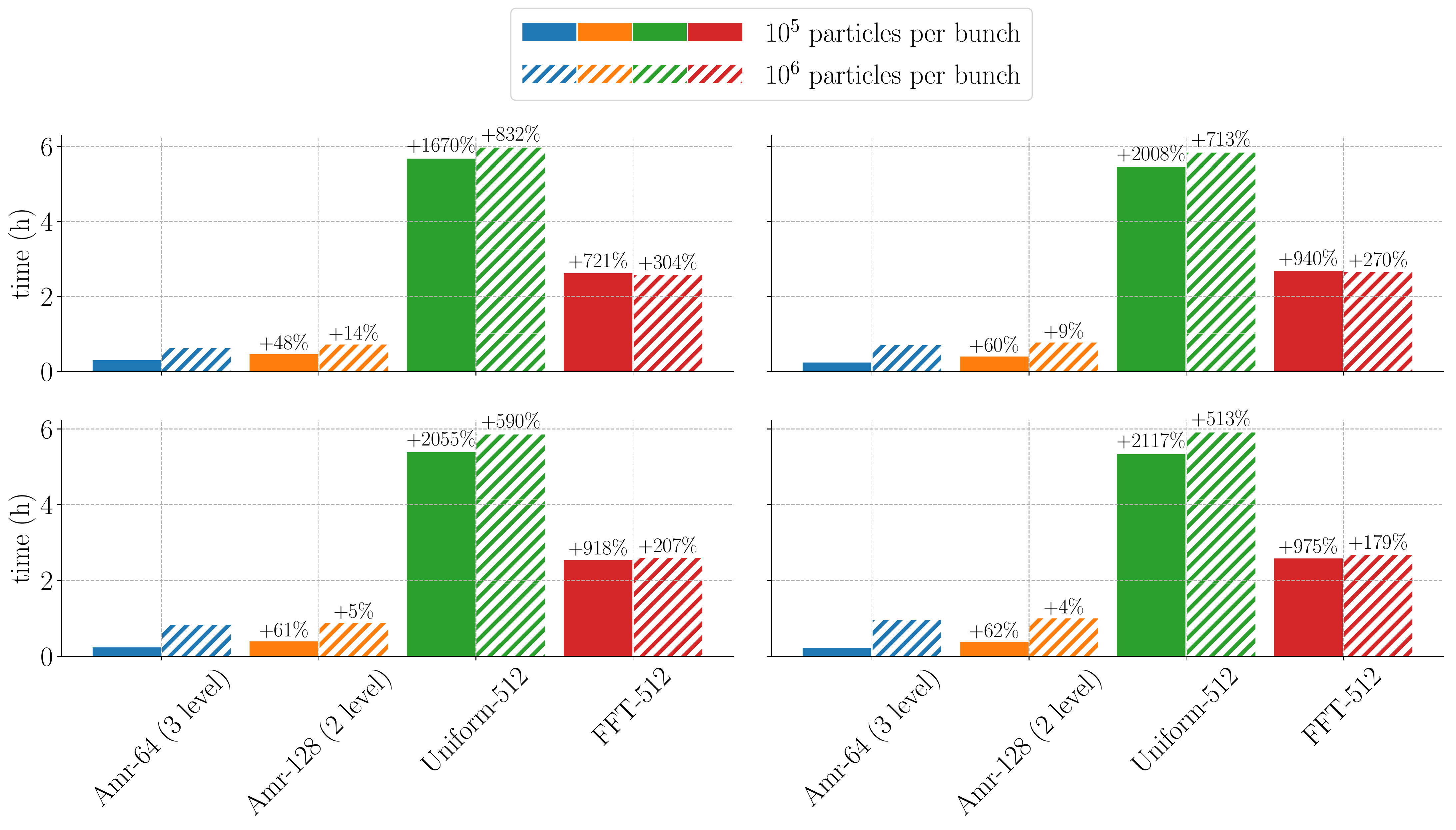}
        \caption{Total simulation CPU time with 5 (top left), 7 (top right), 9
        (bottom left) and 11 (bottom right) neighbouring bunches.\ A bunch consists either of $10^5$ or $10^6$
        macro particles.\ All simulations were run with $36$ MPI-processes.\ The percentages on top of the bars
        are w.r.t.\ the Amr-64 (3 level) timings.}
        \label{fig:timing}
    \end{figure}
    
    \begin{figure}[htp]
        \centering
        \includegraphics[width=\textwidth]{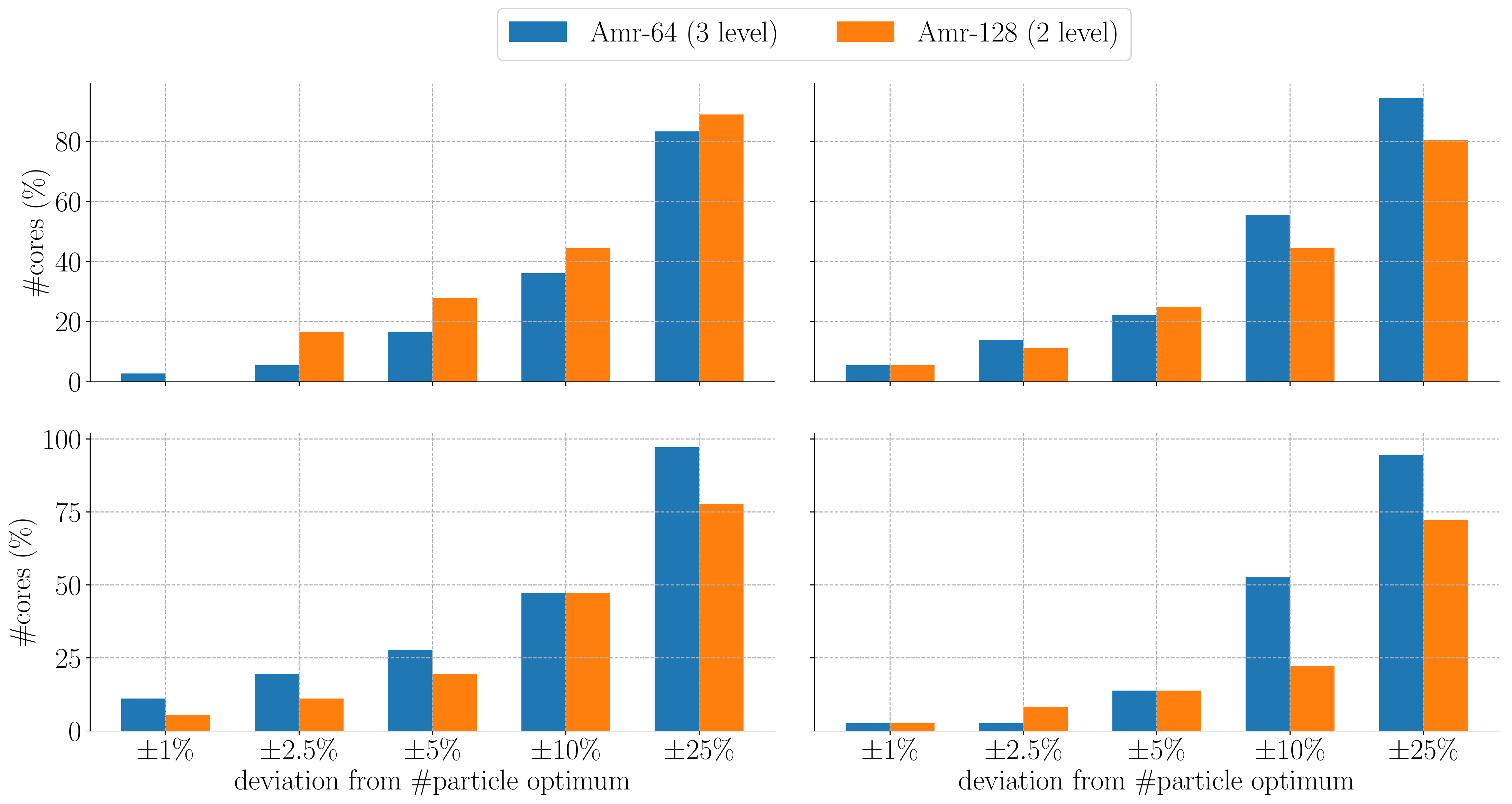}
        \caption{Particle load balancing for 36 MPI-processes and 5 (top left), 7 (top right), 9 (bottom left) or 11 (bottom right)
        neighbouring bunches.\ Each bunch has $10^5$ macro particles.\ The optimum is evaluated as the total number of particles 
        divided by the number of MPI-processes.}
        \label{fig:particle_lbal_1e5}
    \end{figure}
    
    \begin{figure}[htp]
        \centering
        \includegraphics[width=\textwidth]{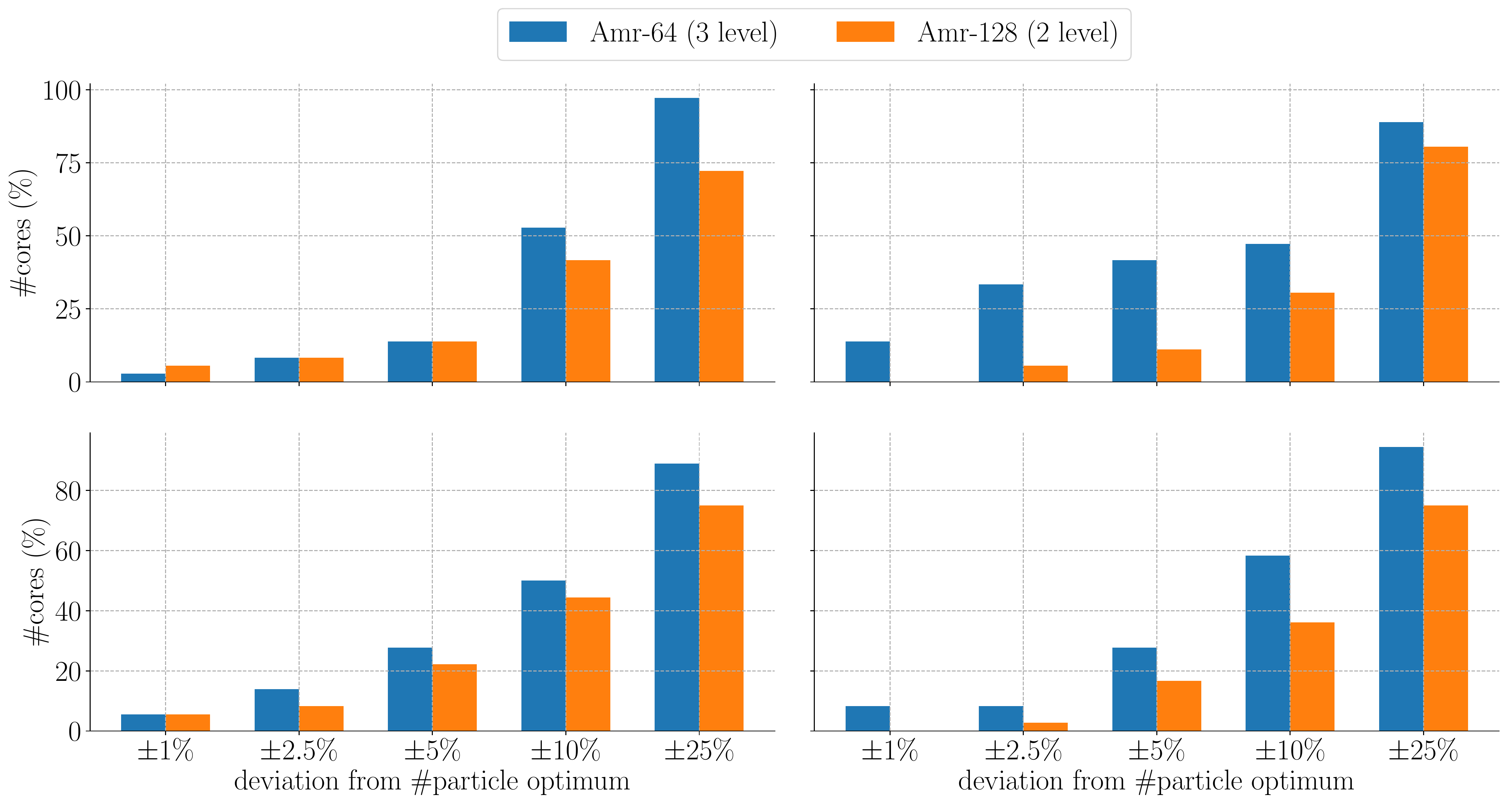}
        \caption{Particle load balancing for 36 MPI-processes and 5 (top left), 7 (top right), 9 (bottom left) or 11 (bottom right)
        neighbouring bunches.\ Each bunch has $10^6$ macro particles.\ The optimum is evaluated as the total number of particles 
        divided by the number of MPI-processes.}
        \label{fig:particle_lbal_1e6}
    \end{figure}
    
    \begin{figure}[htp]
        \centering
        \includegraphics[width=\textwidth]{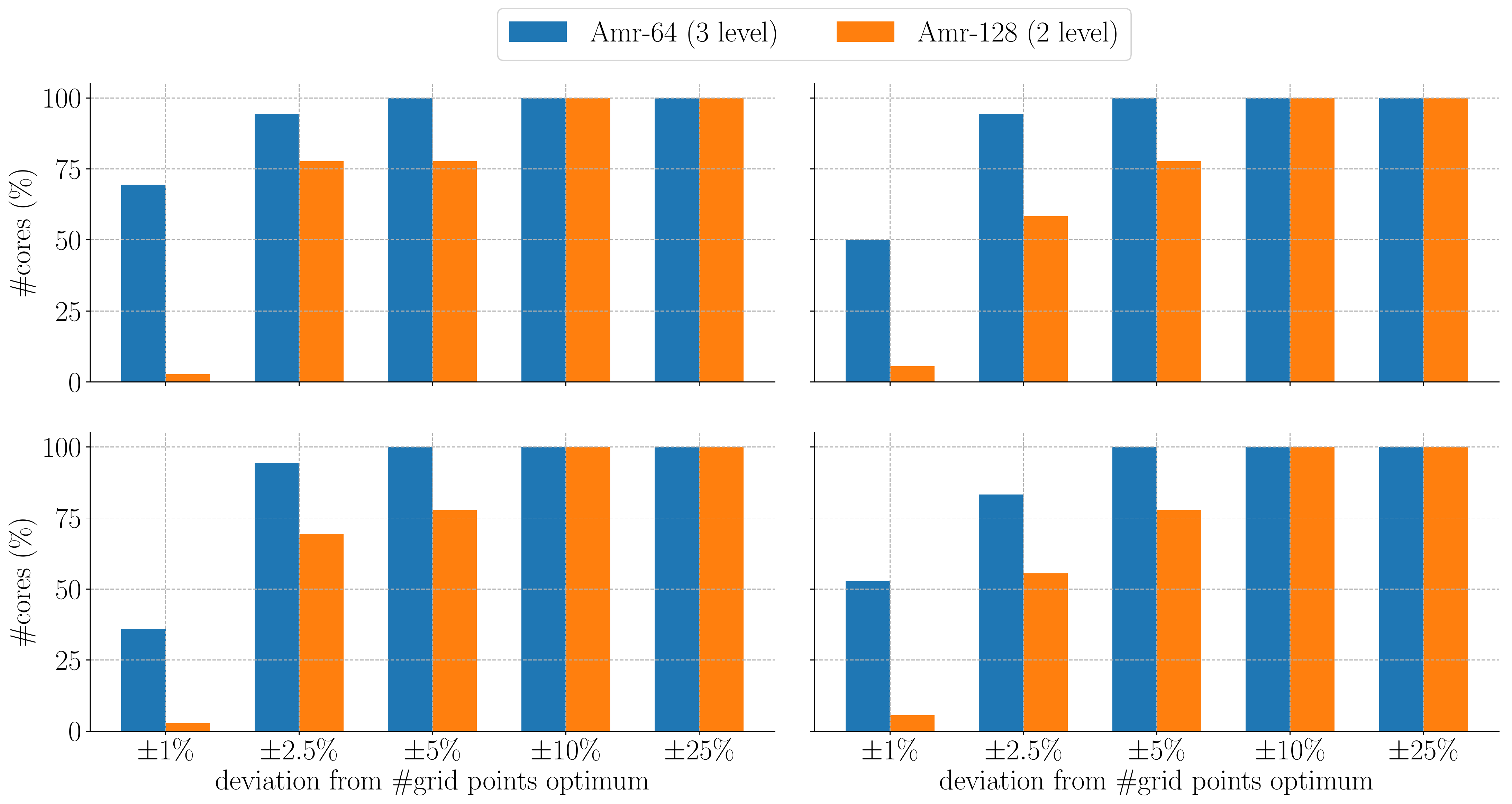}
        \caption{Grid point load balancing for 36 MPI-processes and 5 (top left), 7 (top right), 9 (bottom left) or 11
        (bottom right)
        neighbouring bunches.\ Each bunch has $10^5$ macro particles.\ The optimum is evaluated as the total number of grid points
        per step divided by the number of MPI-processes.}
        \label{fig:grid_lbal_1e5}
    \end{figure}
    
    \begin{figure}[htp]
        \centering
        \includegraphics[width=\textwidth]{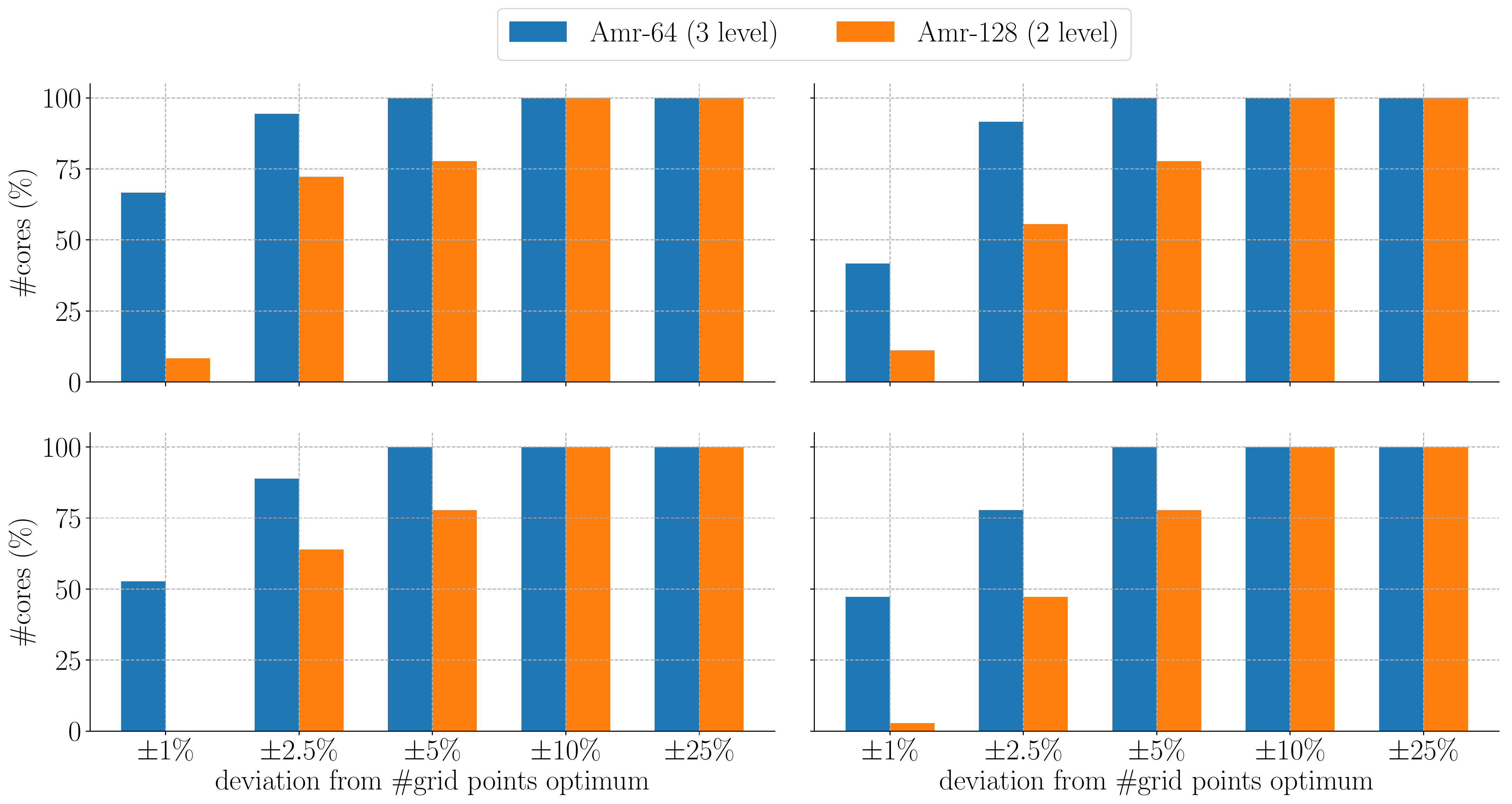}
        \caption{Grid point load balancing for 36 MPI-processes and 5 (top left), 7 (top right), 9 (bottom left) or 11 (bottom 
        right)
        neighbouring bunches.\ Each bunch has $10^6$ macro particles.\ The optimum is evaluated as the total number of grid points
        per step divided by the number of MPI-processes.}
        \label{fig:grid_lbal_1e6}
    \end{figure}

    \section{Performance Benchmark}
    \label{sec:scaling}
    The performance benchmark is done on the multicore partition of Piz Daint, a supercomputer at the Swiss National 
    Supercomputing Centre (CSCS).\ The nodes on the multicore partition consist of two Intel Xeon E5-2695 v4 $@ 2.10\ \si{GHz}$ ($2 
    \times 18$ cores, 64/128 GB RAM) processors \cite{PizDaint}.\
    The benchmark on the GPU partition of Piz Daint confirmed the hardware portability of the new solver.\ However, the data
    transfer between CPU (Central Processing Unit) and GPU as well as the launching of single GPU kernels for
    each matrix-vector or matrix-matrix operation of \tpetra{} showed a performance bottleneck which is why the performance study
    presents a CPU benchmark only.\
    
    The test initialises 11 Gaussian-shaped bunches as described in \Secref{sec:N_Gaussian_Shaped_Bunches} with
    $10^{6}$ macro particles of charge \SI{0.1}{fC} per bunch.\ The Poisson problem is solved 100 times on a three level
    hierarchy (two 
    levels of refinement) with $576^3$ grid points on level zero.\ The particles are randomly displaced within
    $\left[-10^{-3}, 10^{-3}\right]$ after every iteration.\ This represents a
    realistic setup for beam dynamics simulations since the particle distribution in the bunch rest frame changes only marginally
    from one integration time step to another.\ Therefore, it is not necessary to re-mesh the AMR hierarchy and thus rebuild the
    matrices after every time step which gives rise to computational savings.\ The optimal update frequency of the grids
    for neighbouring bunch simulations is currently unknown and is not subject in this article.\ Nevertheless, the computational
    saving is shown with two strong scalings.\ The first benchmark updates the AMR hierarchy after every computation of the
    electrostatic potential while the latter performs a regrid step after every tenth step.\ Since a constant workload per
    MPI-process during an upscaling that is necessary in a fair weak scaling can't be guaranteed, the presented benchmark consists
    of a strong scaling only.\
    
    The blue line in \Figref{fig:strong_scaling} shows the total solver time of the 100 executions.\ As indicated in
    \Tabref{tab:summarised_timing_regrid_100}, the setup of the matrices (violet line), i.e.\ porting the \amrex{} mesh information to 
    \trilinos{}, as well as the evaluation of the linear system of equations on the bottom level (grey line) with the algebraic 
    multigrid solver of \muelu{} consume together more than $77\ \%$ of the time on \num{14400} cores.\ However, the setup time can 
    easily be reduced with a lower regrid frequency as previously mentioned.\ The matrix setup cost in
    the second timing is only $14\ \%$ of the setup cost observed by the first timing.\ Furthermore, the use of an algebraic
    multrigrid solver for the linear system of equations on the bottom level is not an optimal
    choice.\ More suitable would be a geometric multigrid that keeps the structure of the problem which is
    planned for a future paper.\
    
    The parallel efficiency of the strong scaling of \Figref{fig:strong_scaling} is shown in 
    \Figref{fig:parallel_efficiency}.\ The efficiency of the total solve time (blue line) drops below
    \SI{50}{\percent} for 120 or 160 computing nodes.\ In case the AMR hierarchy is updated after every solve, the efficiency is 
    dominated by the bottom solver and the
    matrix setup time.\ However, reducing the regrid frequency shifts the dependency towards the bottom solver.\
    For both regriding configurations we observe an increase in efficiency in case of 400 nodes.\
    Since the maximum number of grid points per dimension on level zero is set to $24$, all cores have
    the same amount of grid points on this level with \num{13824} cores (i.e.\ 384 nodes) that causes the bottom solver to be more 
    efficient.
    
    \begin{table}[!ht]
        \centering
        \begin{tabular}{l
                        S[round-mode=places, round-precision=2, table-format=3.2]
                        S[round-mode=places, round-precision=2, table-format=3.2]
                        S[round-mode=places, round-precision=2, table-format=3.2]
                        S[round-mode=places, round-precision=2, table-format=3.2]}
            \toprule
            \multirow{1}{*}{\bf timing} & \bf CPU avg (s) & \bf fraction ($\boldmath\%$)
                                        & \bf CPU avg (s) & \bf fraction ($\boldmath\%$) \\
                                       & \multicolumn{1}{c}{$100\times$ regriding} & & \multicolumn{1}{c}{$10\times$ regriding} & \\
            \midrule
            total solve         & 378.715     & 100.00   & 343.66     & 100.00 \\
            \midrule
            bottom solver       & 133.792     & 35.3278  & 110.085    & 32.0331 \\
            matrix setup        & 159.633     & 42.1512  &  22.4917   &  6.5447 \\
            smoothing           &  23.2338    &  6.1349  &  17.5876   &  5.1177 \\
            restriction         &  10.4939    &  2.7709  &   7.5509   &  2.1972 \\
            bottom solver setup &   2.26424   &  0.5978  &   1.4371   &  0.4181 \\
            prolongation        &   2.08805   &  0.5513  &   1.59763  &  0.4648 \\
            E-field             &   0.670398  &  0.1770  &   0.610209 &  0.1775 \\
            \midrule
            others              &  46.539612  & 12.2888  &  58.809861 & 53.0469 \\
            \bottomrule
        \end{tabular}
        \caption{Summarised AGMG timings solving Poisson's equation 100 times on \num{14400} cores (400 nodes).\
        It shows the timing results of two configurations.\ The first updates the grids after every ($100\times$ regriding) and the
        second after every tenth ($10\times$ regriding) computation.}
        \label{tab:summarised_timing_regrid_100}
    \end{table}\FloatBarrier\parindent 0pt

    \begin{figure}[!ht]
        \centering
        \includegraphics[width=\textwidth]{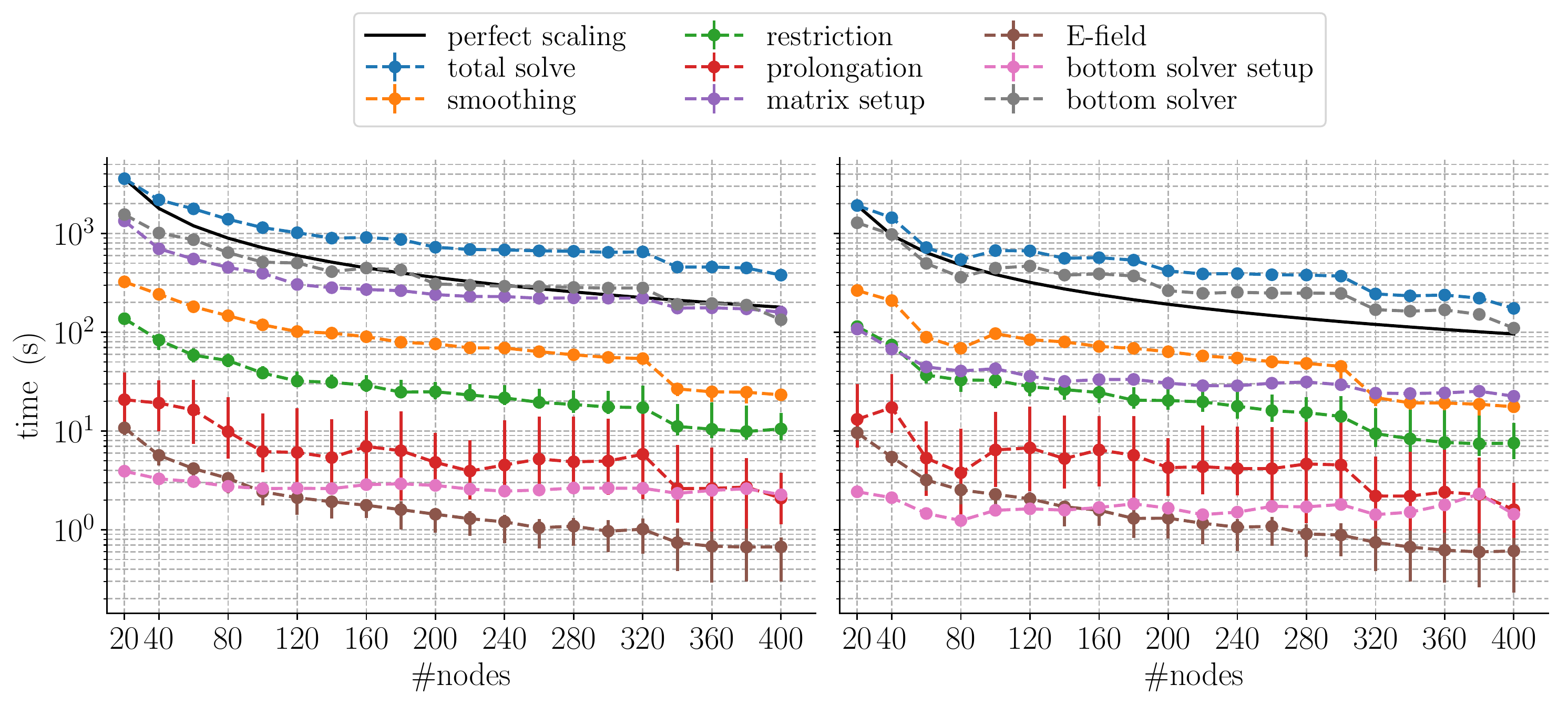}
        \caption{Strong scaling performed on the multicore partition of Piz Daint (Cray XC40)
        with 36 cores per node (without hyperthreading).\ The perfect scaling (black line) uses the total solve time
        with 20 nodes as reference.\ Left: scaling with $100\times$ regriding; right: scaling with $10\times$ regriding.\
        Each marker indicates the average CPU time per operation where the vertical line denotes
        the range by minimum and maximum.}
        \label{fig:strong_scaling}
    \end{figure}\FloatBarrier\parindent 0pt
    
    \begin{figure}[!ht]
        \centering
        \includegraphics[width=\textwidth]{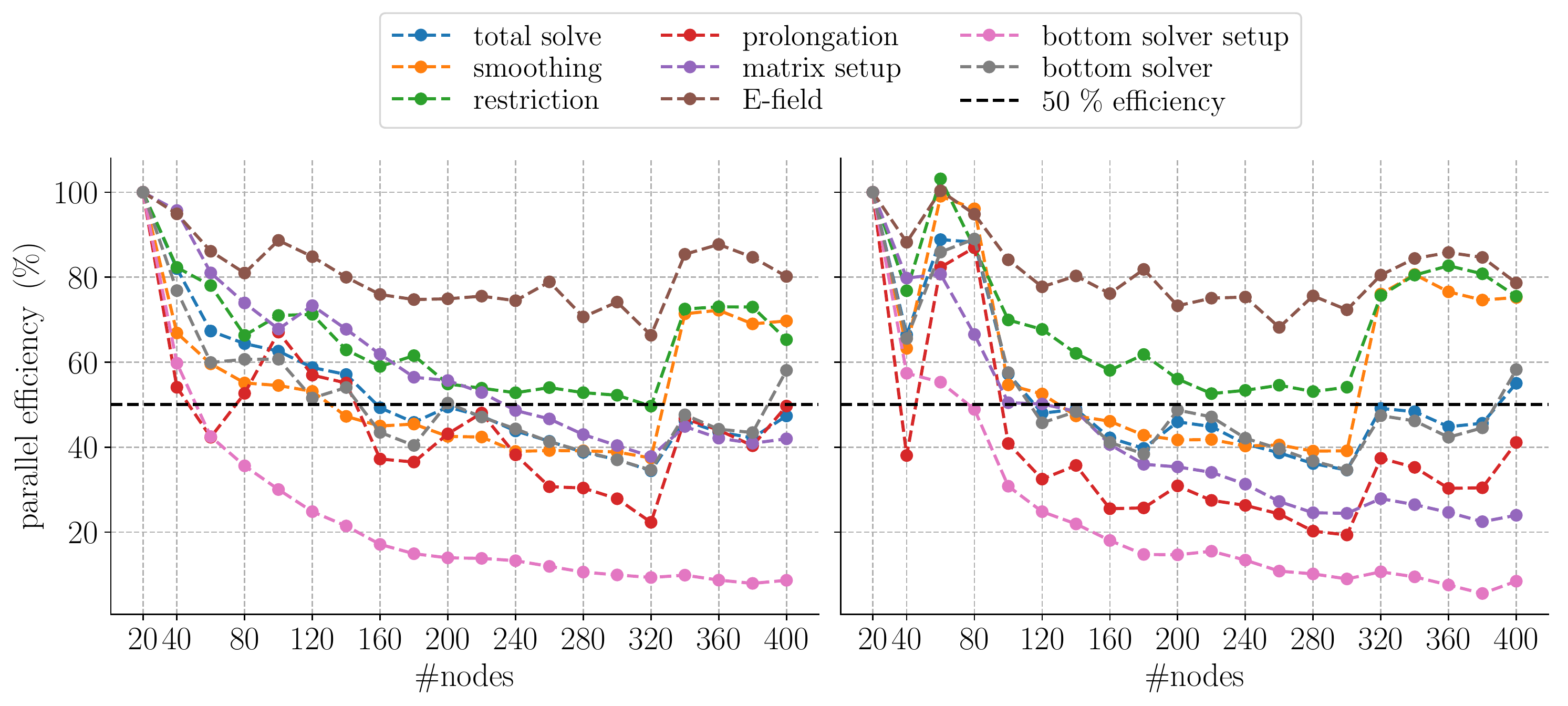}
        \caption{Parallel efficiency.\ The increase of efficiency from 380 to 400 nodes is due
        to an optimal workload on the coarsest level with \num{13824} cores.\
        Left: efficiency with $100\times$ regriding; right: efficiency with $10\times$ regriding.}
        \label{fig:parallel_efficiency}
    \end{figure}\FloatBarrier\parindent 0pt
    
    \section{Conclusion and Outlook}
    \label{sec:conclusion}
    In this article we presented the new adaptive mesh refinement capability of the open-source beam dynamics code \opal{} which has 
    been enhanced by \amrex{}.\ The new feature is supplemented with a hardware architecture independent implementation of a multigrid 
    Poisson solver based on second generation \trilinos{} packages.\ Beside an artificial problem illustrating symmetry
    preservation and a comparison with an analytically solvable problem, the Poisson solver was validated with the built-in \amrex{}
    multi-level solver.\ Although the structure of the mesh is lost when going to the matrix representation,
    the solver shows good scalability on CPUs with a parallel efficiency between \SI{50}{\percent} and \SI{60}{\percent} on
    14'400 cores depending on the AMR regrid frequency.\ The timings indicate that the matrix setup
    and the bottom linear system solver require \SI{77}{\percent} of the total solver time.\ The former
    can be reduced by updating the mesh less frequently.\ The latter might be decreased by replacing the smoothed aggregation
    algebraic multigrid solver of \muelu{} with a structured aggregation procedure, a real geometric multigrid solver or a
    FFT solver which is subject to future research.\
    Thanks to the hardware portability the solver runs on any backend that is supported by \kokkos{}.\ However, due to
    single kernel launches for each matrix-vector operation, the solver is not yet competitive on GPUs.
    
    A small example of the PSI Ring cyclotron demonstrated the benefit of AMR over regular PIC models w.r.t.\
    time to solution and memory consumption at a given accuracy.\
    The presented benchmark shows that AMR requires about four times less memory and the time to solution is at least
    \SI{62.5}{\percent} times shorter than a comparable simulation with the integrated FFT solver of \opal.\ Therefore, the technique 
    of 
    adaptive mesh
    refinement will enable large-scale multi-bunch simulations in high 
    intensity cyclotrons at higher grid resolution in order to more accurately quantify the effect of radially neighbouring
    bunches on halo formation and evolution.

    \section*{Acknowledgment}
    Many thanks to the Center for Computational Sciences and Engineering (CCSE) at Lawrence Berkeley National Laboratory (LBNL), in 
    particular A.\ S.\ Almgren, A.\ Myers and W.\ Zhang.\ The authors appreciate also the support of P.\ Arbenz, Dan.\ F.\ Martin,
    K.~D.~Devine and C.~Siefert in the development of the multigrid solver.\ Furthermore, we'd like to thank the Swiss National 
    Supercomputing Centre (CSCS) for providing the necessary computer time on Piz Daint.\ This project is funded by the Swiss National
    Science Foundation (SNSF) under contract number 200021\_159936.
    

\begin{thebibliography}{35}
\expandafter\ifx\csname natexlab\endcsname\relax\def\natexlab#1{#1}\fi
\providecommand{\url}[1]{\texttt{#1}}
\providecommand{\href}[2]{#2}
\providecommand{\path}[1]{#1}
\providecommand{\DOIprefix}{doi:}
\providecommand{\ArXivprefix}{arXiv:}
\providecommand{\URLprefix}{URL: }
\providecommand{\Pubmedprefix}{pmid:}
\providecommand{\doi}[1]{\href{http://dx.doi.org/#1}{\path{#1}}}
\providecommand{\Pubmed}[1]{\href{pmid:#1}{\path{#1}}}
\providecommand{\bibinfo}[2]{#2}
\ifx\xfnm\relax \def\xfnm[#1]{\unskip,\space#1}\fi
\bibitem[{Hockney and Eastwood(1988)}]{Hockney:1988:CSU:62815}
\bibinfo{author}{R.~W. Hockney}, \bibinfo{author}{J.~W. Eastwood},
  \bibinfo{title}{Computer Simulation Using Particles},
  \bibinfo{publisher}{Taylor \& Francis, Inc.}, \bibinfo{address}{Bristol, PA,
  USA}, \bibinfo{year}{1988}.
\bibitem[{Adelmann et~al.(2016)Adelmann, Locans, and Suter}]{ADELMANN201683}
\bibinfo{author}{A.~Adelmann}, \bibinfo{author}{U.~Locans},
  \bibinfo{author}{A.~Suter},
\newblock \bibinfo{title}{The dynamic kernel scheduler—part 1},
\newblock \bibinfo{journal}{Computer Physics Communications}
  \bibinfo{volume}{207} (\bibinfo{year}{2016}) \bibinfo{pages}{83 -- 90}.
\bibitem[{Berger and Oliger(1984)}]{BERGER1984484}
\bibinfo{author}{M.~J. Berger}, \bibinfo{author}{J.~Oliger},
\newblock \bibinfo{title}{Adaptive mesh refinement for hyperbolic partial
  differential equations},
\newblock \bibinfo{journal}{Journal of Computational Physics}
  \bibinfo{volume}{53} (\bibinfo{year}{1984}) \bibinfo{pages}{484 -- 512}.
\bibitem[{Berger and Colella(1989)}]{BERGER198964}
\bibinfo{author}{M.~Berger}, \bibinfo{author}{P.~Colella},
\newblock \bibinfo{title}{Local adaptive mesh refinement for shock
  hydrodynamics},
\newblock \bibinfo{journal}{Journal of Computational Physics}
  \bibinfo{volume}{82} (\bibinfo{year}{1989}) \bibinfo{pages}{64 -- 84}.
\bibitem[{Hittinger and Banks(2013)}]{HITTINGER2013118}
\bibinfo{author}{J.~Hittinger}, \bibinfo{author}{J.~Banks},
\newblock \bibinfo{title}{Block-structured adaptive mesh refinement algorithms
  for vlasov simulation},
\newblock \bibinfo{journal}{Journal of Computational Physics}
  \bibinfo{volume}{241} (\bibinfo{year}{2013}) \bibinfo{pages}{118 -- 140}.
\bibitem[{Kolobov and Arslanbekov(2016)}]{Kolobov}
\bibinfo{author}{V.~Kolobov}, \bibinfo{author}{R.~Arslanbekov},
\newblock \bibinfo{title}{Electrostatic pic with adaptive cartesian mesh},
\newblock \bibinfo{journal}{Journal of Physics: Conference Series}
  \bibinfo{volume}{719} (\bibinfo{year}{2016}) \bibinfo{pages}{012020}.
\bibitem[{Vay et~al.(2018)Vay, Almgren, Bell, Ge, Grote, Hogan, Kononenko,
  Lehe, Myers, Ng, Park, Ryne, Shapoval, Thévenet, and Zhang}]{VAY2018}
\bibinfo{author}{J.-L. Vay}, \bibinfo{author}{A.~Almgren},
  \bibinfo{author}{J.~Bell}, \bibinfo{author}{L.~Ge},
  \bibinfo{author}{D.~Grote}, \bibinfo{author}{M.~Hogan},
  \bibinfo{author}{O.~Kononenko}, \bibinfo{author}{R.~Lehe},
  \bibinfo{author}{A.~Myers}, \bibinfo{author}{C.~Ng},
  \bibinfo{author}{J.~Park}, \bibinfo{author}{R.~Ryne},
  \bibinfo{author}{O.~Shapoval}, \bibinfo{author}{M.~Thévenet},
  \bibinfo{author}{W.~Zhang},
\newblock \bibinfo{title}{Warp-x: A new exascale computing platform for
  beam–plasma simulations},
\newblock \bibinfo{journal}{Nuclear Instruments and Methods in Physics Research
  Section A: Accelerators, Spectrometers, Detectors and Associated Equipment}
  (\bibinfo{year}{2018}).
\bibitem[{Edwards et~al.(2014)Edwards, Trott, and
  Sunderland}]{CarterEdwards20143202}
\bibinfo{author}{H.~C. Edwards}, \bibinfo{author}{C.~R. Trott},
  \bibinfo{author}{D.~Sunderland},
\newblock \bibinfo{title}{Kokkos: Enabling manycore performance portability
  through polymorphic memory access patterns},
\newblock \bibinfo{journal}{Journal of Parallel and Distributed Computing}
  \bibinfo{volume}{74} (\bibinfo{year}{2014}) \bibinfo{pages}{3202 -- 3216}.
  \bibinfo{note}{Domain-Specific Languages and High-Level Frameworks for
  High-Performance Computing}.
\bibitem[{Edwards et~al.(2012)Edwards, Sunderland, Porter, Amsler, and
  Mish}]{Kokkos}
\bibinfo{author}{H.~C. Edwards}, \bibinfo{author}{D.~Sunderland},
  \bibinfo{author}{V.~Porter}, \bibinfo{author}{C.~Amsler},
  \bibinfo{author}{S.~Mish},
\newblock \bibinfo{title}{Manycore performance-portability: Kokkos
  multidimensional array library},
\newblock \bibinfo{journal}{Scientific Programming} \bibinfo{volume}{20}
  (\bibinfo{year}{2012}) \bibinfo{pages}{89--114}.
\bibitem[{Adelmann et~al.(2017)Adelmann, Baumgarten, Frey, Gsell,
  Valeria~Rizzoglio, Metzger-Kraus, Ineichen, (LANL), (CIAE), Sheehy, (RAL),
  and (MIT)}]{opal:1}
\bibinfo{author}{A.~Adelmann}, \bibinfo{author}{C.~Baumgarten},
  \bibinfo{author}{M.~Frey}, \bibinfo{author}{A.~Gsell},
  \bibinfo{author}{J.~S.~P. Valeria~Rizzoglio},
  \bibinfo{author}{C.~Metzger-Kraus}, \bibinfo{author}{Y.~Ineichen},
  \bibinfo{author}{S.~R. (LANL)}, \bibinfo{author}{C.~W. (CIAE)},
  \bibinfo{author}{S.~Sheehy}, \bibinfo{author}{C.~R. (RAL)},
  \bibinfo{author}{D.~W. (MIT)}, \bibinfo{title}{{The OPAL (Object Oriented
  Parallel Accelerator Library) Framework}}, \bibinfo{type}{Technical Report}
  \bibinfo{number}{PSI-PR-08-02}, Paul Scherrer Institut,
  \bibinfo{year}{(2008-2017)}.
\bibitem[{{Adelmann} et~al.(2019){Adelmann}, {Calvo}, {Frey}, {Gsell},
  {Locans}, {Metzger-Kraus}, {Neveu}, {Rogers}, {Russell}, {Sheehy},
  {Snuverink}, and {Winklehner}}]{2019arXiv190506654A}
\bibinfo{author}{A.~{Adelmann}}, \bibinfo{author}{P.~{Calvo}},
  \bibinfo{author}{M.~{Frey}}, \bibinfo{author}{A.~{Gsell}},
  \bibinfo{author}{U.~{Locans}}, \bibinfo{author}{C.~{Metzger-Kraus}},
  \bibinfo{author}{N.~{Neveu}}, \bibinfo{author}{C.~{Rogers}},
  \bibinfo{author}{S.~{Russell}}, \bibinfo{author}{S.~{Sheehy}},
  \bibinfo{author}{J.~{Snuverink}}, \bibinfo{author}{D.~{Winklehner}},
\newblock \bibinfo{title}{{OPAL a Versatile Tool for Charged Particle
  Accelerator Simulations}},
\newblock \bibinfo{journal}{arXiv e-prints}  (\bibinfo{year}{2019})
  \bibinfo{pages}{arXiv:1905.06654}.
\bibitem[{AMR(2019)}]{AMReX}
\bibinfo{title}{{AMReX}}, \bibinfo{year}{2019}.
  \bibinfo{note}{\url{https://ccse.lbl.gov/AMReX}, release: 18.07}.
\bibitem[{Martin(1998)}]{MartinPhdThesis}
\bibinfo{author}{D.~F. Martin}, \bibinfo{title}{An Adaptive Cell-centered
  Projection Method for the Incompressible Euler Equations}, Ph.D. thesis,
  University of California at Berkeley, \bibinfo{year}{1998}.
\bibitem[{Baker and Heroux(2012)}]{Tpetra}
\bibinfo{author}{C.~G. Baker}, \bibinfo{author}{M.~A. Heroux},
\newblock \bibinfo{title}{{Tpetra, and the use of generic programming in
  scientific computing}},
\newblock \bibinfo{journal}{Scientific Programming} \bibinfo{volume}{20}
  (\bibinfo{year}{2012}) \bibinfo{pages}{115--128}.
\bibitem[{Bavier et~al.(2012)Bavier, Hoemmen, Rajamanickam, and
  Thornquist}]{AmesosBelos}
\bibinfo{author}{E.~Bavier}, \bibinfo{author}{M.~Hoemmen},
  \bibinfo{author}{S.~Rajamanickam}, \bibinfo{author}{H.~Thornquist},
\newblock \bibinfo{title}{{Amesos2 and Belos: Direct and iterative solvers for
  large sparse linear systems}},
\newblock \bibinfo{journal}{Scientific Programming} \bibinfo{volume}{20}
  (\bibinfo{year}{2012}). \bibinfo{note}{Issue 3}.
\bibitem[{Prokopenko et~al.(2014)Prokopenko, Hu, Wiesner, Siefert, and
  Tuminaro}]{MueLu}
\bibinfo{author}{A.~Prokopenko}, \bibinfo{author}{J.~J. Hu},
  \bibinfo{author}{T.~A. Wiesner}, \bibinfo{author}{C.~M. Siefert},
  \bibinfo{author}{R.~S. Tuminaro}, \bibinfo{title}{Mue{L}u User's Guide 1.0},
  \bibinfo{type}{Technical Report} \bibinfo{number}{SAND2014-18874}, Sandia
  National Labs, \bibinfo{year}{2014}.
\bibitem[{Hu et~al.(2014)Hu, Prokopenko, Siefert, Tuminaro, and
  Wiesner}]{MueLuURL}
\bibinfo{author}{J.~J. Hu}, \bibinfo{author}{A.~Prokopenko},
  \bibinfo{author}{C.~M. Siefert}, \bibinfo{author}{R.~S. Tuminaro},
  \bibinfo{author}{T.~A. Wiesner}, \bibinfo{title}{{M}ue{L}u multigrid
  framework}, \bibinfo{howpublished}{\url{http://trilinos.org/packages/muelu}},
  \bibinfo{year}{2014}.
\bibitem[{Prokopenko et~al.(2016)Prokopenko, Siefert, Hu, Hoemmen, and
  Klinvex}]{Ifpack2}
\bibinfo{author}{A.~Prokopenko}, \bibinfo{author}{C.~M. Siefert},
  \bibinfo{author}{J.~J. Hu}, \bibinfo{author}{M.~Hoemmen},
  \bibinfo{author}{A.~Klinvex}, \bibinfo{title}{Ifpack2 {U}ser’s {G}uide
  1.0}, \bibinfo{type}{Technical Report} \bibinfo{number}{SAND2016-5338},
  Sandia National Labs, \bibinfo{year}{2016}.
\bibitem[{Yang et~al.(2010)Yang, Adelmann, Humbel, Seidel, and
  Zhang}]{PhysRevSTAB.13.064201}
\bibinfo{author}{J.~J. Yang}, \bibinfo{author}{A.~Adelmann},
  \bibinfo{author}{M.~Humbel}, \bibinfo{author}{M.~Seidel},
  \bibinfo{author}{T.~J. Zhang},
\newblock \bibinfo{title}{Beam dynamics in high intensity cyclotrons including
  neighboring bunch effects: Model, implementation, and application},
\newblock \bibinfo{journal}{Phys. Rev. ST Accel. Beams} \bibinfo{volume}{13}
  (\bibinfo{year}{2010}) \bibinfo{pages}{064201}.
\bibitem[{Rizzoglio et~al.(2017)Rizzoglio, Adelmann, Baumgarten, Frey,
  Gerbershagen, Meer, and Schippers}]{PhysRevAccelBeams.20.124702}
\bibinfo{author}{V.~Rizzoglio}, \bibinfo{author}{A.~Adelmann},
  \bibinfo{author}{C.~Baumgarten}, \bibinfo{author}{M.~Frey},
  \bibinfo{author}{A.~Gerbershagen}, \bibinfo{author}{D.~Meer},
  \bibinfo{author}{J.~M. Schippers},
\newblock \bibinfo{title}{Evolution of a beam dynamics model for the transport
  line in a proton therapy facility},
\newblock \bibinfo{journal}{Phys. Rev. Accel. Beams} \bibinfo{volume}{20}
  (\bibinfo{year}{2017}) \bibinfo{pages}{124702}.
\bibitem[{Adelmann et~al.(2010)Adelmann, Arbenz, and
  Ineichen}]{ADELMANN20104554}
\bibinfo{author}{A.~Adelmann}, \bibinfo{author}{P.~Arbenz},
  \bibinfo{author}{Y.~Ineichen},
\newblock \bibinfo{title}{A fast parallel poisson solver on irregular domains
  applied to beam dynamics simulations},
\newblock \bibinfo{journal}{Journal of Computational Physics}
  \bibinfo{volume}{229} (\bibinfo{year}{2010}) \bibinfo{pages}{4554 -- 4566}.
\bibitem[{Toggweiler et~al.(2014)Toggweiler, Adelmann, Arbenz, and
  Yang}]{TOGGWEILER2014255}
\bibinfo{author}{M.~Toggweiler}, \bibinfo{author}{A.~Adelmann},
  \bibinfo{author}{P.~Arbenz}, \bibinfo{author}{J.~Yang},
\newblock \bibinfo{title}{A novel adaptive time stepping variant of the
  {B}oris-{B}uneman integrator for the simulation of particle accelerators with
  space charge},
\newblock \bibinfo{journal}{Journal of Computational Physics}
  \bibinfo{volume}{273} (\bibinfo{year}{2014}) \bibinfo{pages}{255 -- 267}.
\bibitem[{Vay et~al.(2012)Vay, Grote, Cohen, and
  Friedman}]{1749-4699-5-1-014019}
\bibinfo{author}{J.-L. Vay}, \bibinfo{author}{D.~P. Grote},
  \bibinfo{author}{R.~H. Cohen}, \bibinfo{author}{A.~Friedman},
\newblock \bibinfo{title}{Novel methods in the particle-in-cell accelerator
  code-framework warp},
\newblock \bibinfo{journal}{Computational Science \& Discovery}
  \bibinfo{volume}{5} (\bibinfo{year}{2012}) \bibinfo{pages}{014019}.
\bibitem[{Colella and Norgaard(2010)}]{COLELLA2010947}
\bibinfo{author}{P.~Colella}, \bibinfo{author}{P.~C. Norgaard},
\newblock \bibinfo{title}{Controlling self-force errors at refinement
  boundaries for {AMR-PIC}},
\newblock \bibinfo{journal}{Journal of Computational Physics}
  \bibinfo{volume}{229} (\bibinfo{year}{2010}) \bibinfo{pages}{947 -- 957}.
\bibitem[{Jackson(1999)}]{Jackson99}
\bibinfo{author}{J.~D. Jackson}, \bibinfo{title}{Classical Electrodynamics},
  \bibinfo{edition}{3rd} ed., \bibinfo{publisher}{John Wiley \& Sons, Inc.},
  \bibinfo{address}{New York}, \bibinfo{year}{1999}.
\bibitem[{{Turk} et~al.(2011){Turk}, {Smith}, {Oishi}, {Skory}, {Skillman},
  {Abel}, and {Norman}}]{2011ApJS..192....9T}
\bibinfo{author}{M.~J. {Turk}}, \bibinfo{author}{B.~D. {Smith}},
  \bibinfo{author}{J.~S. {Oishi}}, \bibinfo{author}{S.~{Skory}},
  \bibinfo{author}{S.~W. {Skillman}}, \bibinfo{author}{T.~{Abel}},
  \bibinfo{author}{M.~L. {Norman}},
\newblock \bibinfo{title}{{yt: A Multi-code Analysis Toolkit for Astrophysical
  Simulation Data}},
\newblock \bibinfo{journal}{Astrophysical Journal, Supplement}
  \bibinfo{volume}{192} (\bibinfo{year}{2011}) \bibinfo{pages}{9}.
\bibitem[{Martin and Cartwright(1996)}]{Martin96}
\bibinfo{author}{D.~F. Martin}, \bibinfo{author}{K.~L. Cartwright},
  \bibinfo{title}{Solving {P}oisson's Equation using Adaptive Mesh Refinement},
  \bibinfo{type}{Technical Report} \bibinfo{number}{UCB/ERL M96/66}, Univ.
  Calif. Berkeley, \bibinfo{year}{1996}.
\bibitem[{Almgren et~al.(1998)Almgren, Bell, Colella, Howell, and
  Welcome}]{ALMGREN19981}
\bibinfo{author}{A.~S. Almgren}, \bibinfo{author}{J.~B. Bell},
  \bibinfo{author}{P.~Colella}, \bibinfo{author}{L.~H. Howell},
  \bibinfo{author}{M.~L. Welcome},
\newblock \bibinfo{title}{A conservative adaptive projection method for the
  variable density incompressible navier–stokes equations},
\newblock \bibinfo{journal}{Journal of Computational Physics}
  \bibinfo{volume}{142} (\bibinfo{year}{1998}) \bibinfo{pages}{1 -- 46}.
\bibitem[{Alvin and Eli(1980)}]{doi:10.1002/cpa.3160330603}
\bibinfo{author}{B.~Alvin}, \bibinfo{author}{T.~Eli},
\newblock \bibinfo{title}{Radiation boundary conditions for wave-like
  equations},
\newblock \bibinfo{journal}{Communications on Pure and Applied Mathematics}
  \bibinfo{volume}{33} (\bibinfo{year}{1980}) \bibinfo{pages}{707--725}.
\bibitem[{Bayliss et~al.(1982)Bayliss, Gunzburger, and
  Turkel}]{10.2307/2101222}
\bibinfo{author}{A.~Bayliss}, \bibinfo{author}{M.~Gunzburger},
  \bibinfo{author}{E.~Turkel},
\newblock \bibinfo{title}{Boundary conditions for the numerical solution of
  elliptic equations in exterior regions},
\newblock \bibinfo{journal}{SIAM Journal on Applied Mathematics}
  \bibinfo{volume}{42} (\bibinfo{year}{1982}) \bibinfo{pages}{430--451}.
\bibitem[{Khebir et~al.(1990)Khebir, Kouki, and Mittra}]{58681}
\bibinfo{author}{A.~Khebir}, \bibinfo{author}{A.~B. Kouki},
  \bibinfo{author}{R.~Mittra},
\newblock \bibinfo{title}{Asymptotic boundary conditions for finite element
  analysis of three-dimensional transmission line discontinuities},
\newblock \bibinfo{journal}{IEEE Transactions on Microwave Theory and
  Techniques} \bibinfo{volume}{38} (\bibinfo{year}{1990})
  \bibinfo{pages}{1427--1432}.
\bibitem[{Gordon and Fook(1993)}]{241666}
\bibinfo{author}{R.~K. Gordon}, \bibinfo{author}{S.~H. Fook},
\newblock \bibinfo{title}{A finite difference approach that employs an
  asymptotic boundary condition on a rectangular outer boundary for modeling
  two-dimensional transmission line structures},
\newblock \bibinfo{journal}{IEEE Transactions on Microwave Theory and
  Techniques} \bibinfo{volume}{41} (\bibinfo{year}{1993})
  \bibinfo{pages}{1280--1286}.
\bibitem[{Biswas et~al.(2015)Biswas, Singh, and Kumar}]{doi:10.1063/1.4931738}
\bibinfo{author}{D.~Biswas}, \bibinfo{author}{G.~Singh},
  \bibinfo{author}{R.~Kumar},
\newblock \bibinfo{title}{Boundary conditions for the solution of the
  three-dimensional poisson equation in open metallic enclosures},
\newblock \bibinfo{journal}{Physics of Plasmas} \bibinfo{volume}{22}
  (\bibinfo{year}{2015}) \bibinfo{pages}{093119}.
\bibitem[{Wangler and Crandall(2000)}]{Wangler}
\bibinfo{author}{T.~P. Wangler}, \bibinfo{author}{K.~R. Crandall},
\newblock \bibinfo{title}{Beam halo in proton linac beams},
\newblock Number~\bibinfo{number}{20} in \bibinfo{series}{International Linac
  Conference}, \bibinfo{year}{2000}. \URLprefix
  \url{http://accelconf.web.cern.ch/AccelConf/l00/papers/TU202.pdf}.
\bibitem[{CSCS(2018)}]{PizDaint}
\bibinfo{author}{CSCS}, \bibinfo{year}{2018}.
  \bibinfo{note}{\url{https://www.cscs.ch/computers/piz-daint/}, visited: 8.\
  October 2018}.

\end{thebibliography}
    

\end{document}